\documentclass[conference]{IEEEtran}
\usepackage{cite}
\usepackage{amsmath,amssymb,amsfonts}
\usepackage{algorithmic}
\usepackage{graphicx}
\usepackage{textcomp}
\usepackage{xcolor}
\usepackage{color}
\usepackage{subfigure}
\usepackage{multicol}

\def\BibTeX{{\rm B\kern-.05em{\sc i\kern-.025em b}\kern-.08em
    T\kern-.1667em\lower.7ex\hbox{E}\kern-.125emX}}

\renewcommand{\baselinestretch}{.97} 

\begin{document}

\title{Rank and Condition Number Analysis for UAV MIMO Channels Using Ray Tracing\\
\thanks{Identify applicable funding agency here. If none, delete this.}
}

\author{\IEEEauthorblockN{Donggu Lee and Ismail Guvenc}
\IEEEauthorblockA{Department of Electrical and Computer Engineering, North Carolina State University, Raleigh, NC, USA \\
\{dlee42,iguvenc\}@ncsu.edu}
}

\maketitle

\begin{abstract}
Channel rank and condition number of multi-input multi-output (MIMO) channels can be effective indicators of achievable rates with spatial multiplexing in mobile networks. 
%Especially, channel rank and condition number can be directly adopted for spatial multiplexing mode of MIMO systems. 
%Ray tracing technologies-based link analysis with accurate angular information of rays and material properties of obstacles can be an effective tool for time-varying MIMO channels. 
In this paper, we use extensive ray tracing simulations to investigate channel rank, condition number, and signal coverage distribution for air-to-ground MIMO channels. We consider UAV-based user equipment (UE) at altitudes of 3~m, 30~m, 70~m, and 110~m from the ground. Moreover, we also consider their communication link with a cellular base station in urban and rural areas. In particular, Centennial Campus and Lake Wheeler Road Field Labs of NC State University are considered, and their  geographical information extracted from the open street map (OSM) database is incorporated into ray tracing simulations. Our results characterize how the channel rank tends to reduce as a function of UAV altitude, while also providing insights into the effects of geography, building distribution, and  threshold parameters on channel rank and condition number.  
\end{abstract}


\section{Introduction}
Reliable and high data rate connectivity with unmanned aerial vehicles (UAVs) have critical applications for mission-critical and  public safety communications, search and rescue, and surveillance.  
Spatial multiplexing with multi-input multi-output (MIMO) systems makes it possible to achieve higher data rates for UAVs over a wide coverage without allocating additional bandwidth or transmit power. Various performance metrics, including uninformed transmitter capacity, singular value spread, condition numbers, and correlation matrix distance can be considered for MIMO systems \cite{MIMO_performance_metric}. Especially, channel rank and condition number can be rapidly employed for determining whether spatial multiplexing can be adopted for a given MIMO channel realization~\cite{MIMO_white_paper, MIMO_adaptation_1, MIMO_adaptation_2}. 
%\textcolor{red}{Selecting a proper transmission mode in MIMO can play an essential role in mission-oriented systems that have specific data rate requirements over resource-limited environments.} 
 
Fundamental backgrounds and insights into channel rank and condition number in MIMO systems are provided in \cite{MIMO_white_paper},  %Assessing MIMO channels with channel rank and condition number is studied. 
which also evaluates the signal-to-noise ratio (SNR) requirements needed for obtaining a certain MIMO spectral efficiency for a given condition number. 
In \cite{capacity_analysis_paper}, the theoretical lower and upper bounds of channel capacity for high-rank MIMO systems are studied. Here, a constraint for line-of-sight MIMO links is derived with the condition number of the channel matrix. 
%The theoretical upper and lower bounds of channel capacity using condition numbers can be confirmed through numerical analysis.
In \cite{vertical_MIMO_paper}, the effect of vertical separation of MIMO antennas and the size of the array on the channel rank and throughput are studied experimentally. Vertically and horizontally placed MIMO antennas in LTE base stations  and the effects of antenna configuration on spatial multiplexing are studied in~\cite{vertical_MIMO_paper_2} considering 2 and 4 transmit antennas. 
A path loss model for unmanned aerial vehicles (UAVs) in an urban environment is studied in \cite{material_paper, path_loss_arxiv}. Material characteristics such as permittivity and conductivity of buildings and ground are considered to develop the path loss model for UAV communication links. %An accurate path loss model for an air-to-ground channel can be derived with the measurement data. 

\begin{figure}[t!]
    \centering
    \subfigure[]{
    \includegraphics[width=0.45\columnwidth]{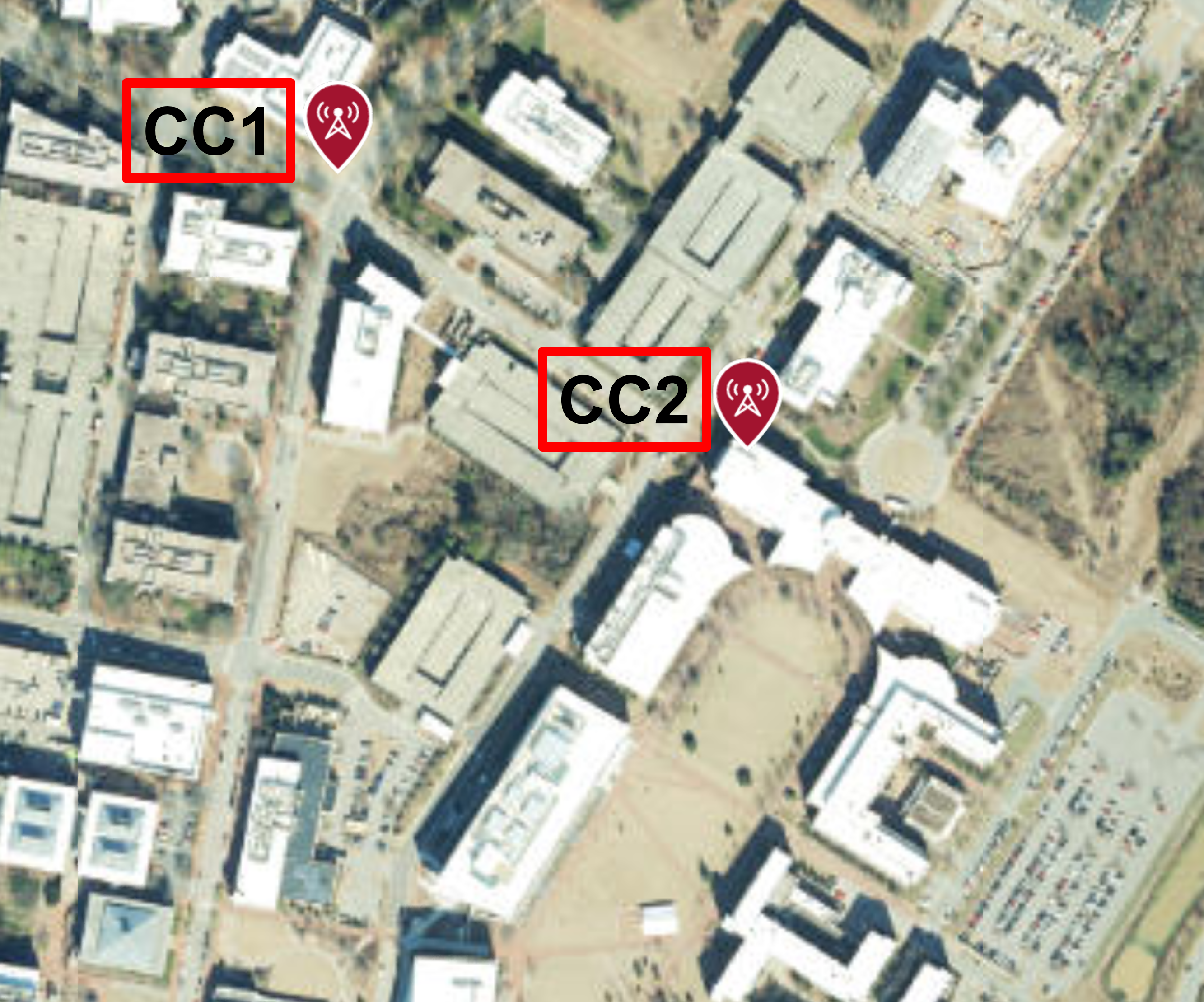}
    \label{fig:map_CC1_CC2}
    }
    \subfigure[]{
    \includegraphics[width=0.45\columnwidth]{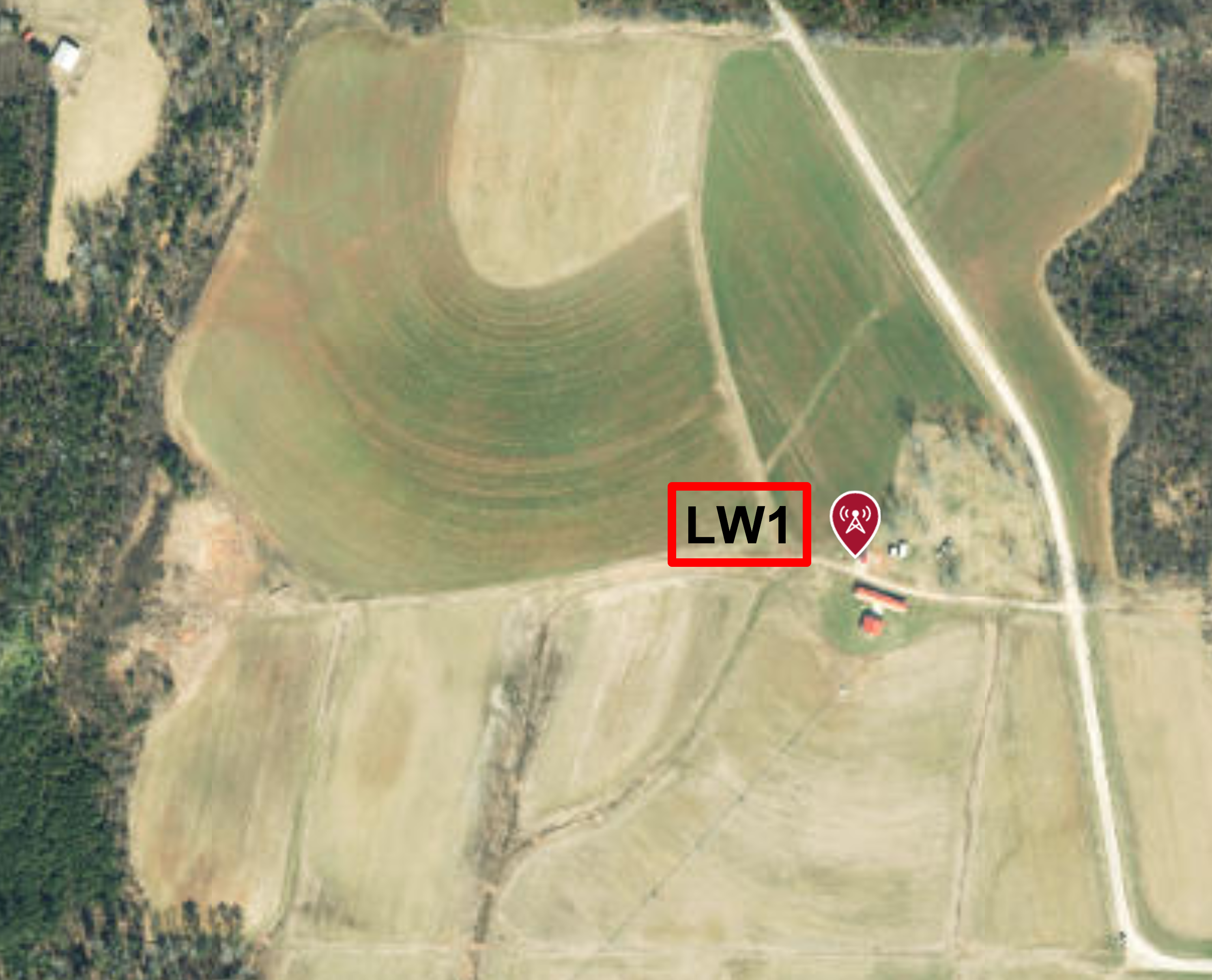}
    \label{fig:map_LW}
    }\vspace{-3mm}
    \caption{Map view of the location of transmitters used in ray tracing simulations: (a) CC1, CC2, and (b) LW1.}
    \label{fig:my_label}\vspace{-6mm}
\end{figure}

To our best knowledge, channel rank and condition number analysis for UAV channels in typical urban and rural scenarios is not available in the literature.  
Ray tracing simulations can serve as a reasonable approach to investigate the distribution of channel rank for different scenarios, and to analyze time-varying MIMO systems \cite{RT_survey}. They can also be used for deciding transmission mode (spatial multiplexing, beam forming, and diversity) for a given channel realization and to evaluate the achievable data rate for each scenario~\cite{RT_mode_paper}. In this paper, we investigate channel rank, condition number,  and received signal strength indicator (RSSI) distribution for UAV MIMO channels using ray tracing.
%, which can provide an accurate trajectory with angular information and material properties. 
Simulations are carried out at different UAV-based user equipment (UE) altitudes, considering the buildings and topography information in the Centennial Campus and Lake Wheeler Road Field Labs of NC  State University (see Fig.~\ref{fig:my_label}). Subsequently, geographical and statistical distributions of channel rank and condition number of a MIMO channel are analyzed for different environments, UAV altitudes, and threshold settings.  

This paper is organized as follows. A description of the system model and ray tracing simulation scenario is provided in Section~\ref{Sec:2}. We simulate, analyze, and discuss RF signal coverage, channel rank detection, and condition number for MIMO UAV links in Section~\ref{Sec:3}, Section~\ref{Sec:4}, and Section~\ref{Sec:5}, respectively, while the last section concludes the paper. 

\section{System Model}\label{Sec:2}

\subsection{MIMO Communication Model}
Received signal vectors of a simple MIMO communication system can be expressed as  
%\begin{equation}\label{eqn: MIMO}
$\boldsymbol{Y} = \boldsymbol{Hx} + \boldsymbol{n}$, 
%\end{equation}
where $\boldsymbol{H}$ denotes the $N_{r} \times N_{t}$ MIMO channel matrix, $\boldsymbol{x}$ is the $N_{t} \times 1$ transmit signal vector, and $\boldsymbol{n}$ represents the $N_{t} \times 1$ noise vector, respectively~\cite{MIMO_white_paper}. Moreover, $N_r$ denotes the number of receiving antennas and $N_t$ denotes the number of transmitting antennas, which are set to be 4 for both receiving and transmitting antennas in this work. The channel coefficients \textit{$h_{ij}$} for the signal from \textit{j}-th transmit antenna to \textit{i}-th receive antenna should be estimated at the receiver side to decode received symbols.
%solve the equation (\ref{eqn: MIMO}). 
%The solvability of the equation can be assessed with the rank and condition of the channel matrix.

For this scenario, the rank of a channel matrix can be an indicator of how many data streams can be supported for the spatial multiplexing \cite{lte_book}. In particular, the channel rank can be determined by the number of non-zero singular values, which can be obtained by the singular value decomposition of the channel matrix as
\begin{equation}\label{eqn: SVD}
    \boldsymbol{H} = \boldsymbol{U}\boldsymbol{\Sigma}\boldsymbol{V^{*}},
\end{equation}
where $\boldsymbol{U}$ is the $\textit{m} \times \textit{m}$ complex unitary matrix, $\boldsymbol{\Sigma}$ is the $\textit{m} \times \textit{n}$ rectangular diagonal matrix, with diagonal elements $\sigma_i$, with $1<i<\min\{m, n\}$, and $\boldsymbol{V}$ represents the $\textit{n} \times \textit{n}$ complex unitary matrix. Without loss of generality, $\sigma_i$ are sorted in descending order as $i$ increases. The rank $R$ of a channel matrix satisfies $R\leq \min\{m, n\}$, and may be calculated as $R=\arg\underset{i}{\max}~ {\rm I}\{\sigma_i>\sigma_{\rm Thr}\}$, where $I\{.\}$ is an indicator function that returns $1$ if its input is valid and $0$ otherwise, and $\sigma_{\rm Thr}$ is a threshold for the minimum singular value. While  more complex approaches are possible for determining the channel rank, we consider a threshold-based approach for simplicity.

The matrix $\boldsymbol{\Sigma}$ obtained from singular value decomposition can be seen to characterize multiple single-input single-output (SISO) channels with zero cross interference \cite{MIMO_white_paper}, from which the condition number can be derived as 
%\begin{equation}
${\rm CN} = \sigma_{\rm max} / \sigma_{\rm min}$, 
%\end{equation}
where $\sigma_{\rm max}$ and $\sigma_{\rm min}$ denote the maximum and minimum singular values of $\boldsymbol{\Sigma}$. Condition number can provide critical information in addition to the channel rank, as it can characterize the link qualities across different spatial data streams and how reliably the data streams can be reconstructed~\cite{MIMO_white_paper}. Since UAV channels typically have strong line-of-sight (LoS) dominance, $\sigma_{\rm min}$ may be very small in many scenarios, and hence we will relax the definition of condition number in Section~\ref{Sec:5} to consider only the strongest two singular values.   
%\noindent \#1. MIMO basic mathematical representation \\
%\#2. Channel rank and condition number

%\begin{figure}[t]
 %   \centering
  %  \subfigure[Receiver altitude: 3 m]{
   % \includegraphics[width=0.42\columnwidth]{%images/1a/CC2_3m_1x4_3GHz_Rank_edited_v10.png}
    %\label{fig:CC2_rank_1}
    %}
    %\subfigure[Receiver altitude: 30 m]{
    %\includegraphics[width=0.42\columnwidth]%{images/1a/CC2_30m_1x4_3GHz_Rank_edited_v10.png}
    %\label{fig:CC2_rank_2}
    %}
    %\subfigure[Receiver altitude: 70 m]{
    %\includegraphics[width=0.42\columnwidth]%{images/1a/CC2_70m_1x4_3GHz_Rank_edited_v10.png}
   % \label{fig:CC2_rank_3}
   % }
    %\subfigure[Receiver altitude: 110 m]{
    %\includegraphics[width=0.42\columnwidth]{images/1a/CC2_110m_1x4_3GHz_Rank_edited_v10.png}
    %\label{fig:CC2_rank_4}
    %}
    
 %   \caption{Channel ranks in  Centennial Campus with CC2 as the cellular BS.}
 %   \label{fig:CC2_rank}
%\end{figure} 

\subsection{Ray Tracing Model and Scenario}
In this paper, the shooting and bouncing ray (SBR) method \cite{ray_tracing_access_paper,sbr_paper}-based ray tracing model has been employed. The basic concept of SBR is to trace all rays launched from a source to decide if the rays arrive at a receiver after they travel over a reflective or diffracting trajectory. The SBR method has three procedures: 1) ray launching, 2) ray tracing, and 3) ray reception. 
In the ray launching process, it is assumed that all the rays launched from the source are uniformly distributed to ensure that all rays carry the same amount of transmitting power. Here, each ray can be represented by a line in 3D space and its trajectory can be expressed as \cite{sbr_paper}
$(x_1, y_1, z_1) = (x_0, y_0, z_0) + (s_x, s_y, s_z)t$,
where $(x_0, y_0, z_0)$ denotes the reference point, $(s_x, s_y, s_z)$ denotes the direction vector, and $t$ is the time of the trajectory, respectively. 

%\begin{figure}[t!]
   % \centering
 %   \subfigure[Receiver altitude: 3 m]{
%    \includegraphics[width=0.45\columnwidth]{images/1a/LW_3m_1x4_3GHz_Rank_edited_v10.png}
%    \label{fig:LW_rank_1}
 %   }
 %   \subfigure[Receiver altitude: 30 m]{
 %   \includegraphics[width=0.45\columnwidth]%{images/1a/LW_30m_1x4_3GHz_Rank_edited_v10.png}
%    \label{fig:LW_rank_2}
 %   }
%    \subfigure[Receiver altitude: 70 m]{
%    \includegraphics[width=0.45\columnwidth]%{images/1a/LW_70m_1x4_3GHz_Rank_edited_v10.png}
%    \label{fig:LW_rank_3}
 %   }
 %   \subfigure[Receiver altitude: 110 m]{
  %  \includegraphics[width=0.45\columnwidth]{images/1a/LW_110m_1x4_3GHz_Rank_edited_v10.png}
 %   \label{fig:LW_rank_4}
%    }
%    \caption{Channel ranks in  Lake Wheeler with LW1 as the cellular BS.}
 %   \label{fig:LW_rank}
%\end{figure}

After the launching, the reflective and diffracting rays are traced. A reflected or diffracted ray will be generated if the launched rays intersect any object on its path. The electrical field after reflection and diffraction can be obtained as \cite{sbr_paper}:
%\begin{equation}
    $\boldsymbol{E}(x_{i+1}, y_{i+1}, z_{i+1}) = D_i \Gamma_i \boldsymbol{E}(x_i, y_i, z_i) e^{j\theta}$,  
%\end{equation}
where $D_i$ represents the divergence factor that is related to the spreading of the ray tube right after the \textit{i}-th reflection, $\Gamma_i$ denotes the reflection coefficient, $\boldsymbol{E}$$(x_i, y_i, z_i)$ represents the incident electric field, and $\theta$ is the phase shift, respectively.
In the last step, a ray tube indicates a possible area of the ray traveling can be derived. When the ray tube intersects the receiving field, the receiving rays can be determined.   
%\noindent \#1. Basic mathematical representation of shooting and bounce ray model

\begin{table}[t]
    \centering
    \caption{Simulation parameters for ray tracing.}\vspace{-1mm}
    \begin{tabular}{|c|c|c|}
    \hline
      \textbf{Parameters}   & \textbf{Description} & \textbf{Value} \\
      \hline
        $f_{c}$ & Carrier frequency & 3.4 GHz \\
        \hline
        $a_{\rm{BS, CC1}}$ & Height of CC1 base station & 10 m \\
        \hline
        $a_{\rm{BS, CC2}}$ & Height of CC2 base station & 25 m \\
        \hline
        $a_{\rm{BS, LW1}}$ & Height of LW1 base station & 10 m \\
        \hline
        $T_{\rm{Cen}}$ & Target area of Centennial Campus & 580 m $\times$ 460 m \\
        \hline
        $T_{\rm{LW}}$ & Target area of Lake Wheeler & 740 m $\times$ 600 m \\
        \hline
        $N_{\rm{ref}}$ & Maximum number of reflected rays & 2\\
        \hline
        $N_{\rm ele}$ & Number of antenna elements & 4 (TX and RX) \\
        \hline
        $d_{\rm{tx}}$ & Element spacing for TX antennas & $\lambda$ \\
        \hline
        $d_{\rm{rx}}$ & Element spacing for RX antennas & $0.5\lambda$ 
        \\
        \hline
        $\rm{P_{TX}}$ & Transmit power of TX antennas & 10 W
        \\
        \hline
    \end{tabular}    
    \label{tab:sim_param}\vspace{-5mm}
\end{table}

%\section{Ray Tracing  Scenario for Channel Rank and Condition Number Analysis}

We consider ray tracing simulations for modeling the propagation around NSF AERPAW platform's fixed node locations at NC State University~\cite{aerpawWebsite}. Matlab's Ray Tracing tool~\cite{matlab_ray_tracing} is used for carrying out the simulations. Fig.~\ref{fig:my_label} shows the locations of the transmitters in the Centennial Campus of NC State University with CC1 and CC2 fixed nodes and Lake Wheeler Road Field Labs with the LW1 fixed node. Here, the locations of the transmitters are marked with red markers. The Centennial Campus area contains buildings and is a campus urban environment, while the Lake Wheeler area is an open rural area. The geographical information is obtained from the open street map (OSM) database \cite{osm} and the information about buildings such as building heights is obtained from the OSM Buildings database \cite{osm_buildings}. We assume that if a receiver is inside a building, it will not have connectivity. Channel rank,  condition number, and RSSI analysis are conducted within these two areas as will be described as follows. 
%\textcolor{red}{IG: Here you need to explain how you used geographical and building map information and cite the corresponding website. You need to also describe how you placed}

The following assumptions are made for rank and condition number analysis: 1) carrier frequency is set to be 3.4 GHz; 2) 4 element array antenna with antenna spacing of $\lambda$ for the transmitter and half of $\lambda$ for the receiver are employed, where $\lambda$ denotes the wavelength; 3) the heights of the transmitter are set to be 10 m, 25 m, and 10 m from the ground for CC1, CC2, and LW1, respectively; 4) the maximum number of reflection of the rays is set to be 2 with concrete building material, perfect reflector for ground reflection of the Centennial Campus area, and vegetation ground material setting for Lake Wheeler areas \cite{brweb}; 5) the receivers are located on a uniform grid of 20 m resolution for both areas; %6) OFDM is adopted; 
and 6) noise is neglected for the purpose of rank and condition number analysis. The key simulation parameters for ray tracing simulations are summarized in Table~\ref{tab:sim_param}.

%\subsection{AERPAW Testbed}
%\noindent \#1. Description of AERPAW settings (CC1, CC2, and LW)

\begin{figure*}[h!]
    \centering
    \subfigure[UAV altitude: 3 m (CC1)]{
    \includegraphics[width=0.46\columnwidth]{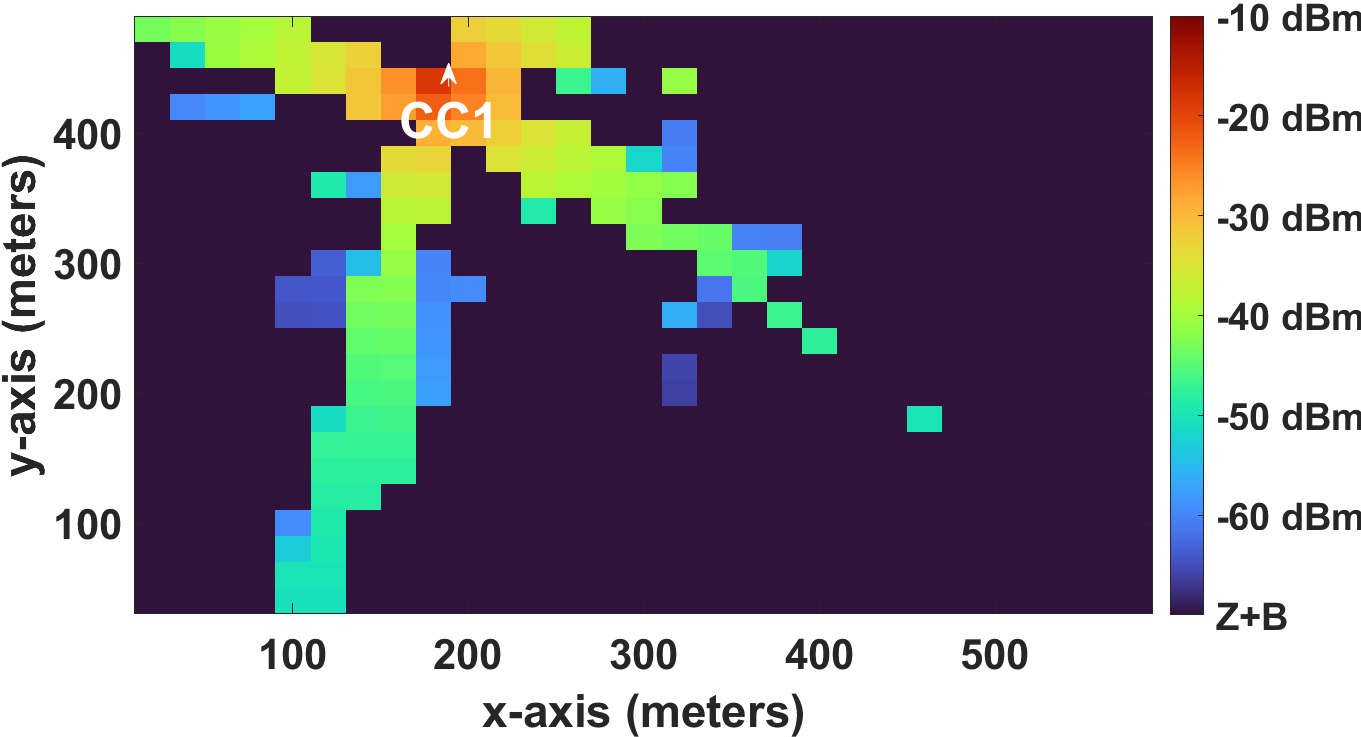}
    \label{fig:CC1_SISO_1}
    }
    \subfigure[UAV altitude: 30 m (CC1)]{
    \includegraphics[width=0.46\columnwidth]{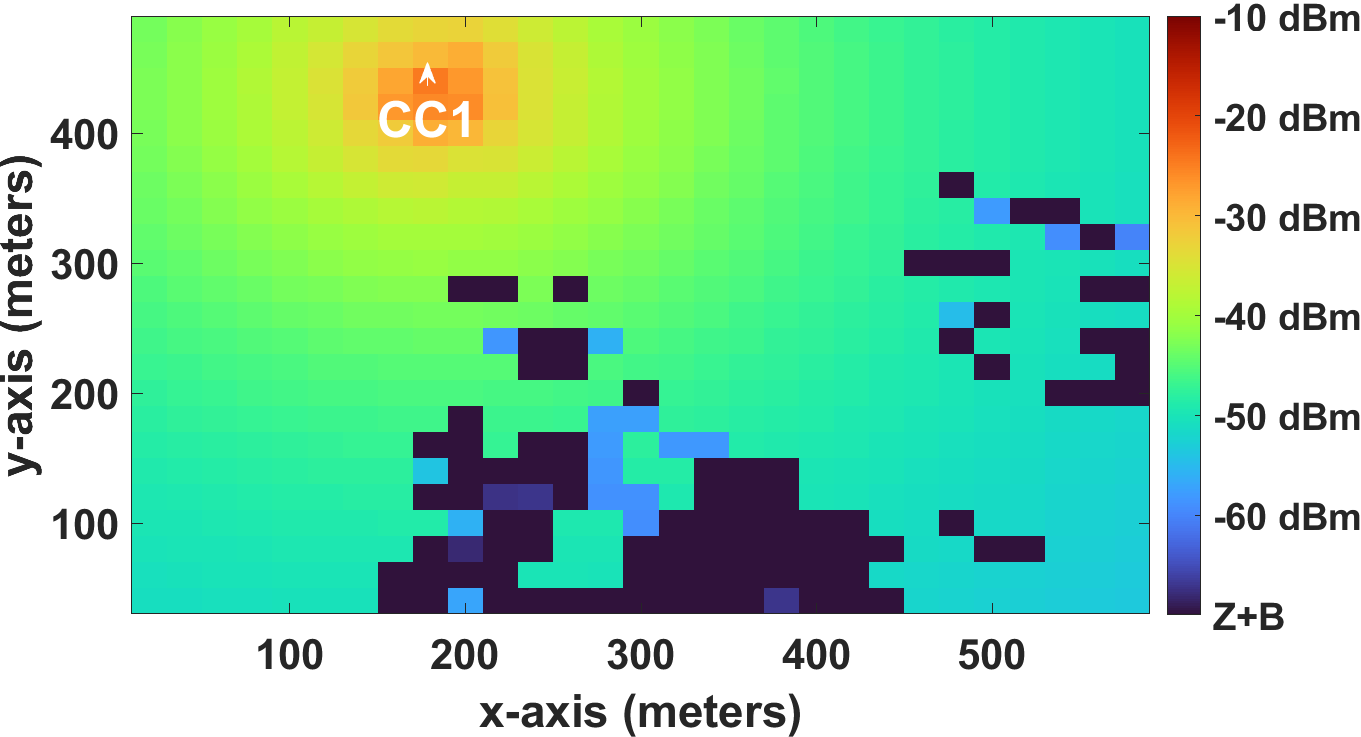}
    \label{fig:CC1_SISO_2}
    }
    \subfigure[UAV altitude: 70 m (CC1)]{
    \includegraphics[width=0.46\columnwidth]{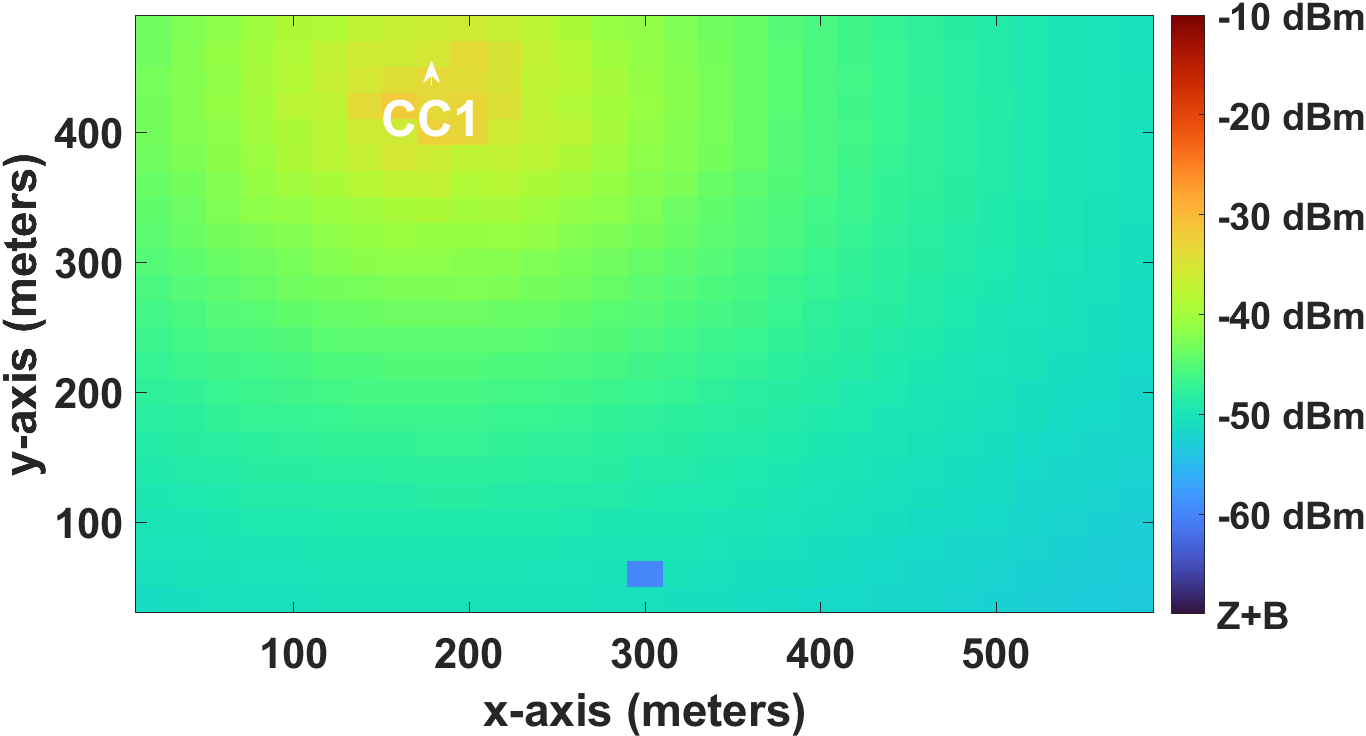}
    \label{fig:CC1_SISO_3}
    }
    \subfigure[UAV altitude: 110 m (CC1)]{
    \includegraphics[width=0.46\columnwidth]{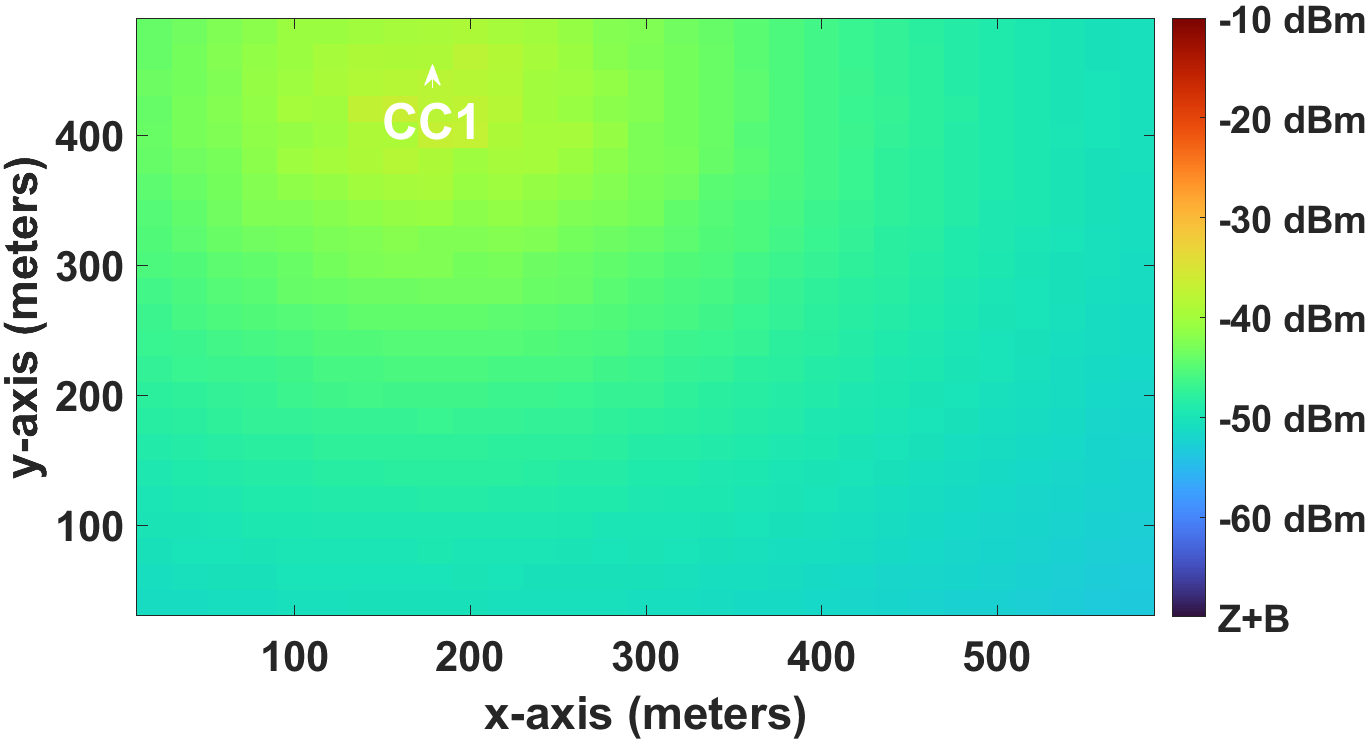}
    \label{fig:CC1_SISO_4}
    }
    \subfigure[UAV altitude: 3 m (CC2)]{
    \includegraphics[width=0.46\columnwidth]{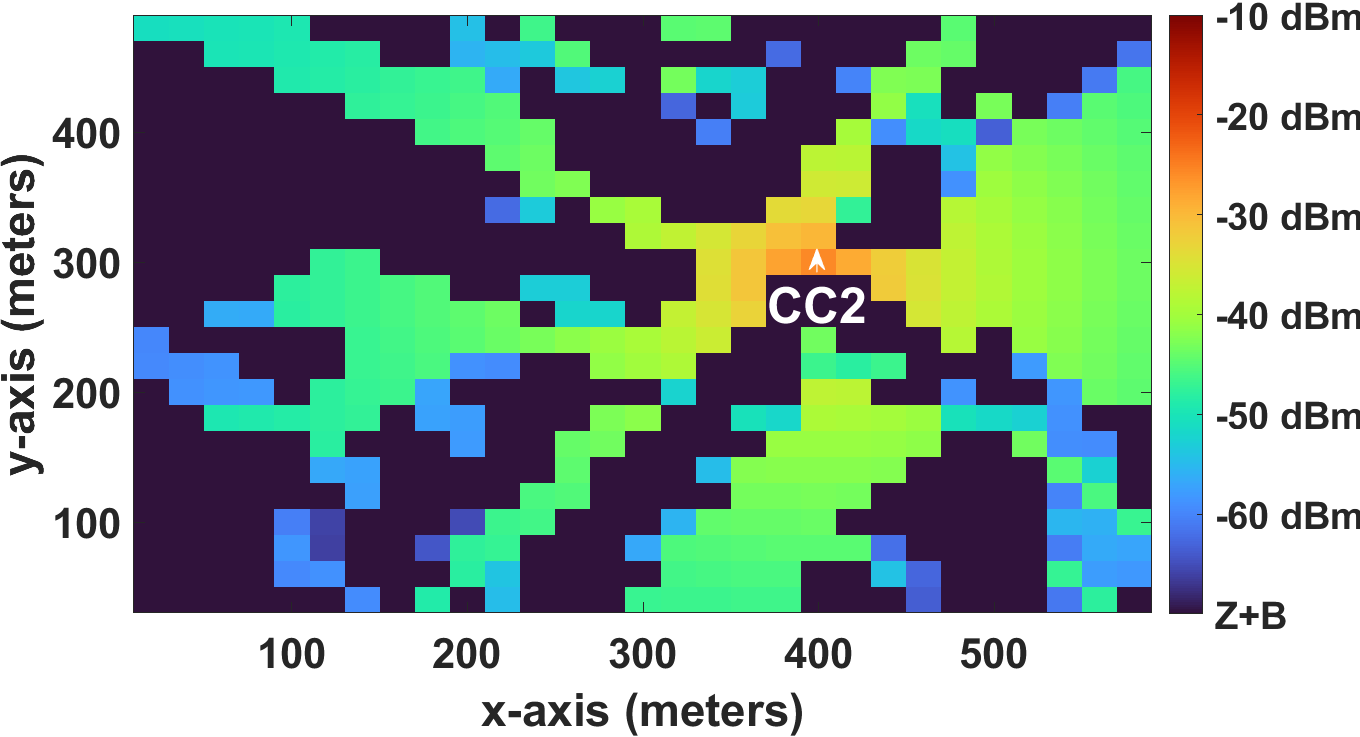}
    \label{fig:CC2_SISO_1}
    }
    \subfigure[UAV altitude: 30 m (CC2)]{
    \includegraphics[width=0.46\columnwidth]{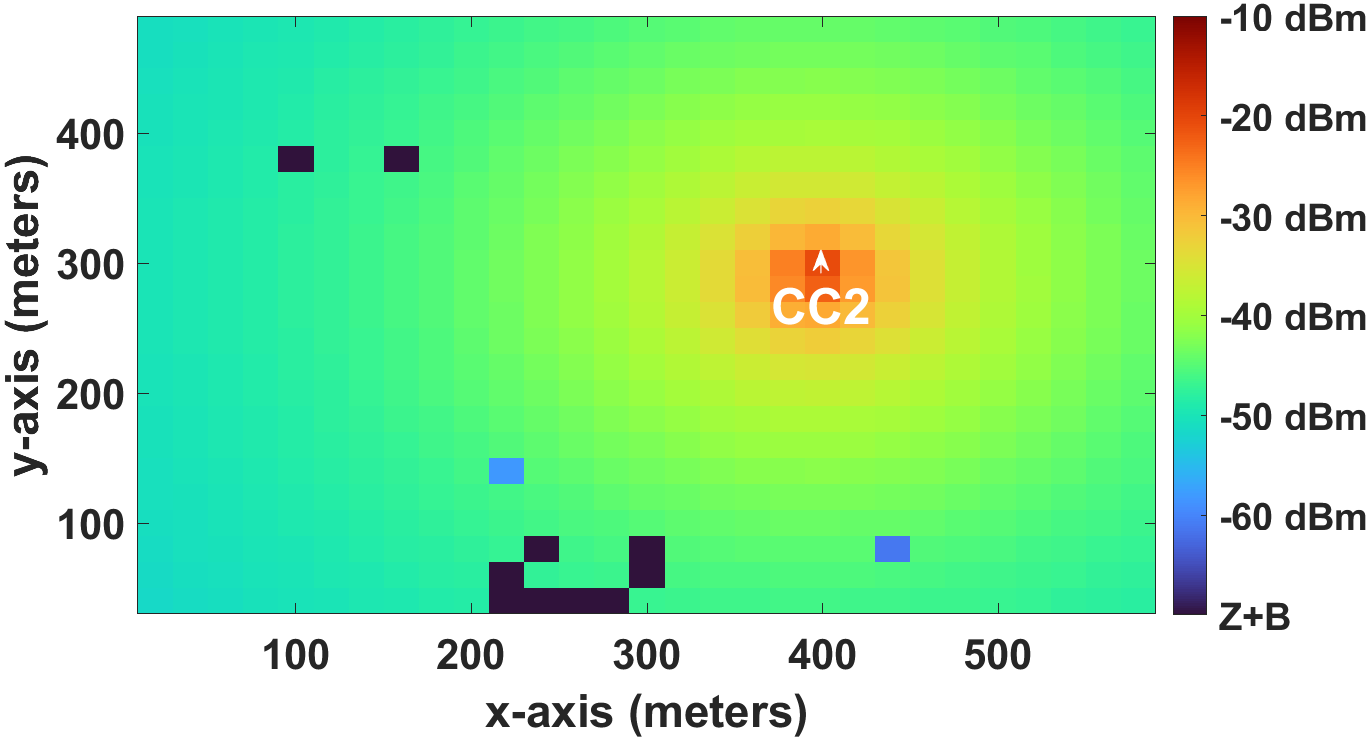}
    \label{fig:CC2_SISO_2}
    }
    \subfigure[UAV altitude: 70 m (CC2)]{
    \includegraphics[width=0.46\columnwidth]{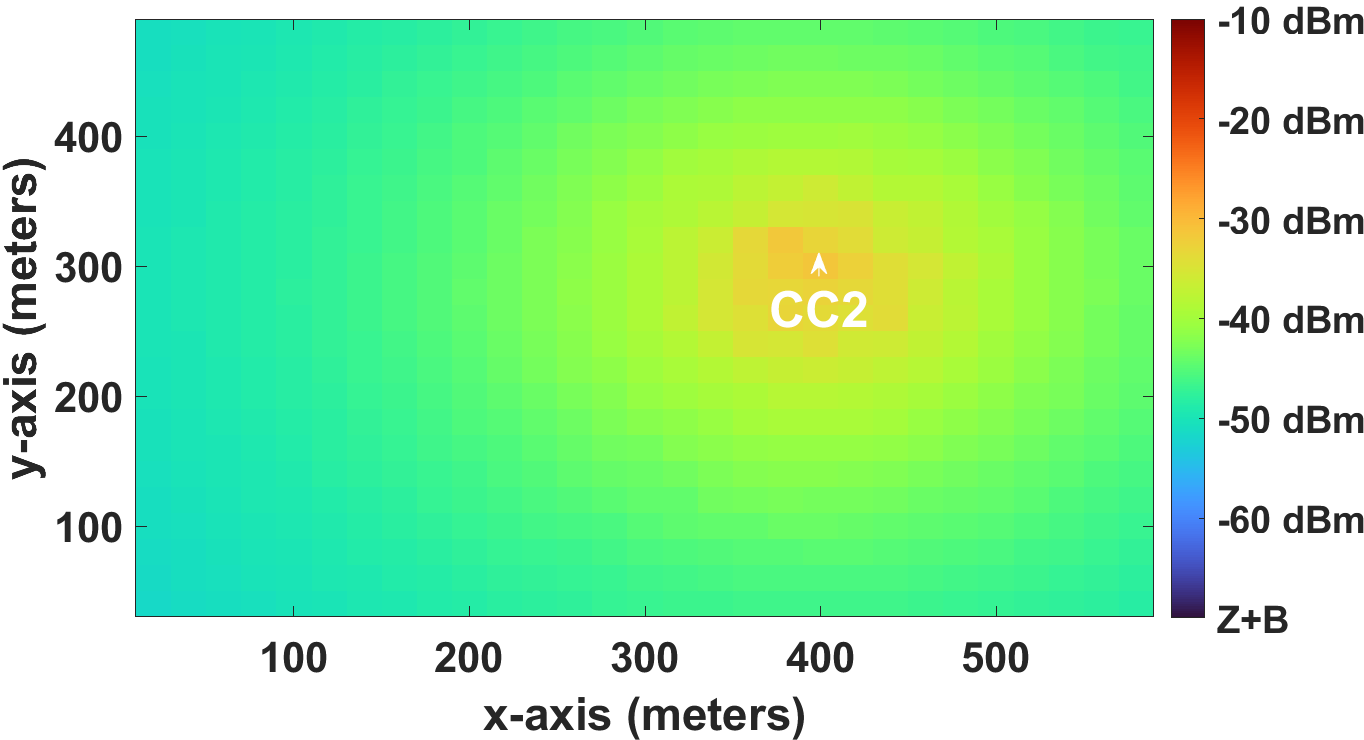}
    \label{fig:CC2_SISO_3}
    }
    \subfigure[UAV altitude: 110 m (CC2)]{
    \includegraphics[width=0.46\columnwidth]{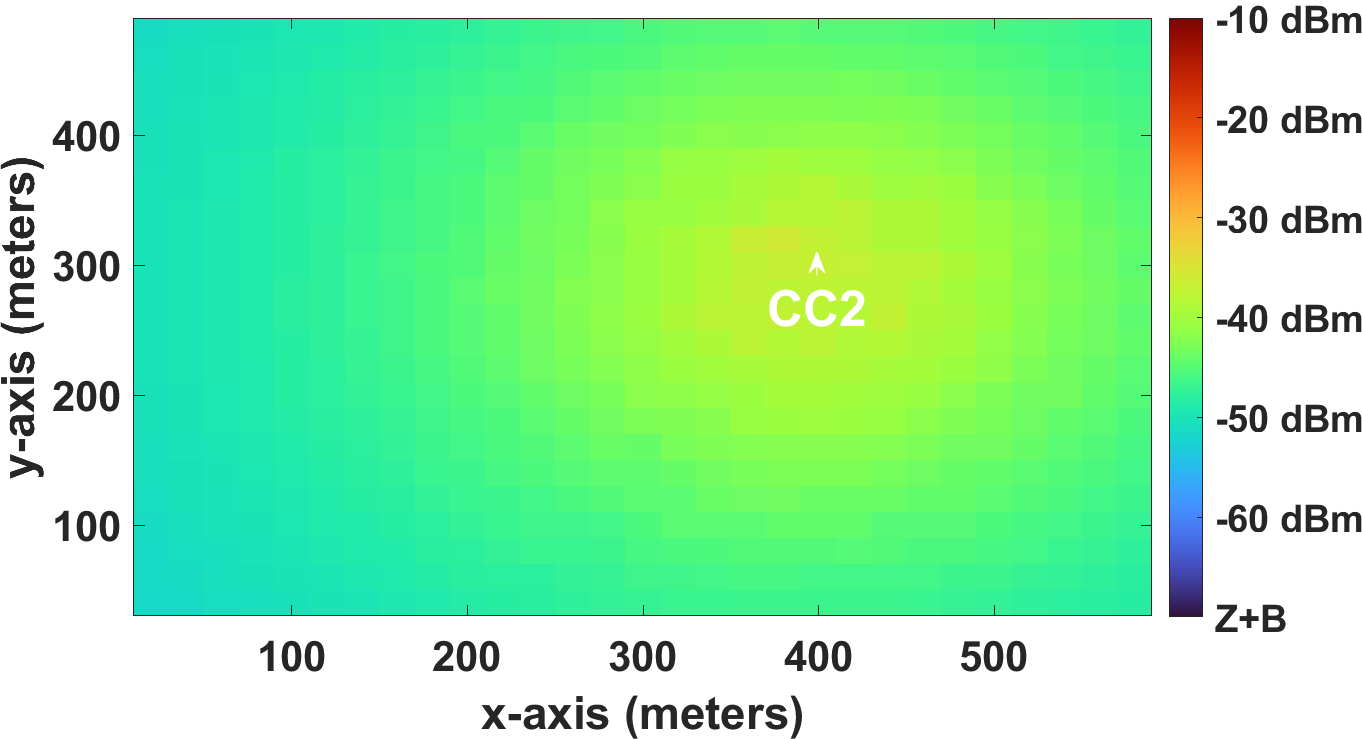}
    \label{fig:CC2_SISO_4}
    }
    \subfigure[UAV altitude: 3 m (LW1)]{
    \includegraphics[width=0.46\columnwidth]{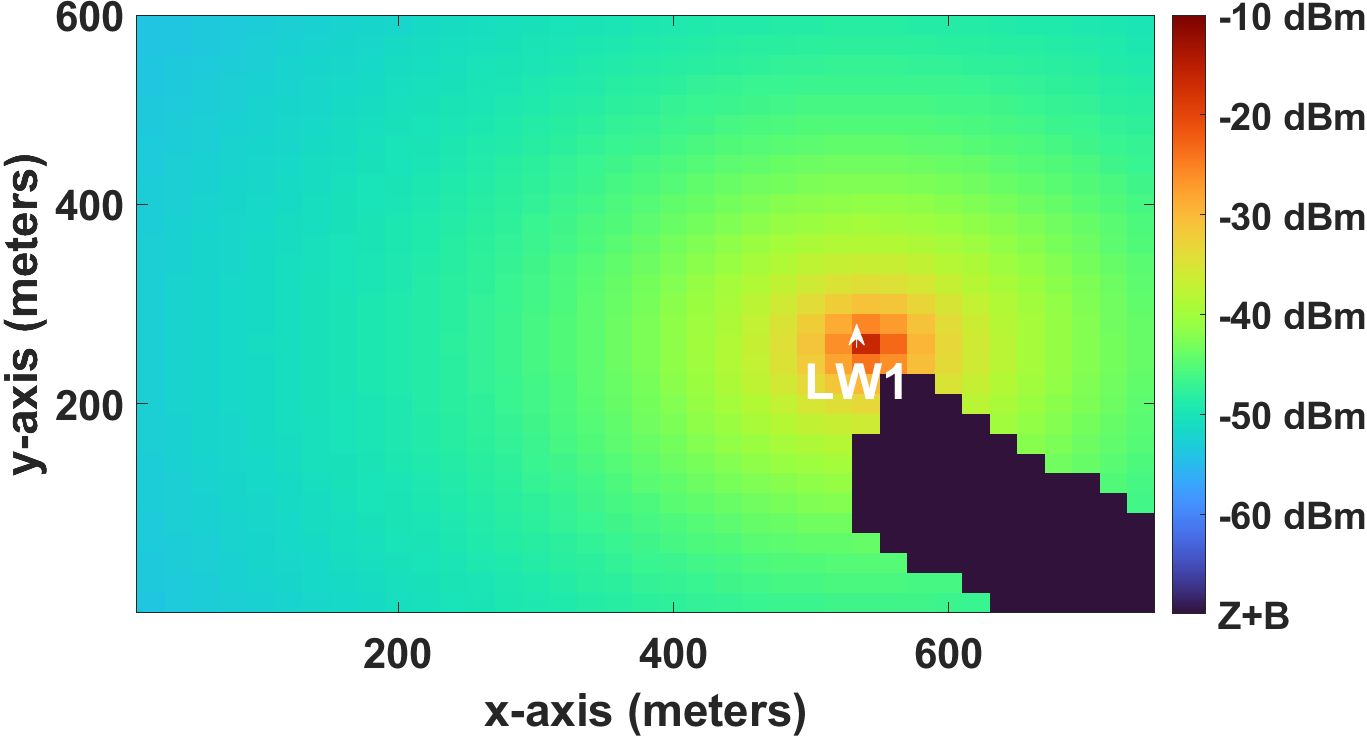}
    \label{fig:LW1_SISO_1}
    }
    \subfigure[UAV altitude: 30 m (LW1)]{
    \includegraphics[width=0.46\columnwidth]{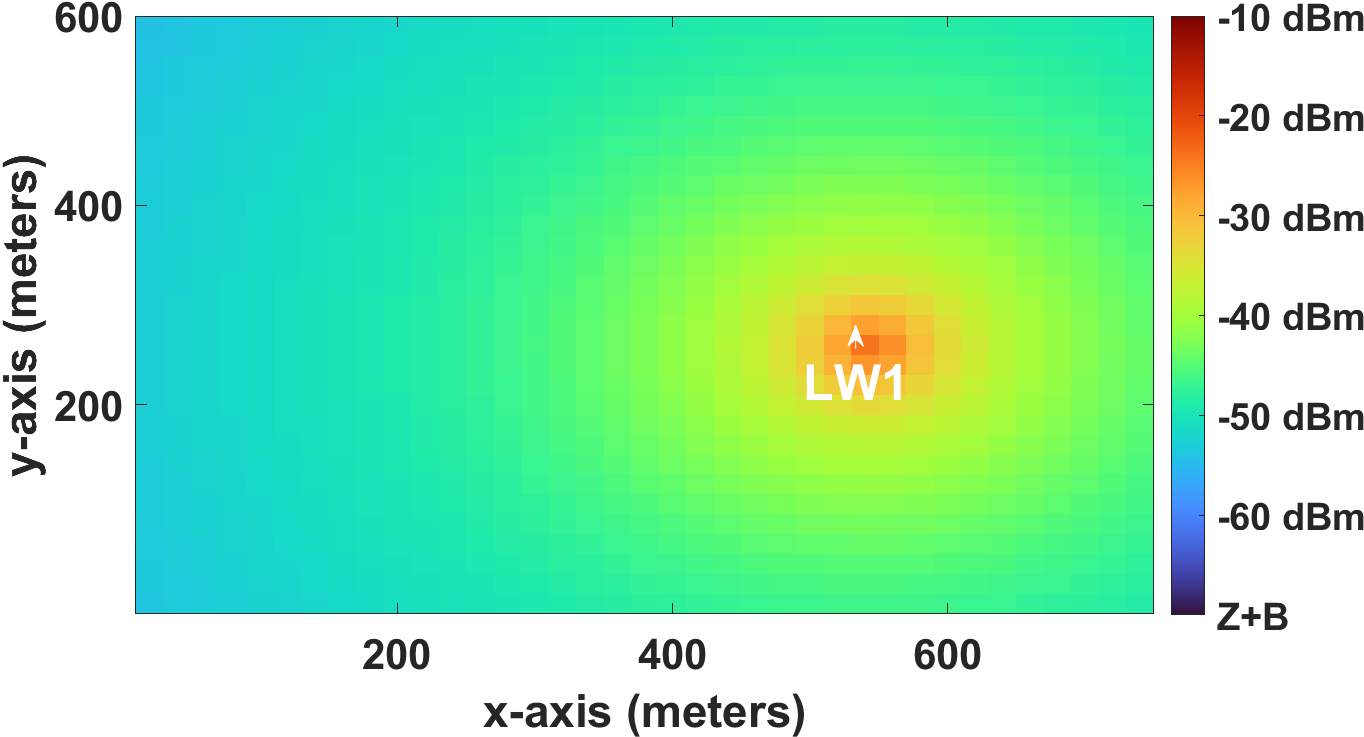}
    \label{fig:LW1_SISO_2}
    }
    \subfigure[UAV altitude: 70 m (LW1)]{
    \includegraphics[width=0.46\columnwidth]{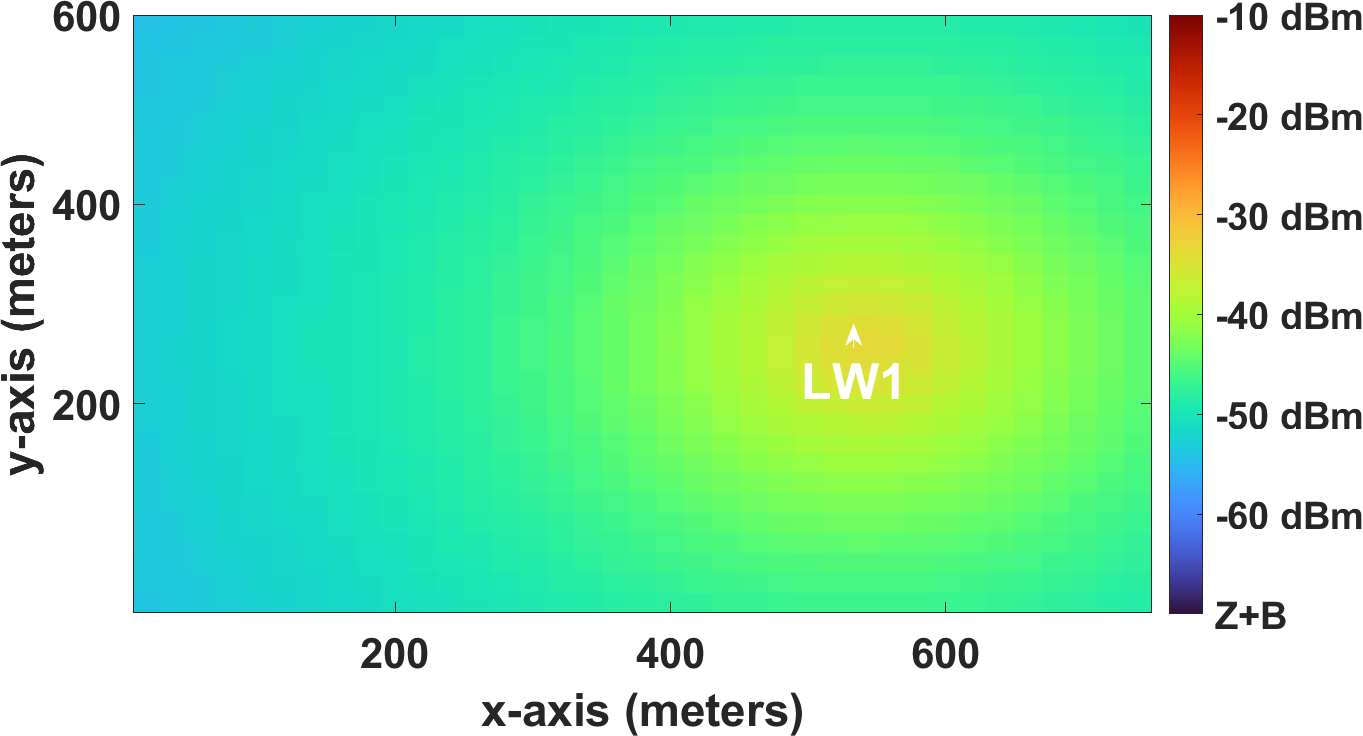}
    \label{fig:LW1_SISO_3}
    }
    \subfigure[UAV altitude: 110 m (LW1)]{
    \includegraphics[width=0.46\columnwidth]{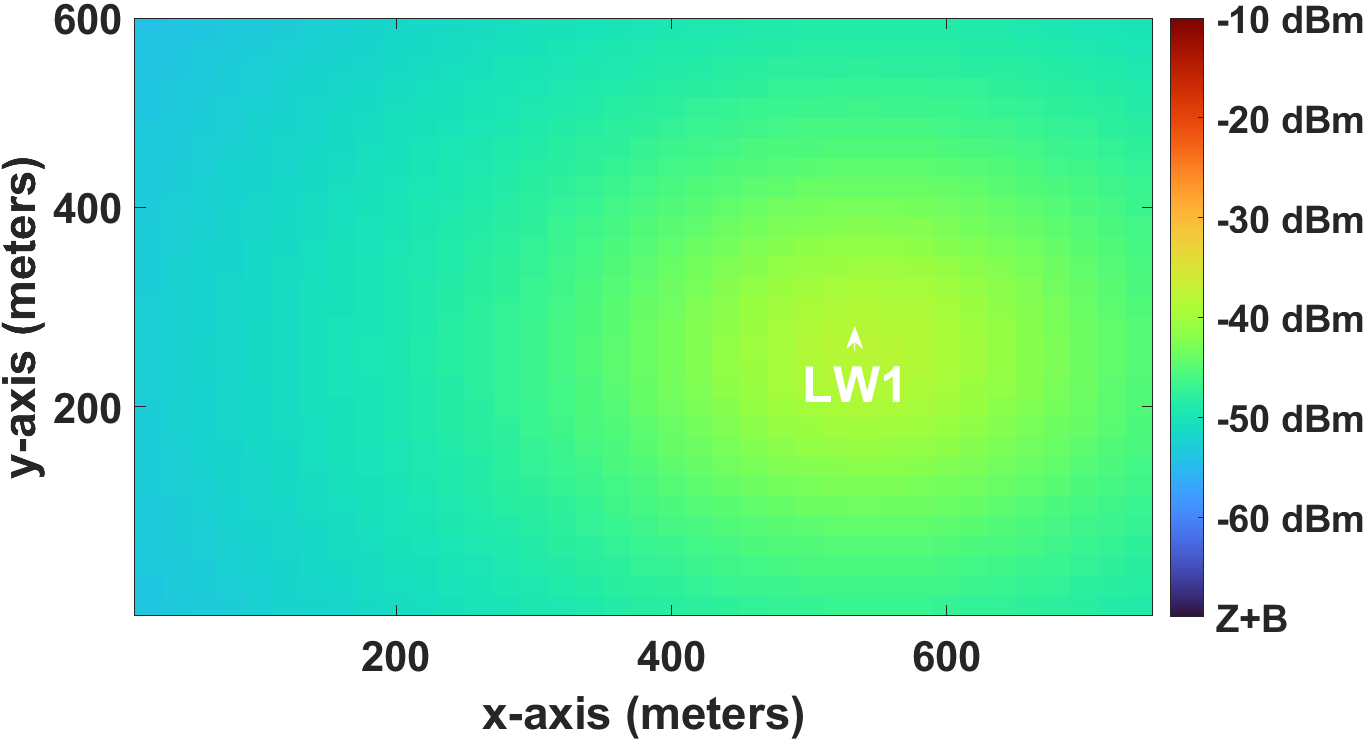}
    \label{fig:LW1_SISO_4}
    }\vspace{-1mm}
    \caption{RSSI in Centennial Campus with CC1 (a)-(d), CC2 (e)-(h), and in Lake Wheeler with LW1 (i)-(l) as the cellular BS (see Fig.~\ref{fig:my_label}). Effects of building blockage are apparent at 3~m and $30$~m UE height, especially for the Centennial Campus.}
    \label{fig:SISO_CC1_CC2_LW1}\vspace{-4mm}
\end{figure*}

\section{RF Signal Coverage Analysis}\label{Sec:3}
In this section, we analyze the RF signal coverage for an air-to-ground link at four different altitudes using ray tracing simulations. The simulation settings from Section~\ref{Sec:2} are adopted except for MIMO settings. Here, SISO antennas are employed for both the transmitter and receiver sides.

\begin{figure}[t!]
    \centering
    \subfigure[CC1]{
    \includegraphics[width=0.45\columnwidth]{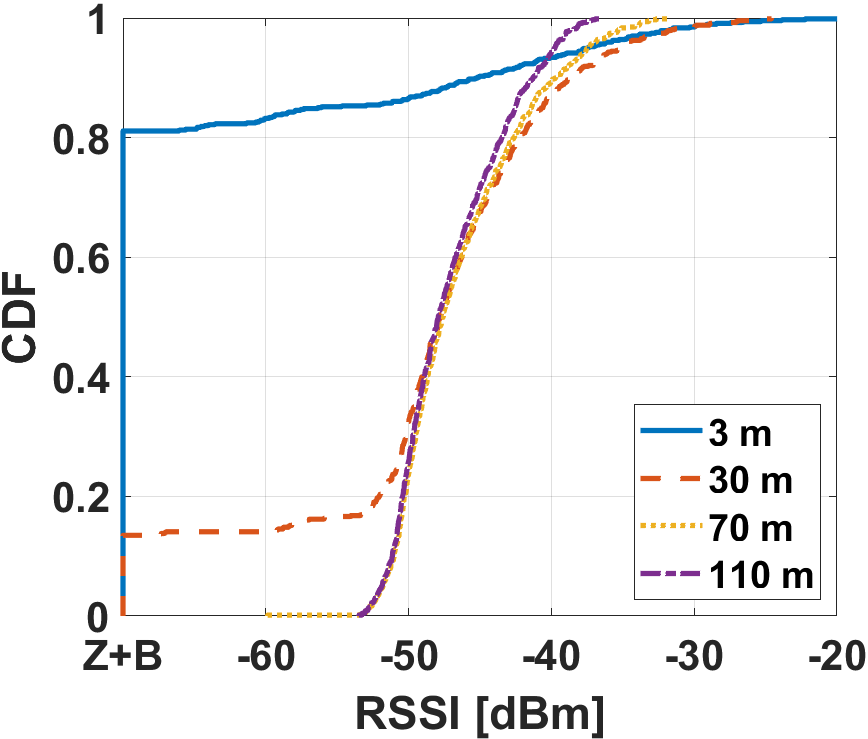}
    \label{fig:CDF_RSSI_CC1}
    }
    \subfigure[CC2]{
    \includegraphics[width=0.45\columnwidth]{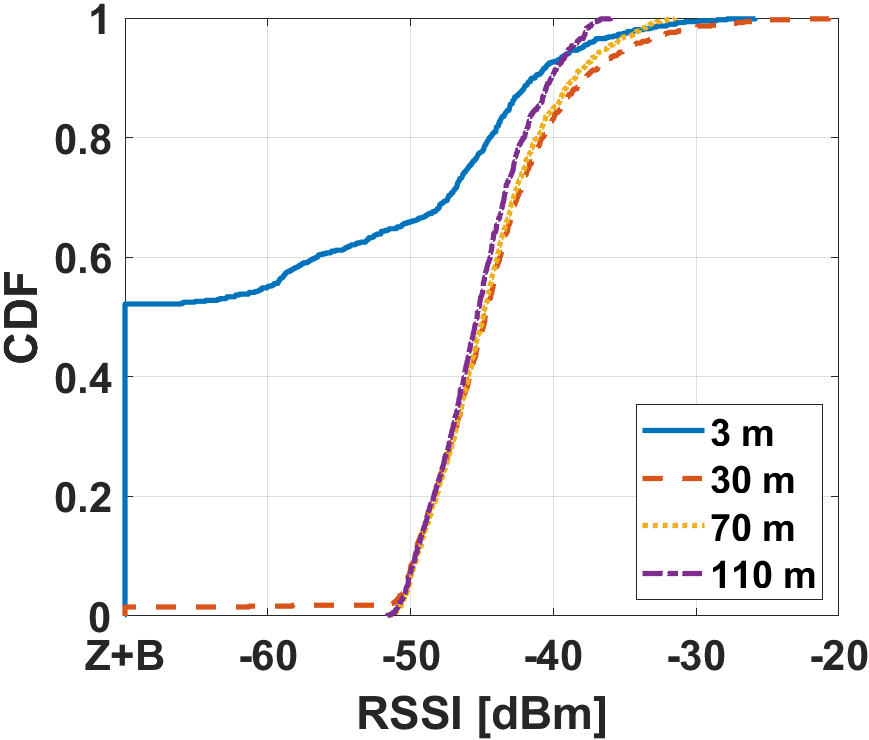}
    \label{fig:CDF_RSSI_CC2}
    }
    \subfigure[LW1]{
    \includegraphics[width=0.45\columnwidth]{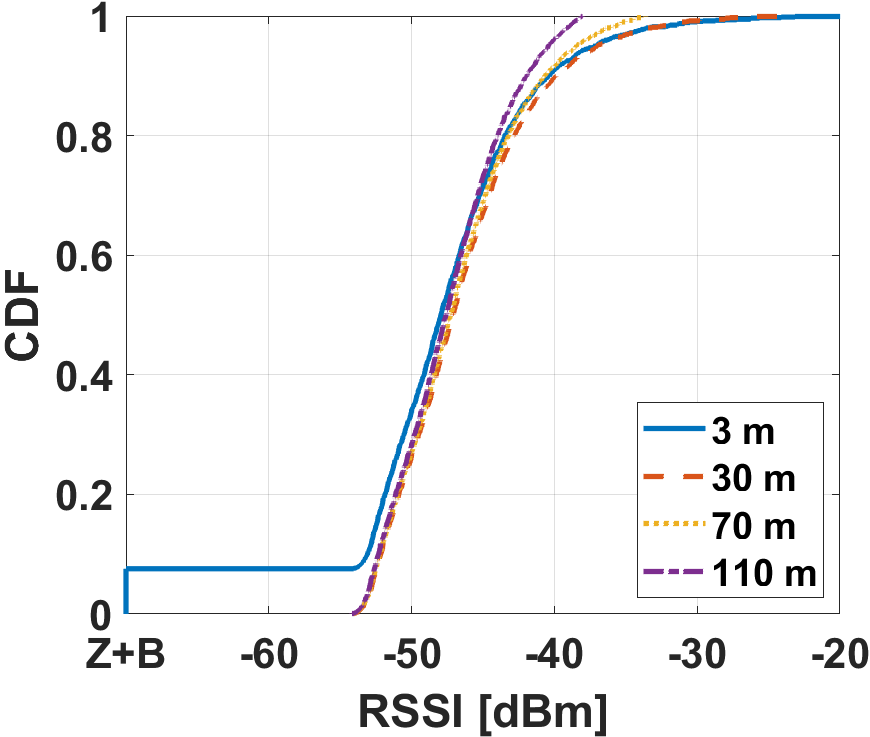}
    \label{fig:CDF_RSSI_LW1}
    }
    \subfigure[$P_Z+P_B$]{
    \includegraphics[width=0.45\columnwidth]{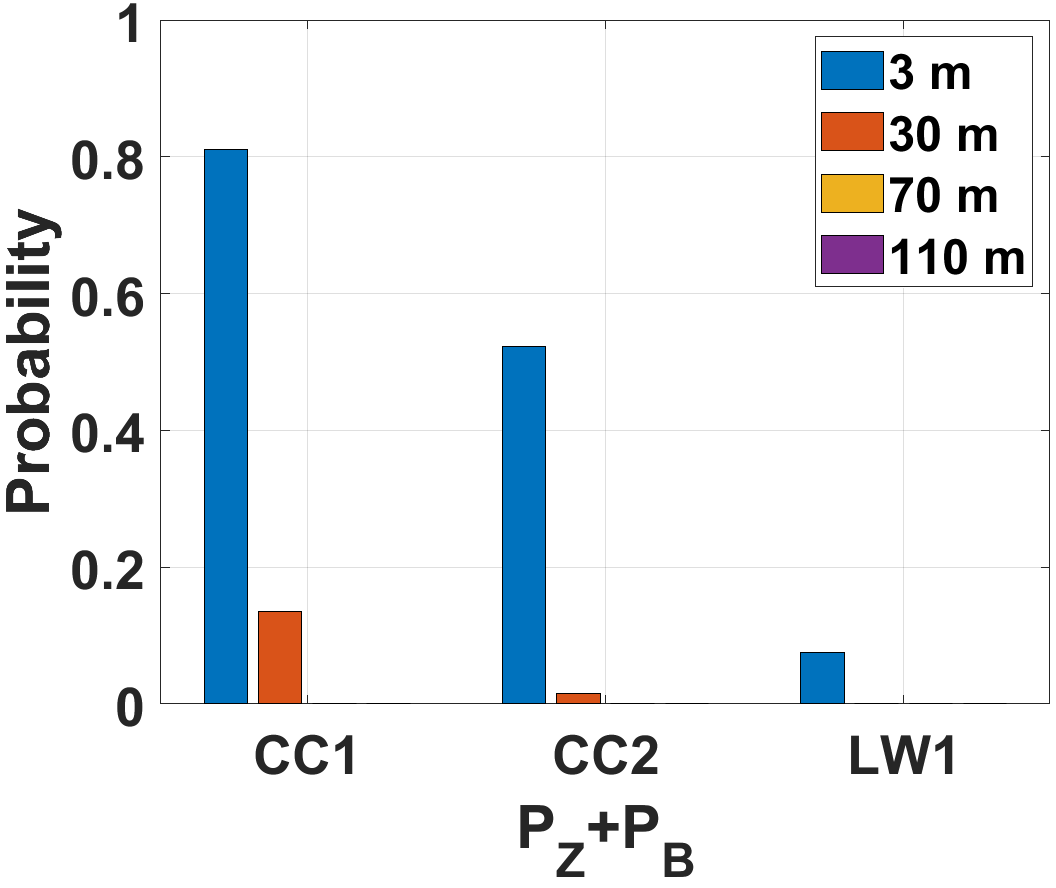}
    \label{fig:CDF_RSSI_Po}
    }     %\vspace{-1mm}
    \caption{CDFs for %altitude-dependent 
    RSSI at CC1, CC2, and LW1 versus altitude.}
    \label{fig:CDFs_RSSI}\vspace{-3mm}
\end{figure}

Simulation results for RSSI for various altitudes of the UAV with CC1, CC2, and LW1 transmitters are shown in Fig.~\ref{fig:SISO_CC1_CC2_LW1}, where $\boldsymbol{Z}$ refers to the outdoor receiver locations with no wireless coverage, and $\boldsymbol{B}$ captures the out-of-coverage receiver locations within buildings, respectively. As seen in Figs.~\ref{fig:CC1_SISO_1}, \ref{fig:CC2_SISO_1}, and \ref{fig:LW1_SISO_1}, there is extensive blockage when the receiver height is 3 m, inclusive of building indoor locations. Moreover,  the distribution of the RSSI depends on the distance from the transmitter as the altitude of the UAV increases. In Fig.~\ref{fig:CDFs_RSSI}, the CDFs for altitude-dependent RSSI at three different transmitters are given. The combined probabilities that there is no link coverage at a given altitude and indoor receiver, $P_{\boldsymbol{Z}}+P_{\boldsymbol{B}}$, are also plotted in Fig.~\ref{fig:CDF_RSSI_Po}.  The RSSI of UAVs is observed to be mostly distributed between -55 dBm to -40 dBm, especially at higher altitudes and for the rural scenario. 

\section{Channel Rank Analysis}\label{Sec:4}

In this section, we analyze the channel rank for an air-to-ground link at multiple altitudes using ray tracing simulations, based on the assumptions provided in Section~\ref{Sec:2}. Once the singular value matrix $\boldsymbol{\Sigma}$ in~\eqref{eqn: SVD} is calculated, the transmitter needs to decide how many spatial layers to use for communication.  It is known that choosing between spatial multiplexing with a larger number of spatial layers versus diversity transmission with fewer spatial layers  is not always trivial \cite[Ch.~8]{tse2005fundamentals}, which depends on the signal-to-noise ratio (SNR) and spatial correlation of received signals at different antennas. It is not worth transmitting data over a spatial layer that has a very small singular value, as the SNR over that spatial layer will be very low, and using diversity transmission may be preferable as it will improve the SNR. In this context, how to set the threshold $\sigma_{\rm Thr}$ for deciding the channel rank is critical as it will determine the number of spatial layers. 

\begin{figure*}[t]
    \centering
    \subfigure[UAV altitude: 3 m (CC1)]{
    \includegraphics[width=0.45\columnwidth]{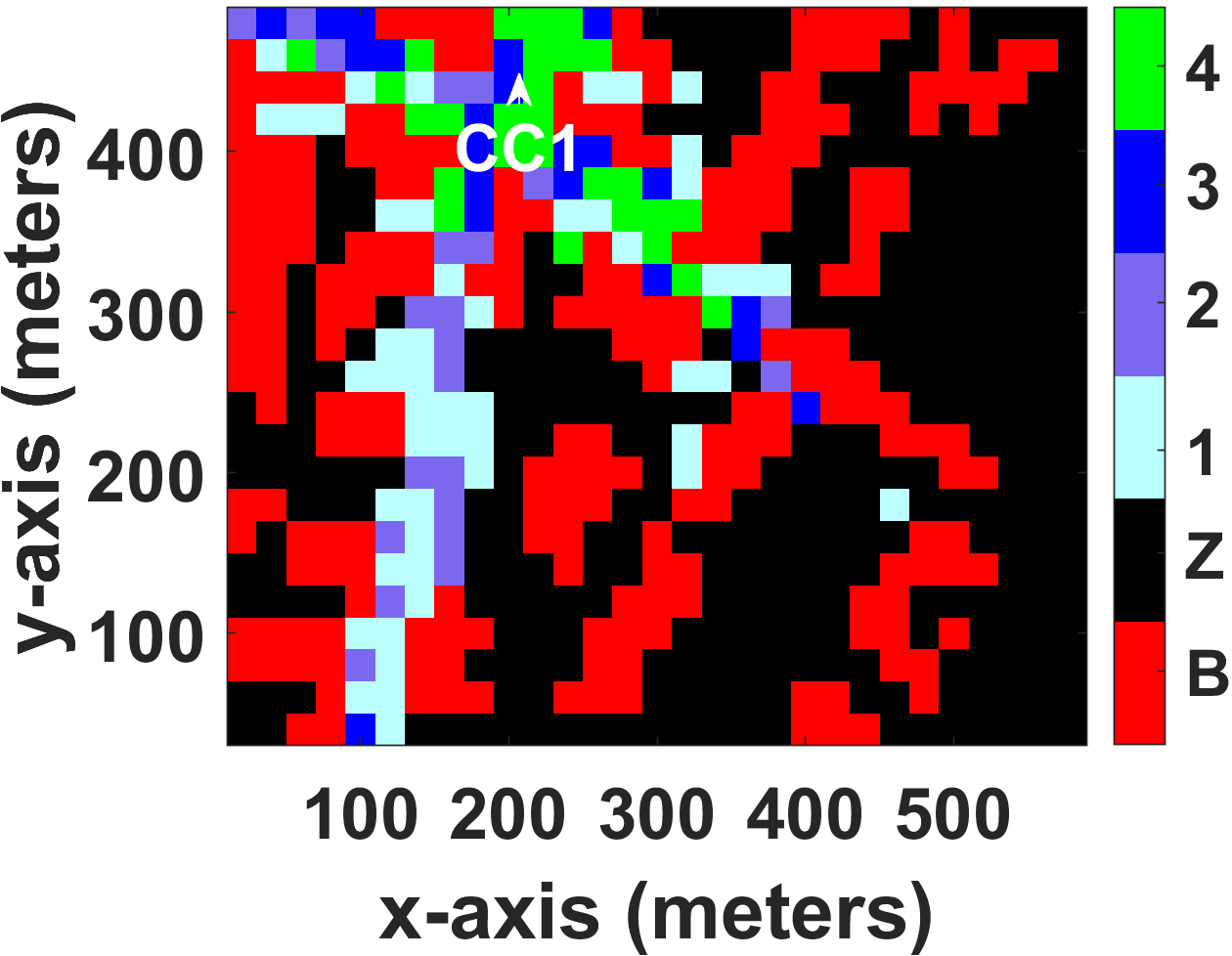}
    \label{fig:CC1_rank_1}
    }
    \subfigure[UAV altitude: 30 m (CC1)]{
    \includegraphics[width=0.45\columnwidth]{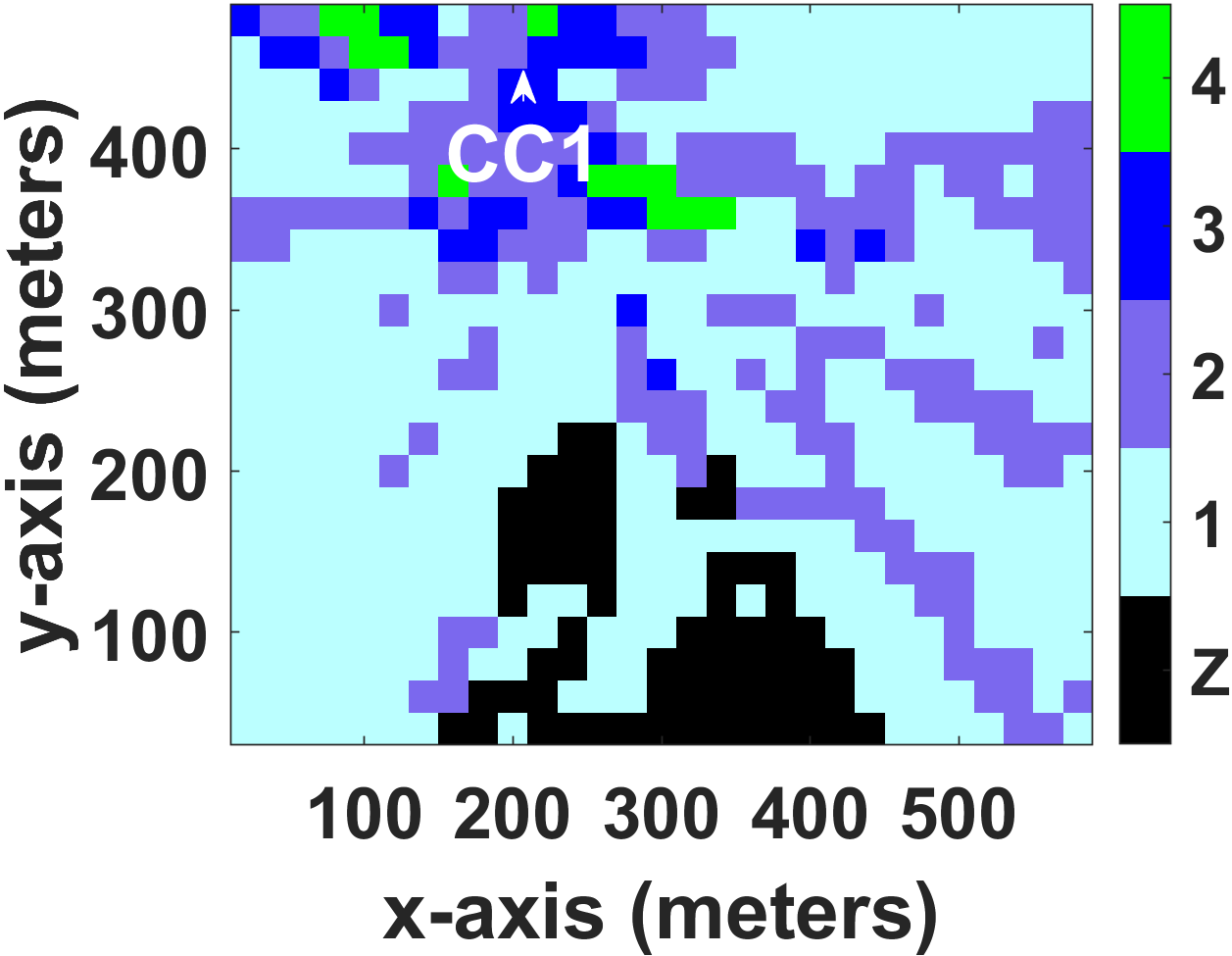}
    \label{fig:CC1_rank_2}
    }
    \subfigure[UAV altitude: 70 m (CC1)]{
    \includegraphics[width=0.45\columnwidth]{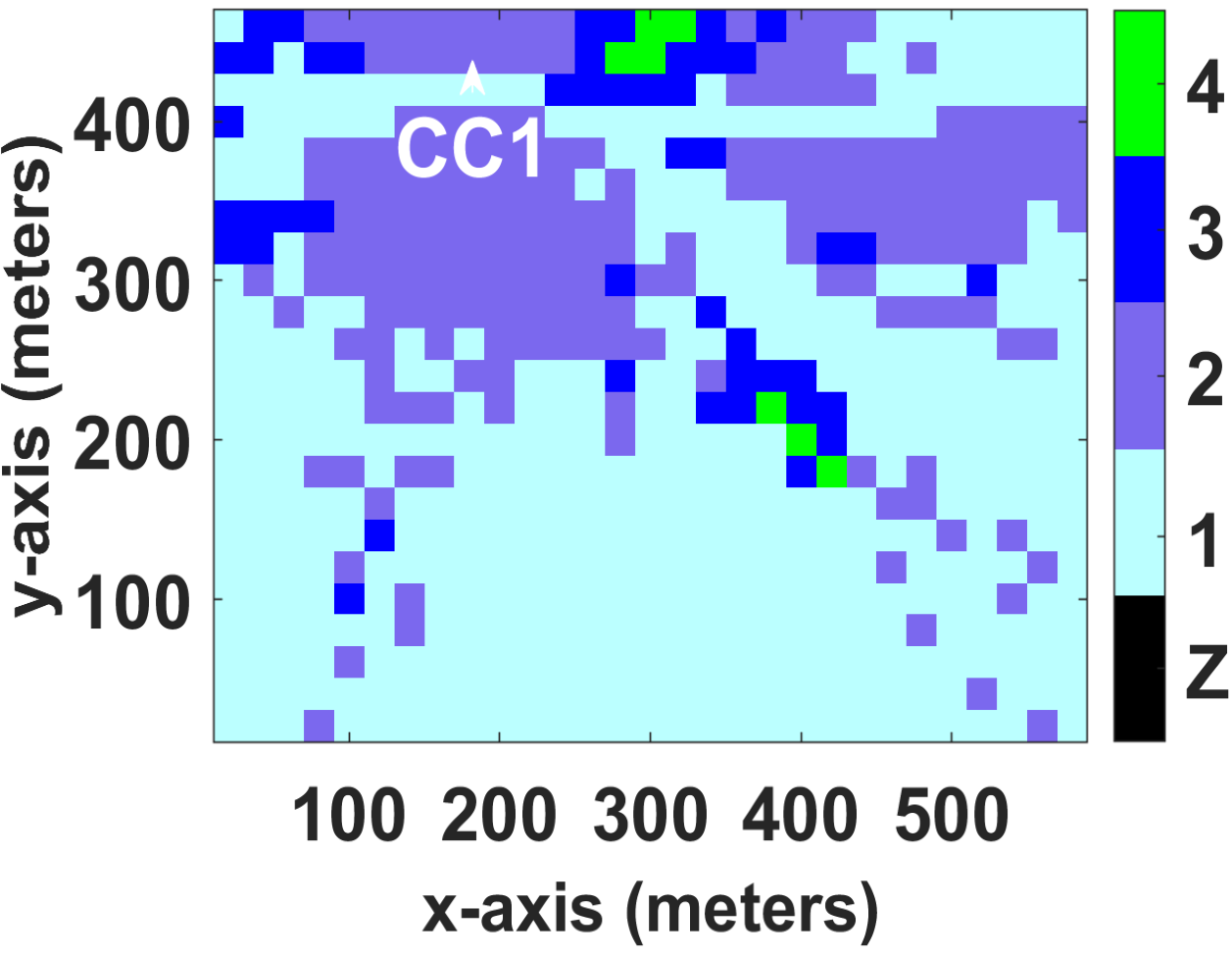}
    \label{fig:CC1_rank_3}
    }
    \subfigure[UAV altitude: 110 m (CC1)]{
    \includegraphics[width=0.45\columnwidth]{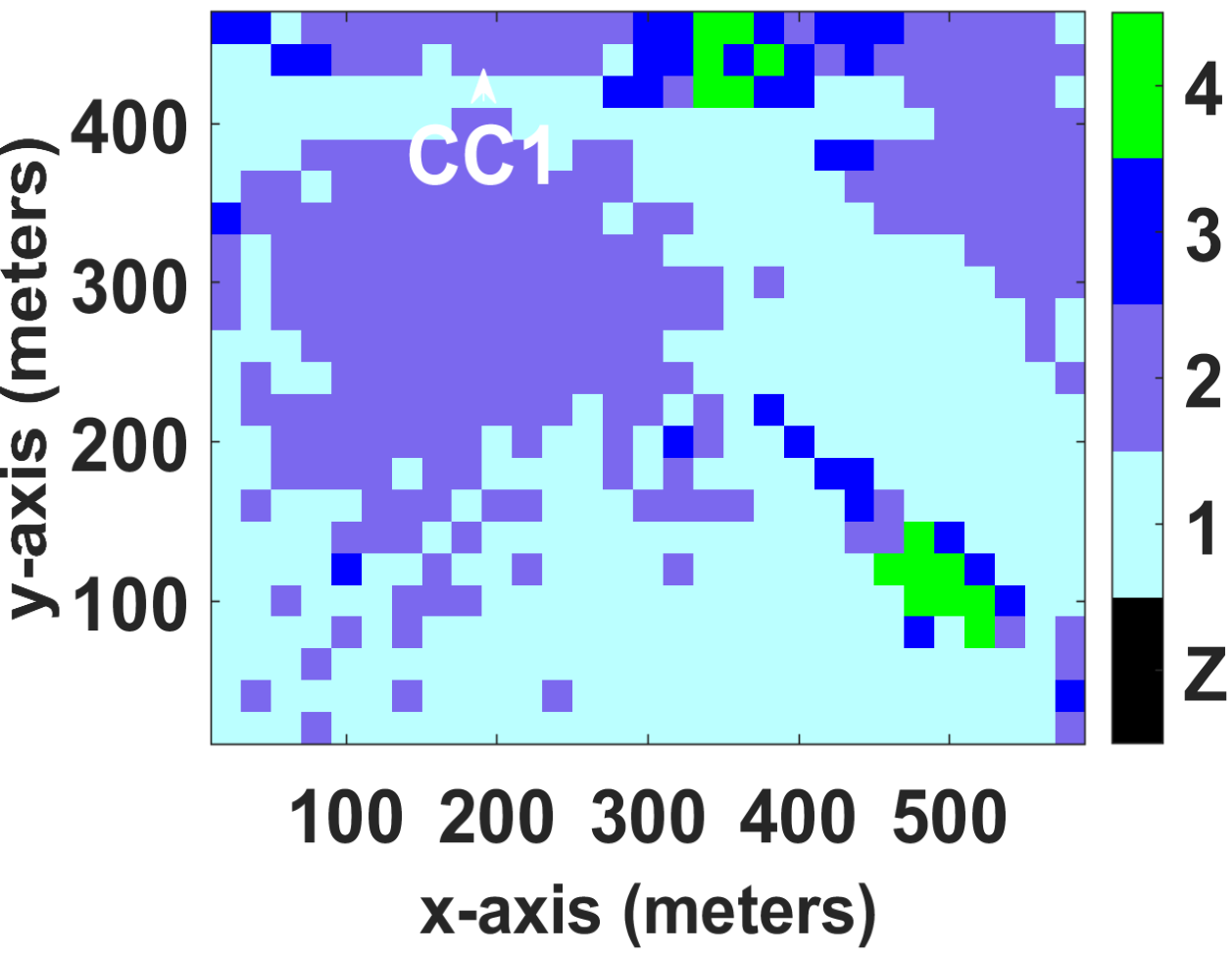}
    \label{fig:CC1_rank_4}
    }
    \subfigure[UAV altitude: 3 m (CC2)]{
    \includegraphics[width=0.45\columnwidth]{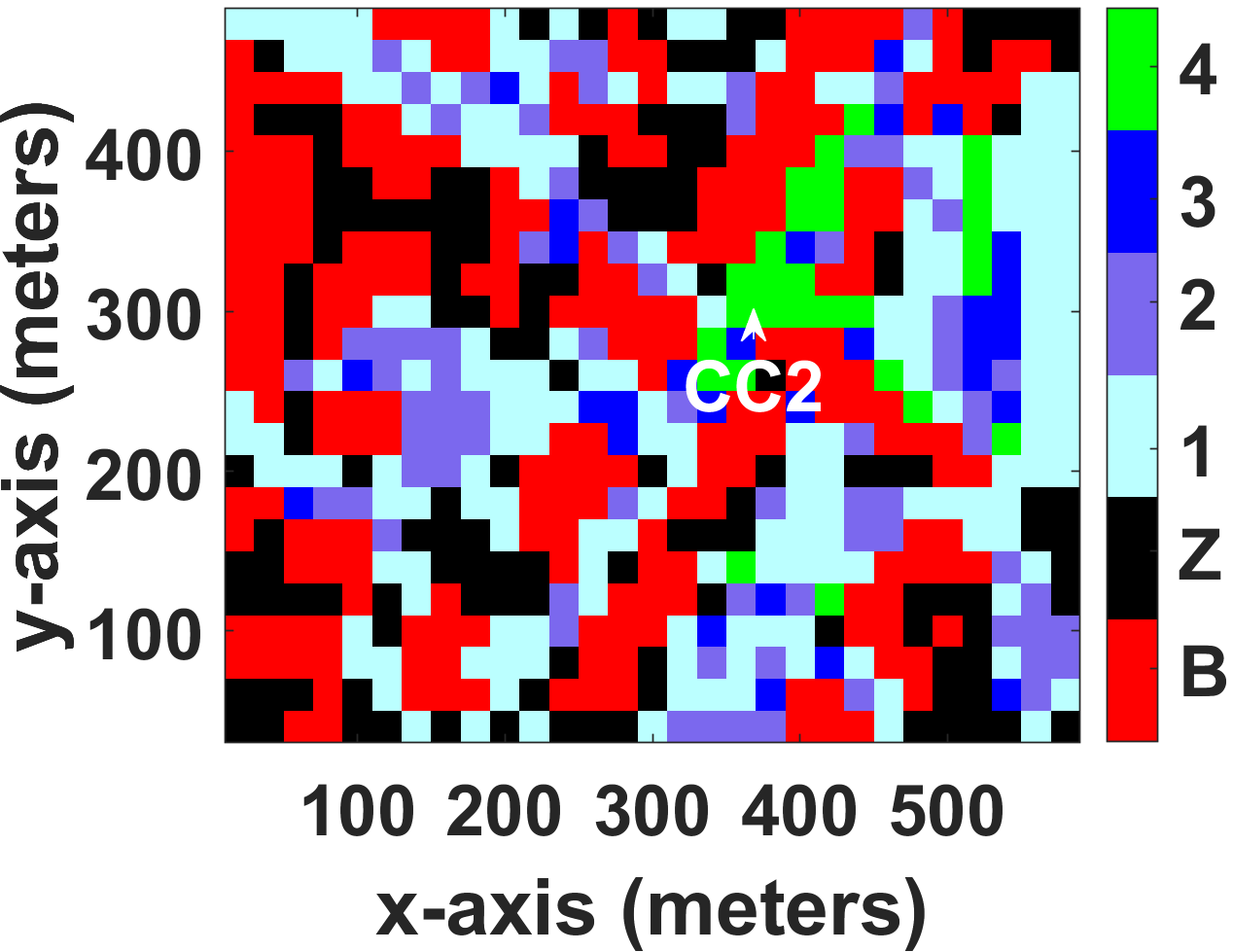}
    \label{fig:CC2_rank_1}
    }
    \subfigure[UAV altitude: 30 m (CC2)]{
    \includegraphics[width=0.45\columnwidth]{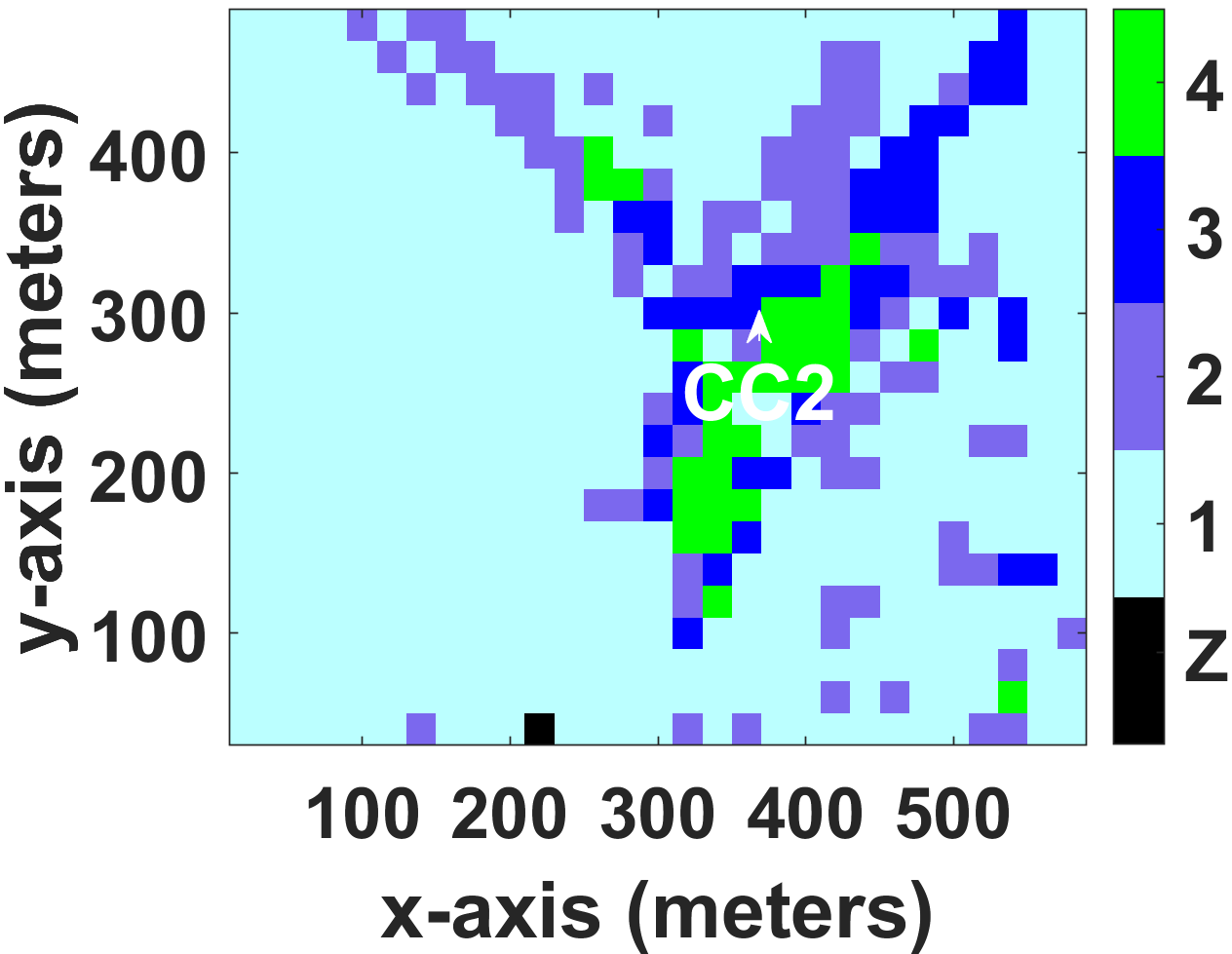}
    \label{fig:CC2_rank_2}
    }
    \subfigure[UAV altitude: 70 m (CC2)]{
    \includegraphics[width=0.45\columnwidth]{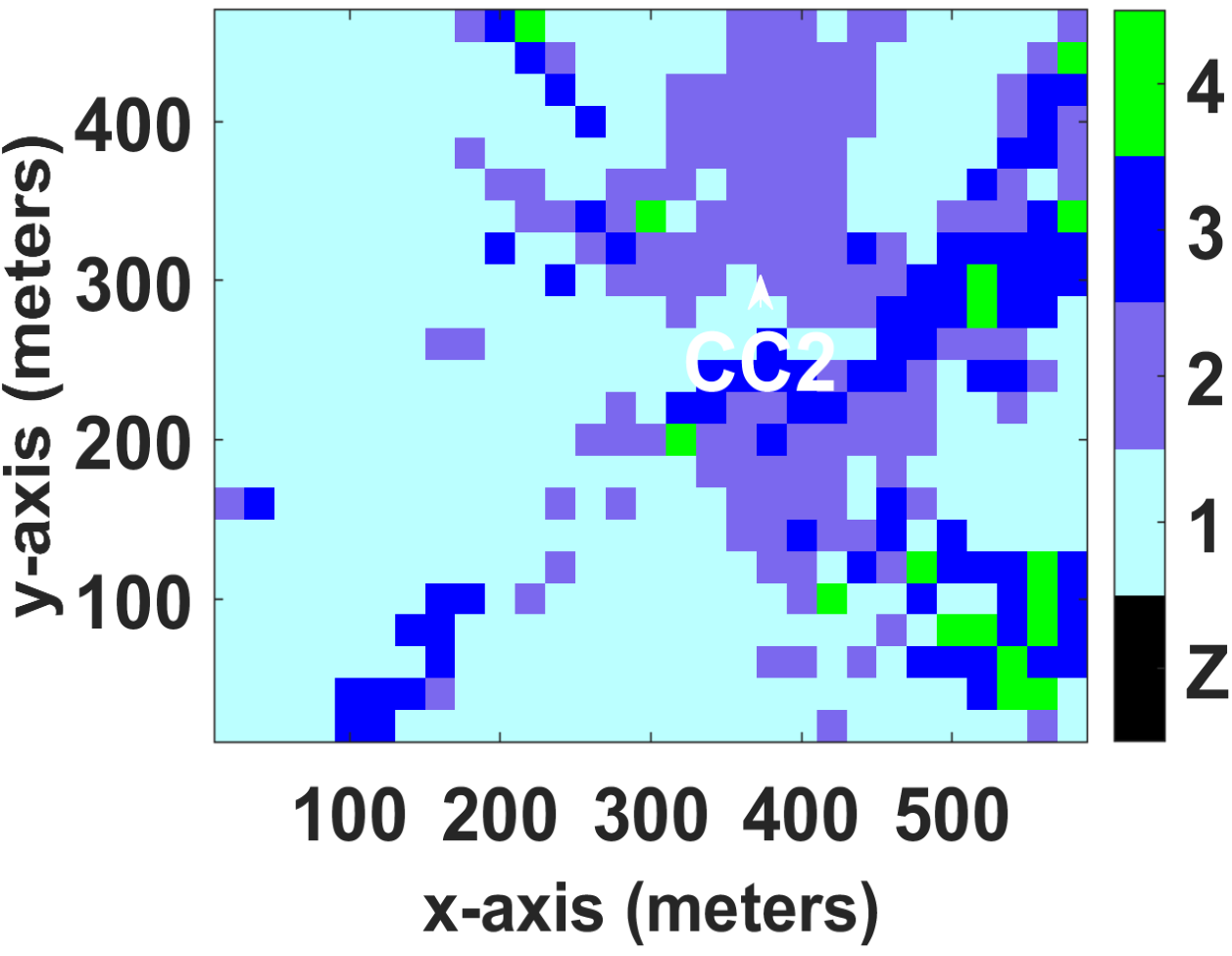}
    \label{fig:CC2_rank_3}
    }
    \subfigure[UAV altitude: 110 m (CC2)]{
    \includegraphics[width=0.45\columnwidth]{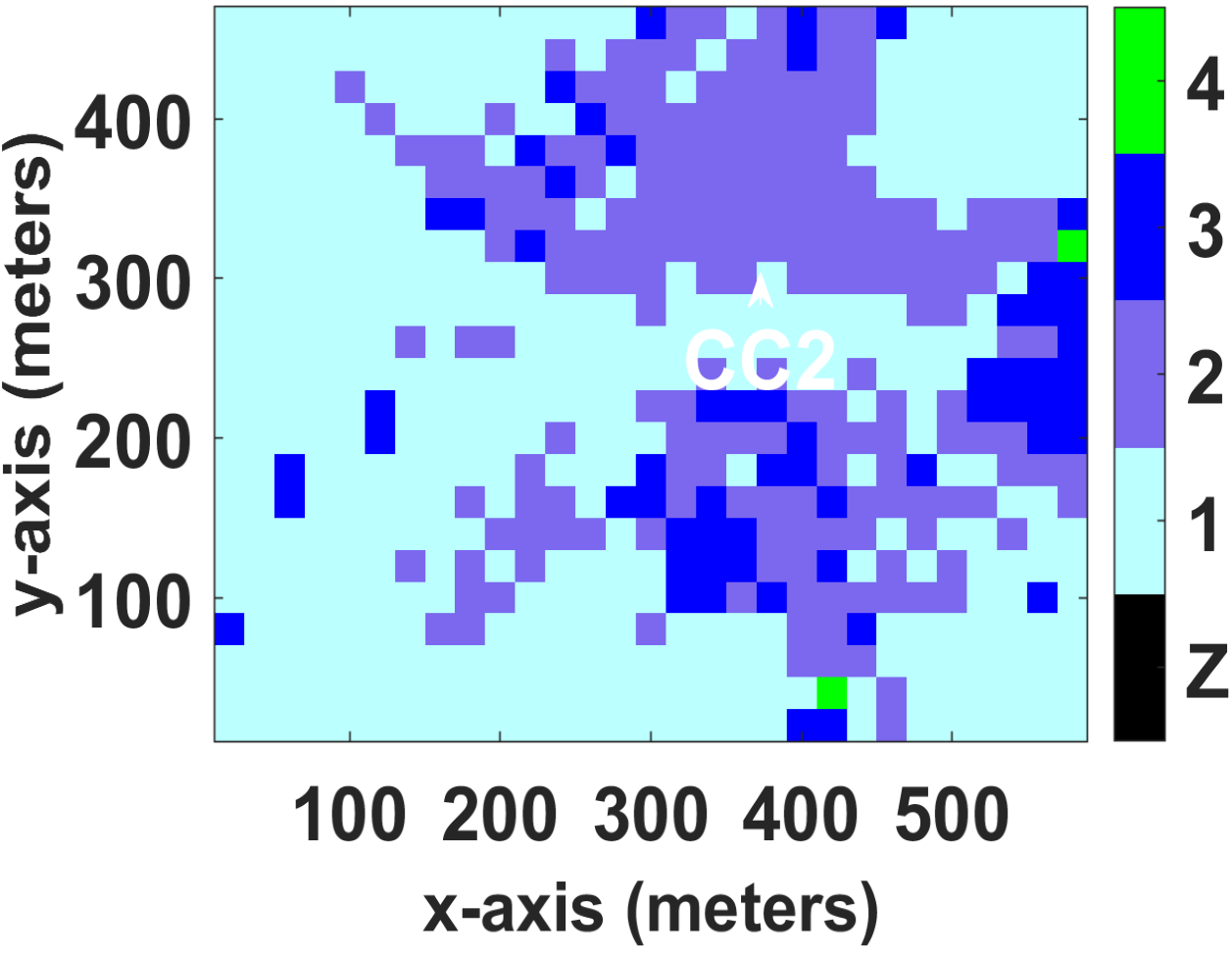}
    \label{fig:CC2_rank_4}
    }
    \subfigure[UAV altitude: 3 m (LW1)]{
    \includegraphics[width=0.45\columnwidth]{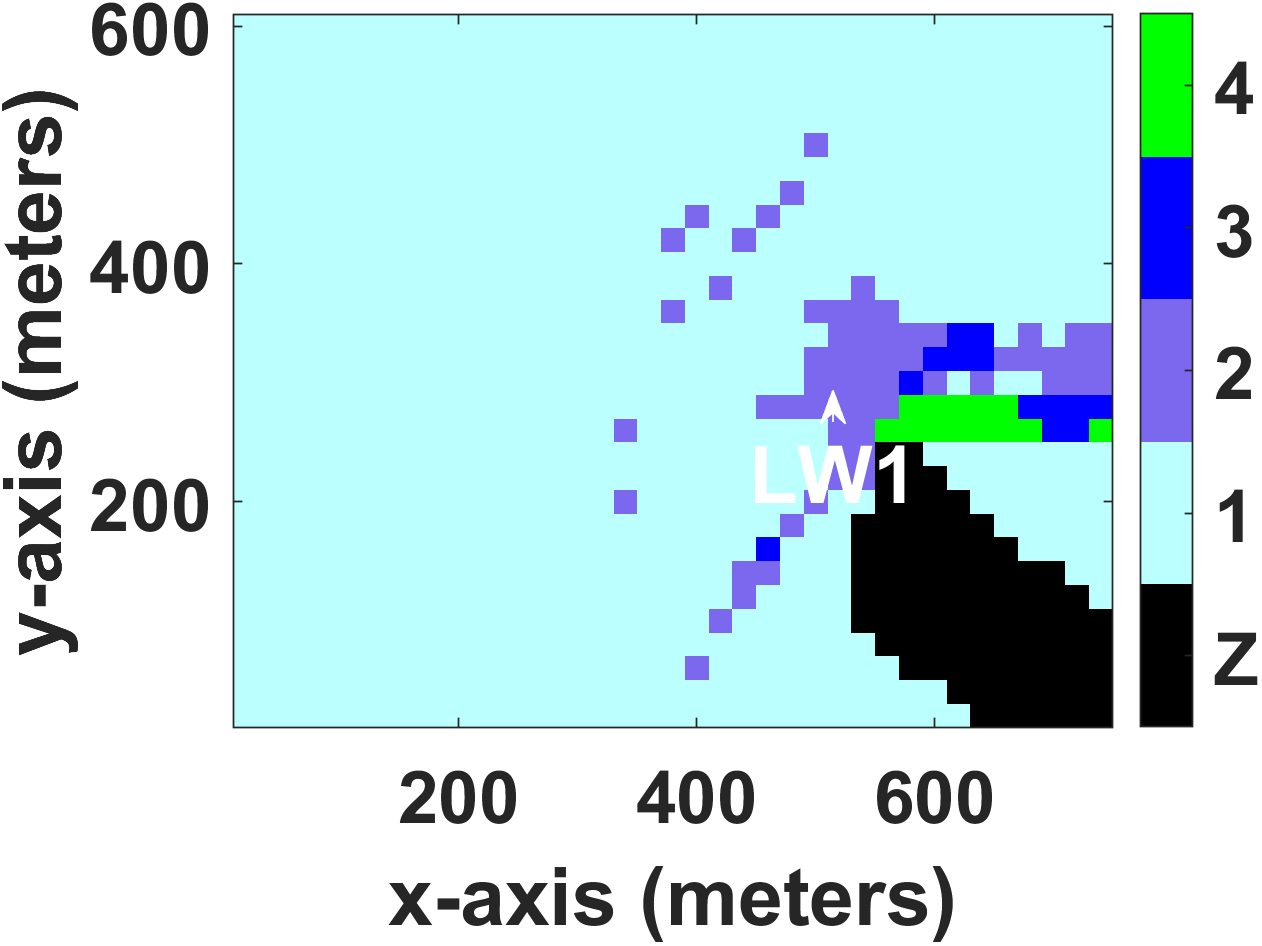}
    \label{fig:LW_rank_1}
    }
    \subfigure[UAV altitude: 30 m (LW1)]{
    \includegraphics[width=0.45\columnwidth]{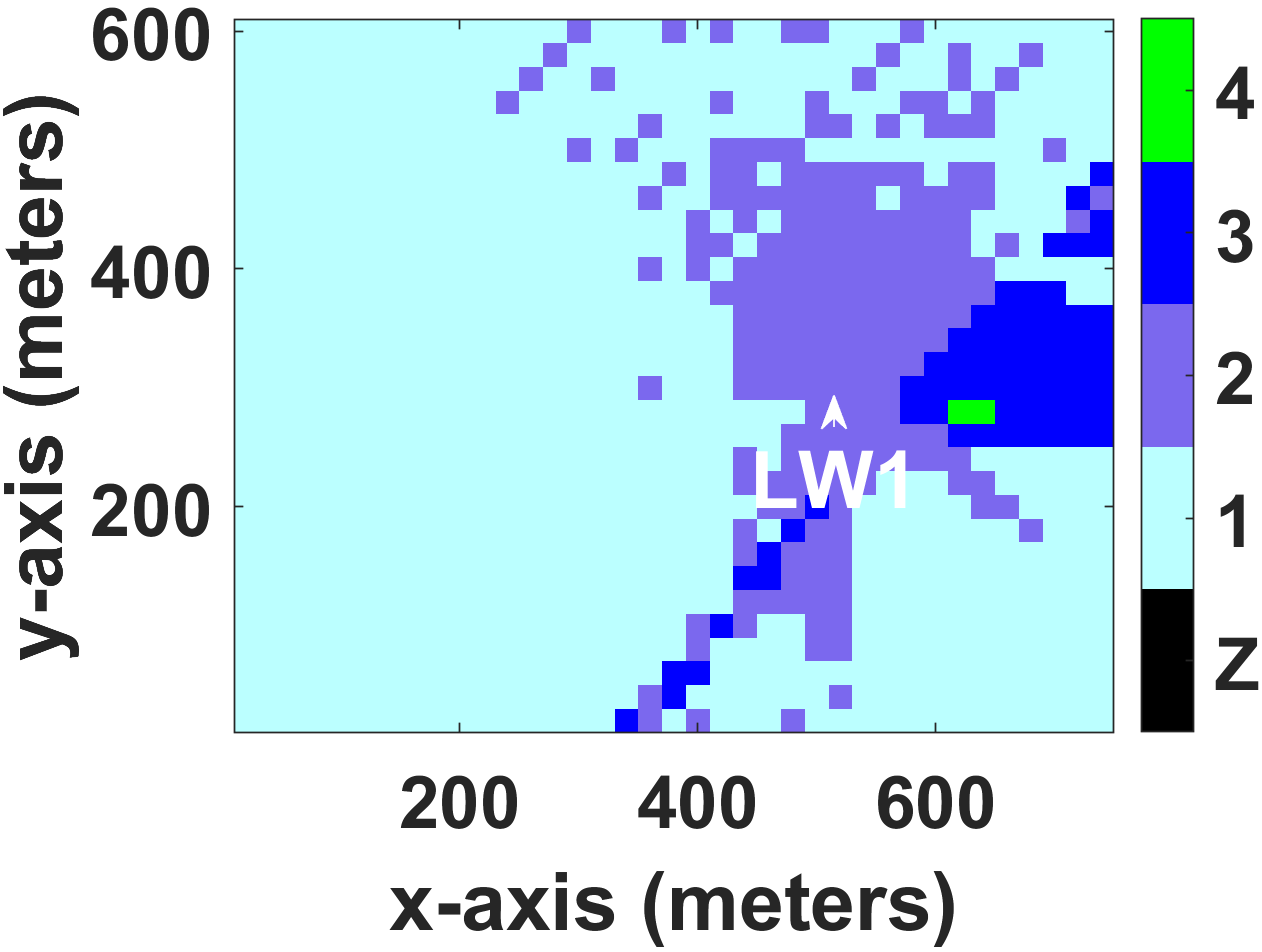}
    \label{fig:LW_rank_2}
    }
    \subfigure[UAV altitude: 70 m (LW1)]{
    \includegraphics[width=0.45\columnwidth]{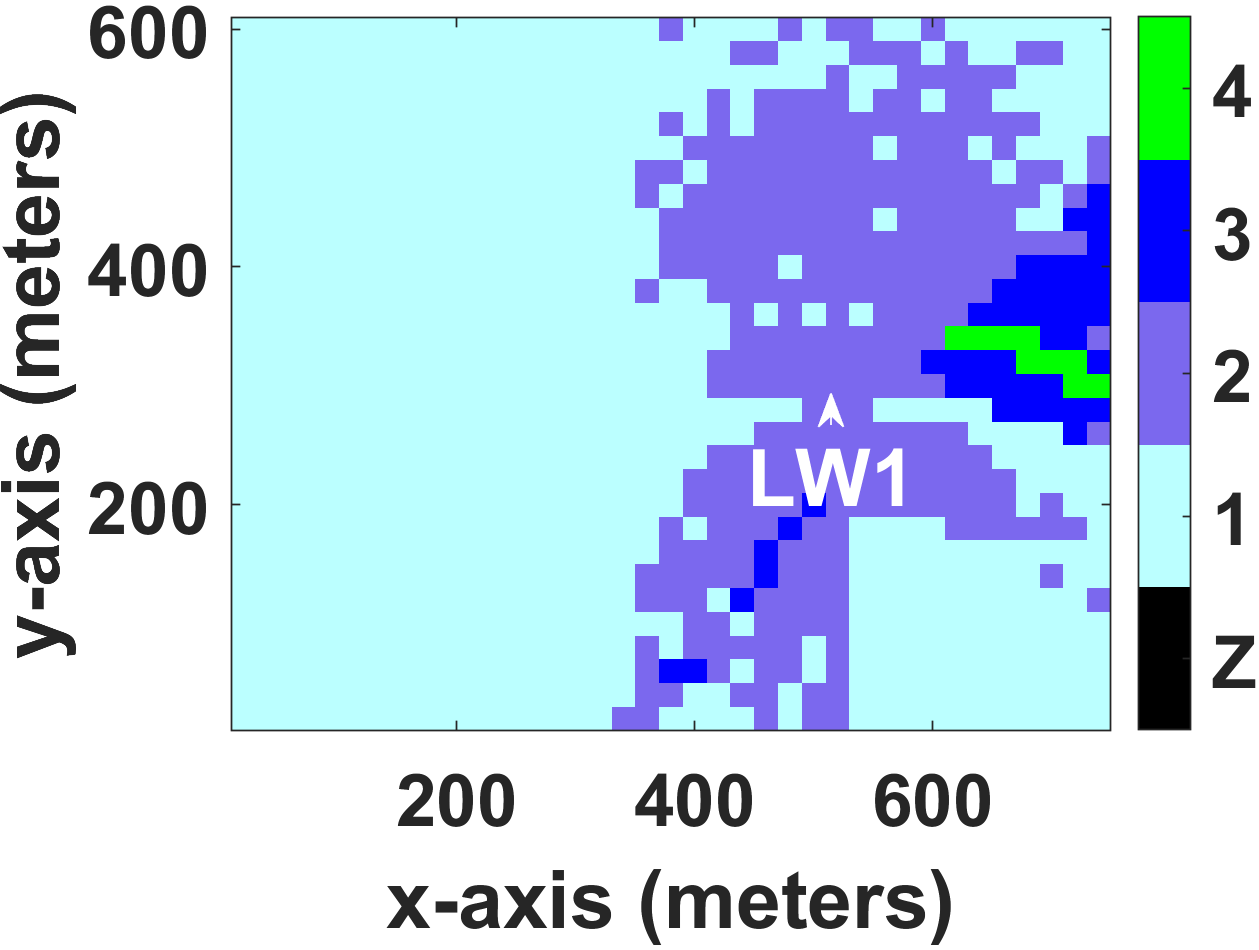}
    \label{fig:LW_rank_3}
    }
    \subfigure[UAV altitude: 110 m (LW1)]{
    \includegraphics[width=0.45\columnwidth]{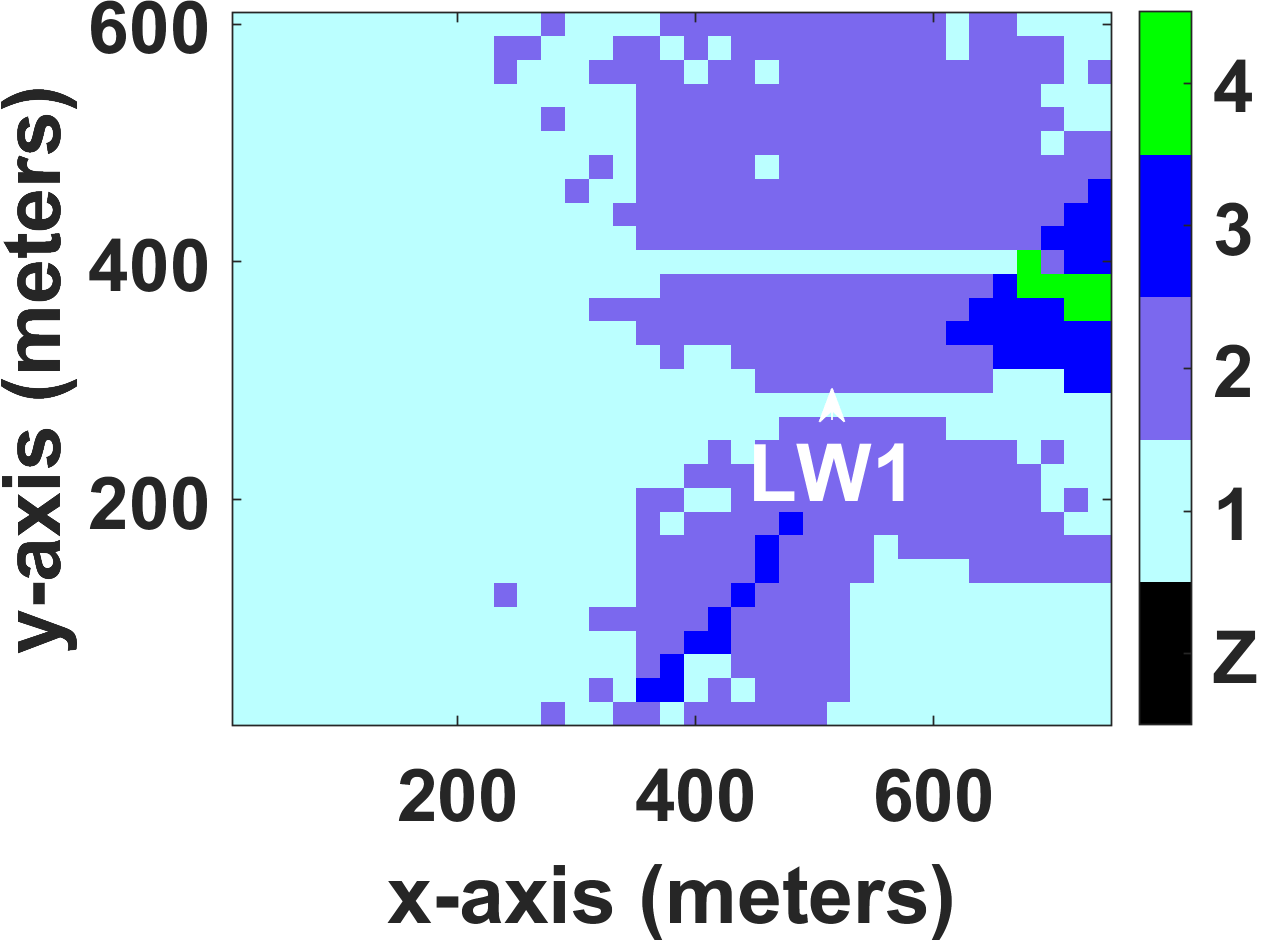}
    \label{fig:LW_rank_4}
    }\vspace{-1mm}
    \caption{Channel ranks in Centennial Campus with CC1 (a)-(d), CC2 (e)-(h), and in Lake Wheeler with LW1 (i)-(l) as the cellular BS (see Fig.~\ref{fig:my_label}).  Channel rank is seen to be correlated across different altitudes for a given transmitter.}
    \label{fig:rank_CC1_CC2_LW1}\vspace{-4.5mm}
\end{figure*} 

%In this paper, 
We consider rounding criteria for finding the singular values of the channel rank matrix. The singular value for the $s$-th strongest singular value  with rounding can be expressed as
\begin{equation}\label{Eq:Thresholding}
    \tilde{\sigma}_{s} = \begin{cases} \sigma_{s}, 
   \; \; \textnormal{if} \; \sigma_{s} \geq \sigma_{\rm{Thr}} \\ 0, \; \; \textnormal{otherwise} \end{cases},
\end{equation}
where $\sigma_{s}$ denotes the $s$-th strongest singular value. Here, we consider that the threshold $\sigma_{\rm{Thr}}$ can be proportional to the strongest singular value for a given $\boldsymbol{\Sigma}$, i.e.,  $\sigma_{\rm{Thr}}=\sigma_1/K$, where $K$ is a  constant value. Alternatively, the threshold for $s$-th strongest singular value can be the average value of the singular value of the corresponding order, i.e.,  $\sigma_{\rm{Thr}}=\sigma_{s, {\rm mean}}=\sum_{s}\sigma_{s}/N$ denotes the mean value of the $s$-th strongest singular value, where $N$ is the  total number of receiver locations at a given UE altitude. The subscript $s$ can be bounded as $\min(m, n)$  based on the dimensions of the rectangular diagonal matrix $\boldsymbol{\Sigma}$.  Then, the channel rank is derived by the number of non-zero singular values after the thresholding and rounding operation in~\eqref{Eq:Thresholding}.
%the singular value that is greater than or equal to the mean value of the corresponding order of singular values.

Under these assumptions, and considering for now that $\sigma_{\rm{Thr}}=\sigma_{s, {\rm mean}}$ is used as a singular value threshold in~\eqref{Eq:Thresholding}, simulation results on UAV channel rank at various receiver altitudes with CC1, CC2, and LW1 serving as transmitters are shown in Fig.~\ref{fig:rank_CC1_CC2_LW1}. 
% In the results, \textcolor{red}{$\boldsymbol{B}$} captures the indoor receiver locations that are within buildings.
% \textcolor{red}{Here, $P_{\rm{o}}$ in Fig. \ref{fig:SISO_CC1_CC2_LW1} is inclusive of $P_{\rm{b}}$ and the zero-rank scenario, which is slightly different because of different antenna settings for RSSI and channel rank analysis.} 
As seen in Fig. \ref{fig:CC1_rank_1}, most of the UEs with 3~m height do not have signal coverage due to the blockage incurred by buildings. The effect of blockage disappears for the 30~m altitude since the height of all the buildings in the area is lower than 30~m. Most receiver locations have ranks of 1 or 2 with 30~m, 70~m, and 110~m UAV altitudes, while only some receivers located at the reflected paths have higher channel ranks. The same trend can be seen in Figs.~\ref{fig:CC2_rank_1}-\ref{fig:CC2_rank_4}. The channel rank distribution is higher with CC2 3~m case compared to the CC1 3~m case, which can be explained by the geographic characteristics and heights of CC1's and CC2's locations. Some correlation of the channel rank at different altitudes for a given transmitter location is apparent, which requires further investigation in the future.

\begin{figure*}[t!]
    \centering
    \subfigure[CC1, $\sigma_{\rm Thr} = \sigma_{s,{\rm mean}}$]{
    \includegraphics[width=0.45\columnwidth]{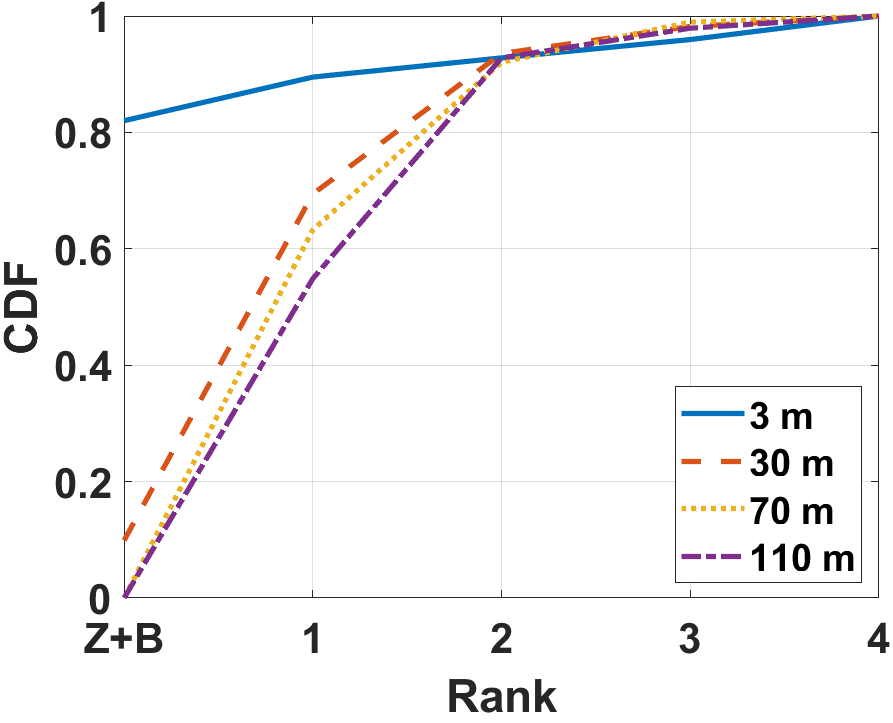}
    \label{fig:CDF_CC1}
    }
    \subfigure[CC1, $\sigma_{\rm Thr} = \sigma_{1}/10$]{
    \includegraphics[width=0.45\columnwidth]{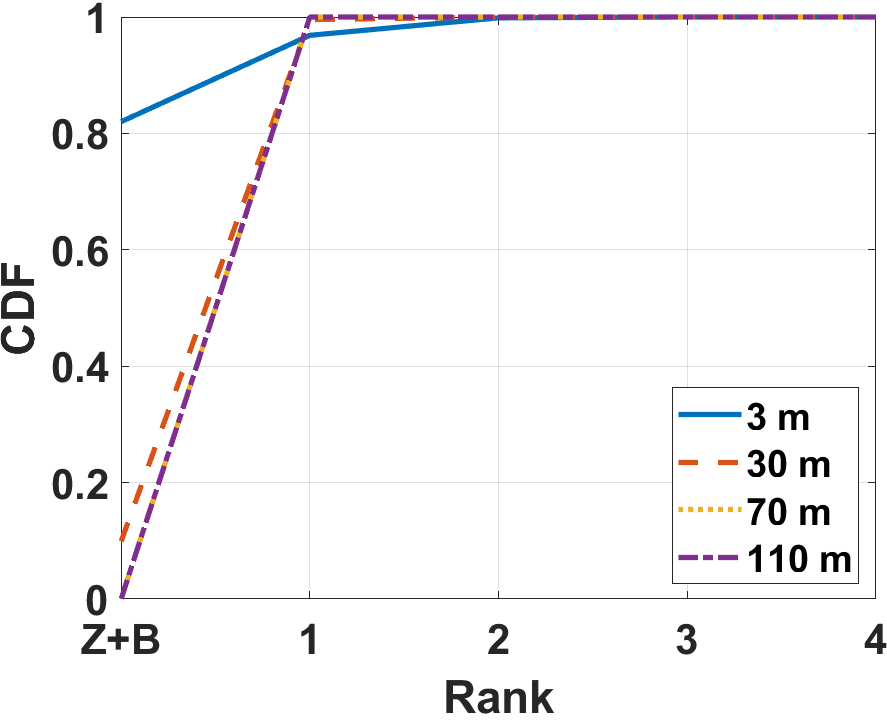}
    \label{fig:CDF_CC1_20dB}
    }
    \subfigure[CC1, $\sigma_{\rm Thr} = \sigma_{1}/10^2$]{
    \includegraphics[width=0.45\columnwidth]{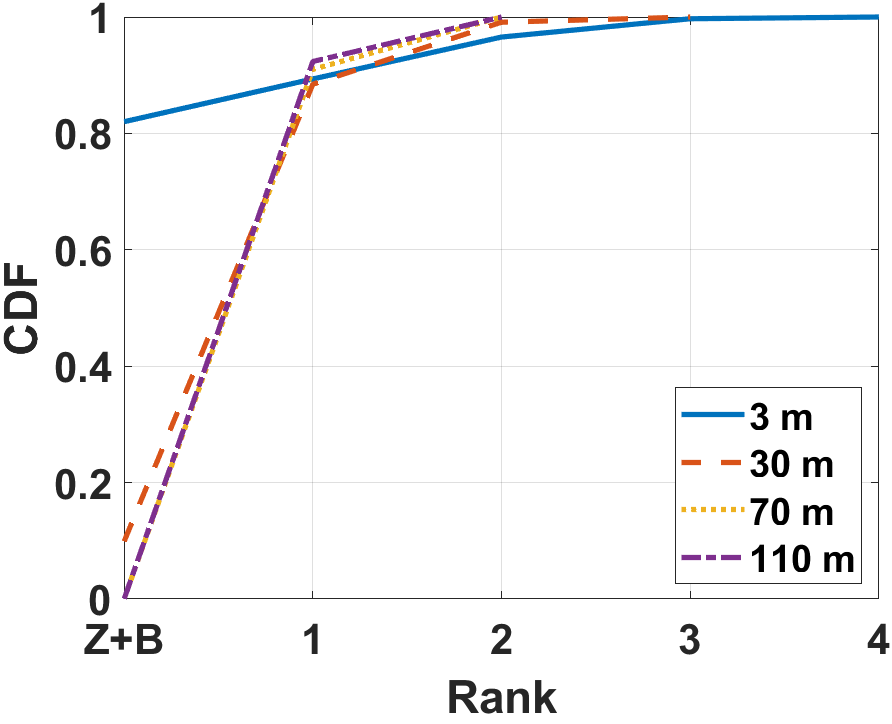}
    \label{fig:CDF_CC1_20dB}
    }
    \subfigure[CC1, $\sigma_{\rm Thr} = \sigma_{1}/10^4$]{
    \includegraphics[width=0.45\columnwidth]{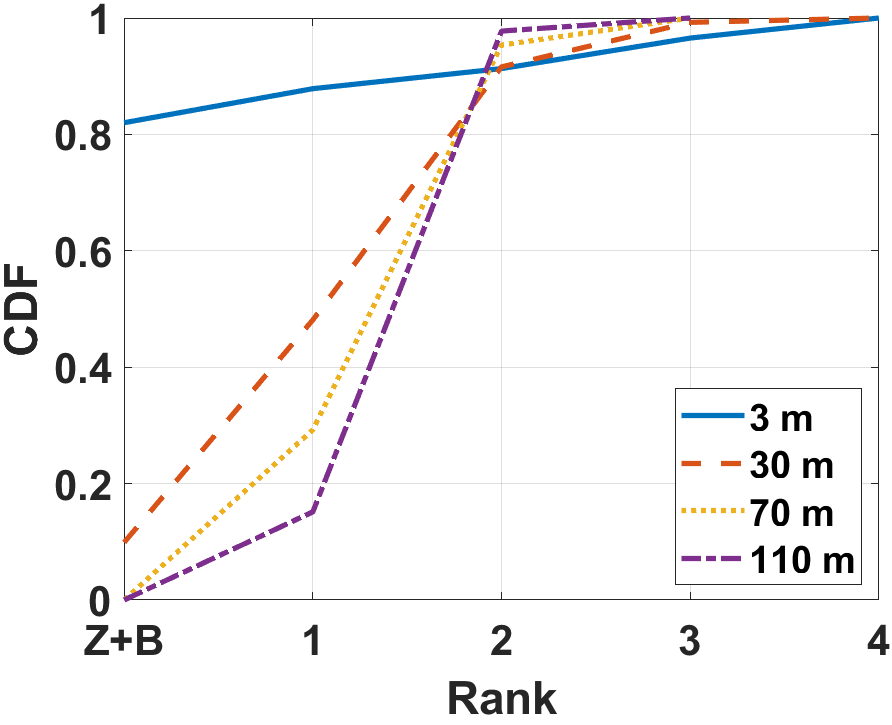}
    \label{fig:CDF_CC1_40dB}
    }
    % \hspace{-1mm}
    \subfigure[CC2, $\sigma_{\rm Thr} = \sigma_{s,{\rm mean}}$]{
    \includegraphics[width=0.45\columnwidth]{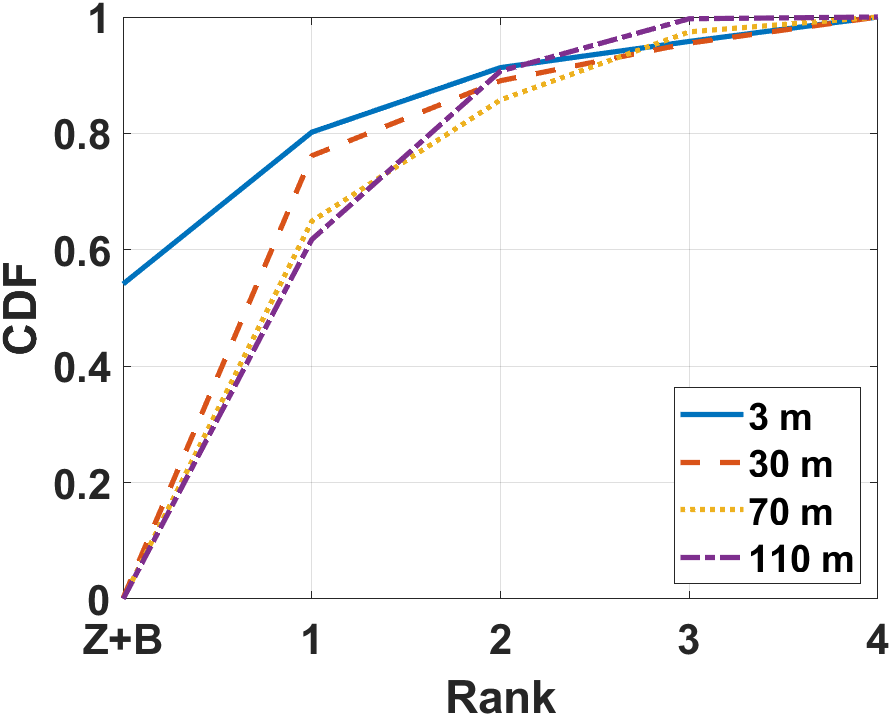}
    \label{fig:CDF_CC2}
    }
    \subfigure[CC2, $\sigma_{\rm Thr} = \sigma_{1}/10$]{
    \includegraphics[width=0.45\columnwidth]{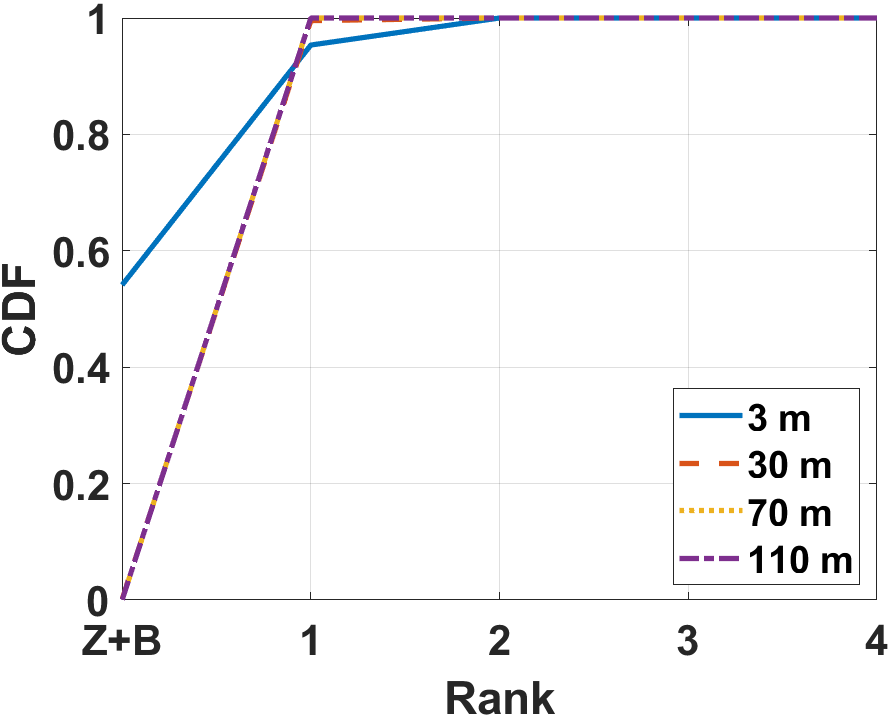}
    \label{fig:CDF_CC2_20dB}
    }
    \subfigure[CC2, $\sigma_{\rm Thr} = \sigma_{1}/10^2$]{
    \includegraphics[width=0.45\columnwidth]{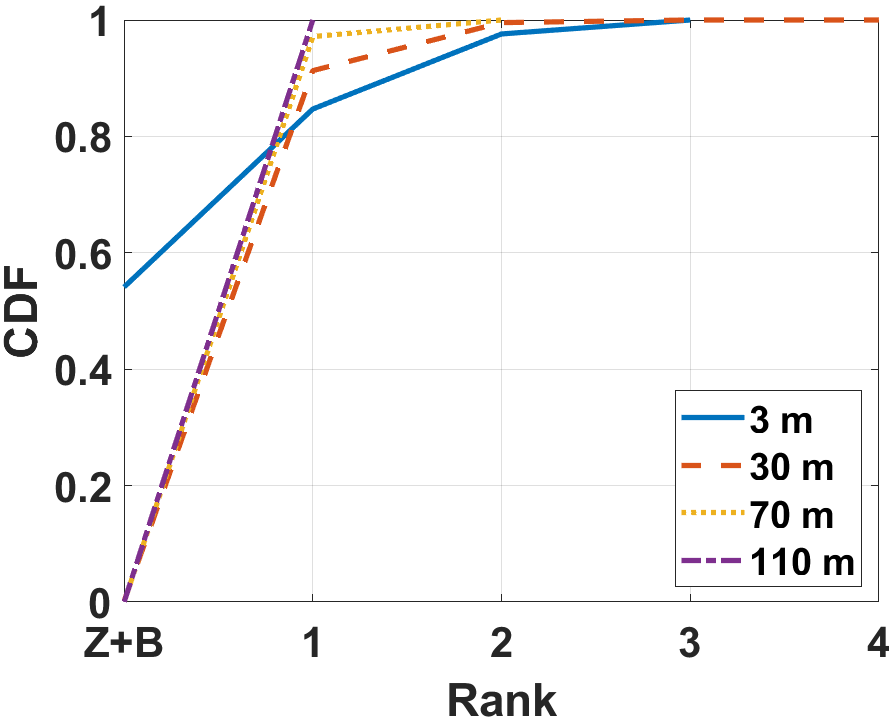}
    \label{fig:CDF_CC2_20dB}
    }
    \subfigure[CC2, $\sigma_{\rm Thr} = \sigma_{1}/10^4$]{
    \includegraphics[width=0.45\columnwidth]{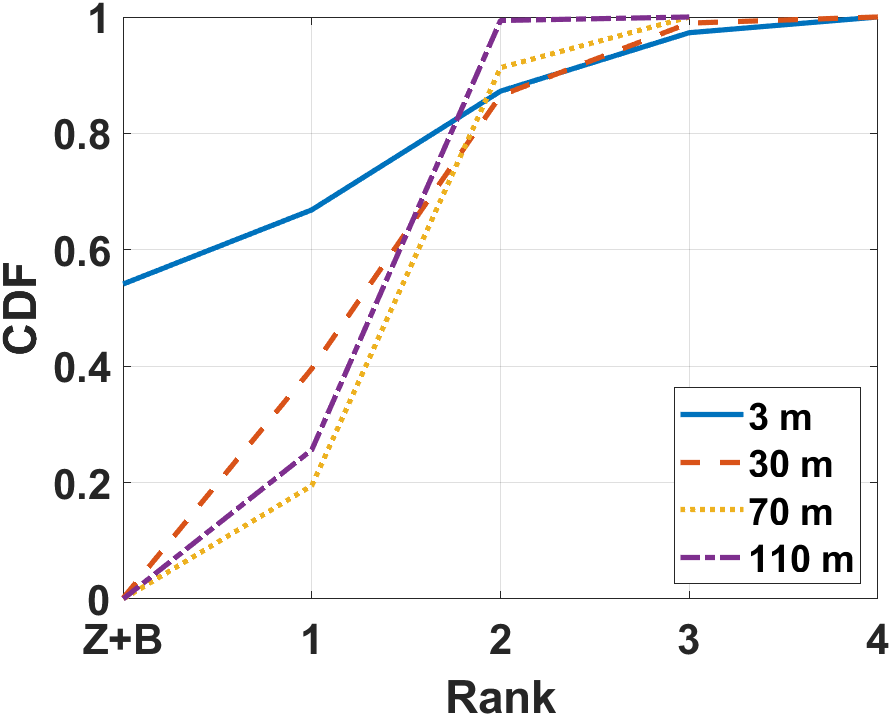}
    \label{fig:CDF_CC2_40dB}
    }
    \subfigure[LW1, $\sigma_{\rm Thr} = \sigma_{s,{\rm mean}}$]{
    \includegraphics[width=0.45\columnwidth]{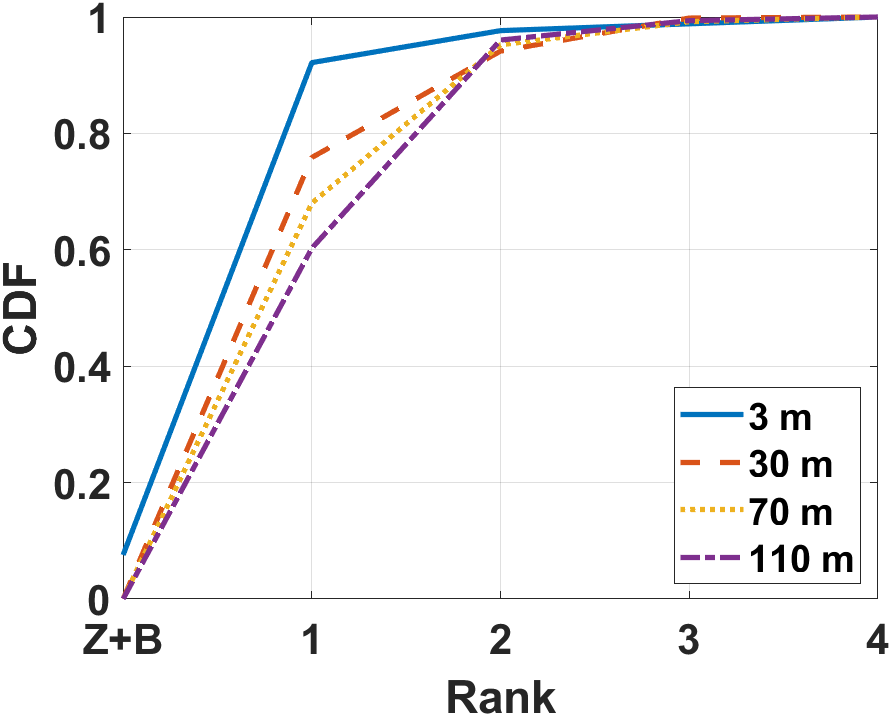}
    \label{fig:CDF_LW}
    }   
    \subfigure[LW1, $\sigma_{\rm Thr} = \sigma_{1}/10$]{
    \includegraphics[width=0.45\columnwidth]{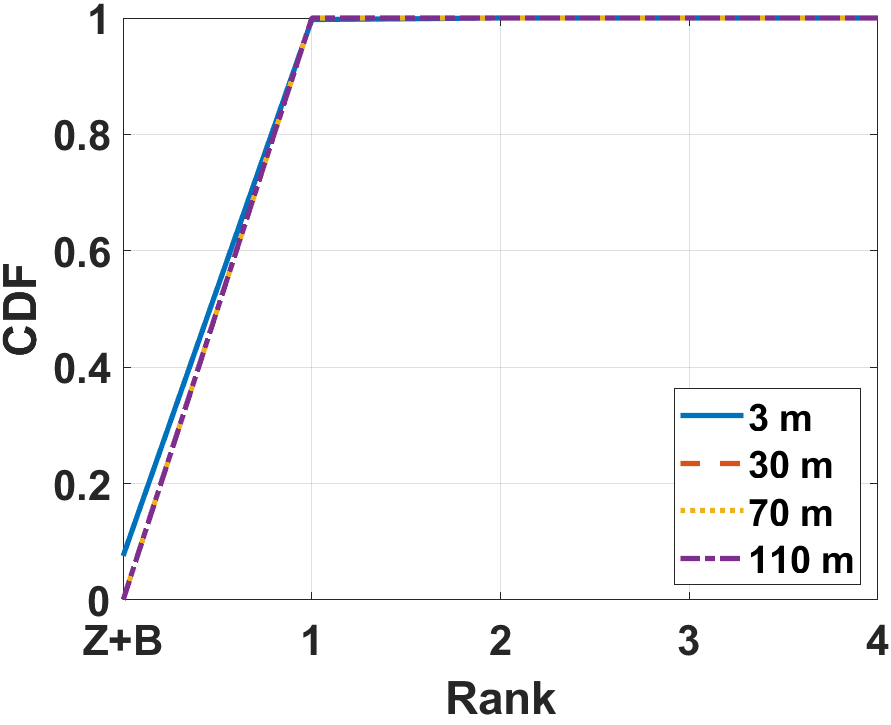}
    \label{fig:CDF_LW_20dB}
    }    
    \subfigure[LW1, $\sigma_{\rm Thr} = \sigma_{1}/10^2$]{
    \includegraphics[width=0.45\columnwidth]{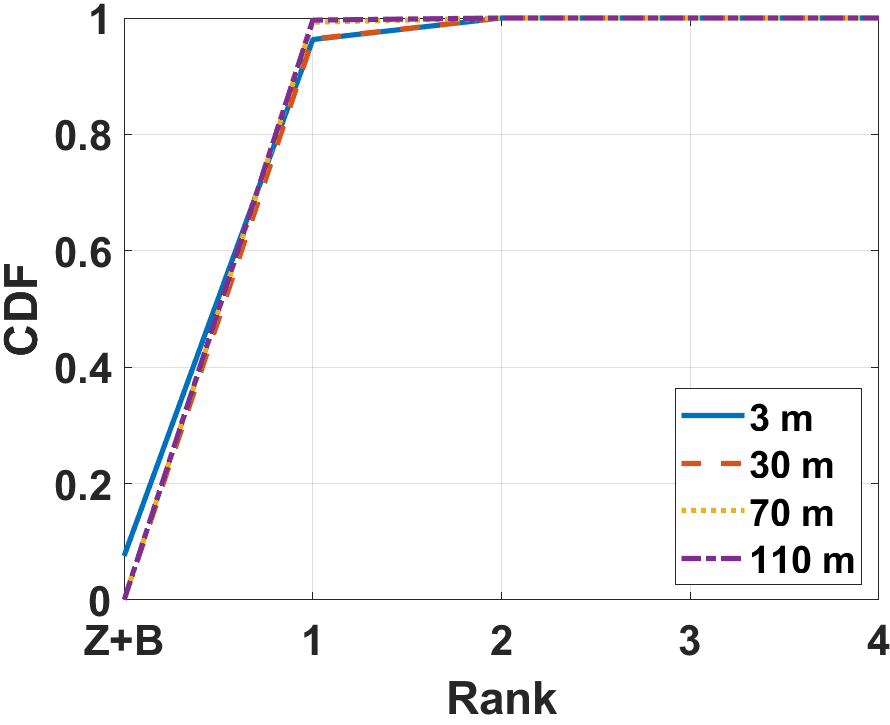}
    \label{fig:CDF_LW_20dB}
    }     
    \subfigure[LW1, $\sigma_{\rm Thr} = \sigma_{1}/10^4$]{
    \includegraphics[width=0.45\columnwidth]{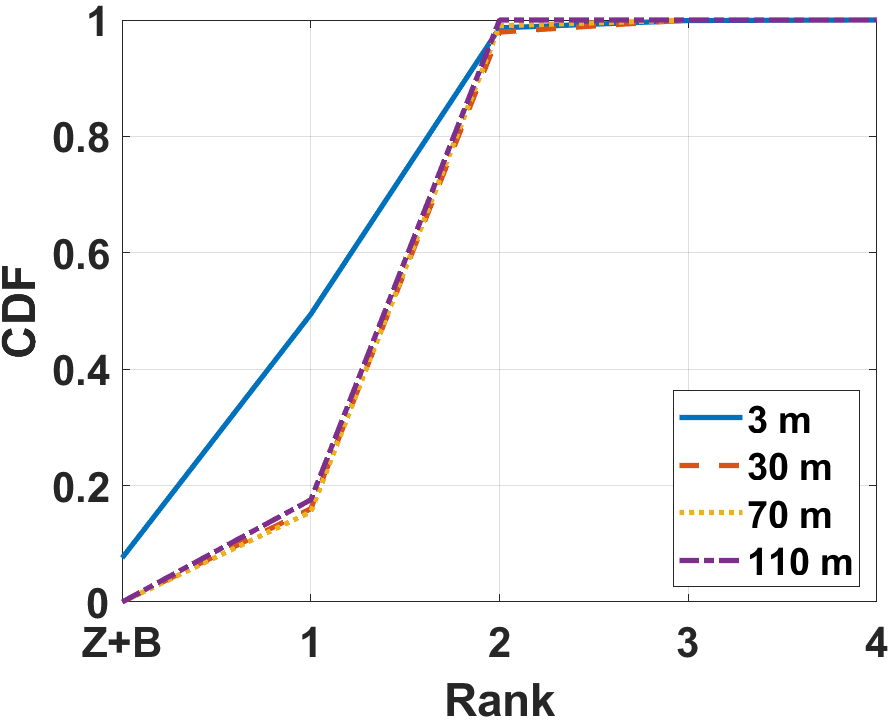}
    \label{fig:CDF_LW_40dB}
    }%\vspace{-2mm}
    \caption{CDFs for altitude-dependent channel rank at CC1, CC2, and LW1, for 3 different singular value thresholding criteria.}
    \label{fig:CDF ranks}\vspace{-0.1cm}
\end{figure*}

In Figs.~\ref{fig:LW_rank_1}-\ref{fig:LW_rank_4}, channel ranks  for various receiver altitudes with LW1 as the transmitter are provided. A blocked signal area due to the presence of a building can be observed in the southeast region in Fig.~\ref{fig:LW_rank_1} for a receiver height of $3$~m. The distribution of the channel rank tends to be higher at receiver locations closer to the LW1 tower, likely due to the presence of paths reflected from the building nearby LW1, as well as the reflected paths from the ground. 

The CDFs for channel rank for CC1, CC2, and LW1 with various different singular value thresholds are shown in Fig.~\ref{fig:CDF ranks}, where $P_{\boldsymbol{Z}}+P_{\boldsymbol{B}}$ denotes the combined probability of the receiver locations with no wireless coverage outside and inside buildings. With $\sigma_{\rm{Thr}}=\sigma_{s, {\rm mean}}$, more than $80\%$  and $50\%$ of receiver locations 
%have 
%$P_{\boldsymbol{Z}}+P_{\boldsymbol{B}}$  
are out of coverage for CC1 and CC2, respectively, while only around $10\%$  of receivers are out of coverage
%have $P_{Z}+P_B$ 
for LW1. %Here, we use a ``zero rank'' to jointly capture indoor receiver locations as well as locations with weak signal coverage. 
We observe that the  probability of having a channel rank of 3 or 4 is very small in all environments and thresholds, especially for the rural LW1 scenario. 
%and that probability is relatively higher when the UAV is being served by CC1 or CC2 (urban) versus being served by LW1 (rural). 
This is as expected since the UAV links typically have a strong LoS multipath component and very few other multipath components, especially for rural scenarios, which results in low-rank channels. 
% The lower altitude receiver locations tend to have a high portion of rank 1. However, this trend slightly vanishes for higher rank, which can be incurred from the geographic properties of the given area. 
%With the constant threshold of $\sigma_{\rm{Thr}}=\sigma/10^2$, most of the cases have a low rank except for LW1 3 m case. 
For $K=10$, close to all channels, excluding $<5\%$ of UE locations around CC1 and CC2 at a UE height of $3$~m, have a channel rank of~$1$ when there is wireless coverage.  
For $K=10^2$, we start observing more UE locations having a channel rank of $2$, but this comes at the cost of observing up to $20$~dB of SNR difference between two spatial layers. 
When $K=10^4$, it is possible to observe channel ranks of up to $4$, but now the range of the SNR over up to four different spatial streams becomes within $40$~dB of the strongest SNR. Urban environments an lower UE altitudes are more likely to observe a higher channel rank.  

\section{Condition Number Analysis}\label{Sec:5}
Simulation settings and assumptions for condition number analysis are identical to the simulation settings in the previous section for rank analysis. In the results,  $\boldsymbol{r1}$ represents the case where the channel rank is one. Due to the use of the singular value threshold in~\eqref{Eq:Thresholding}, the condition number can be undefined when $\sigma_{\rm min}=0$. Since it is less likely to observe a channel rank of 3 and 4 based on the results in Section~\ref{Sec:4}, we investigate the ratio of the first two strongest singular values as a condition number in this section, i.e., $\rm{CN}=\sigma_1/\sigma_2$.
% channel condition number 

\begin{figure*}[h!] 
    \centering
    \subfigure[UAV altitude: 3 m (CC1)]{
    \includegraphics[width=0.45\columnwidth]{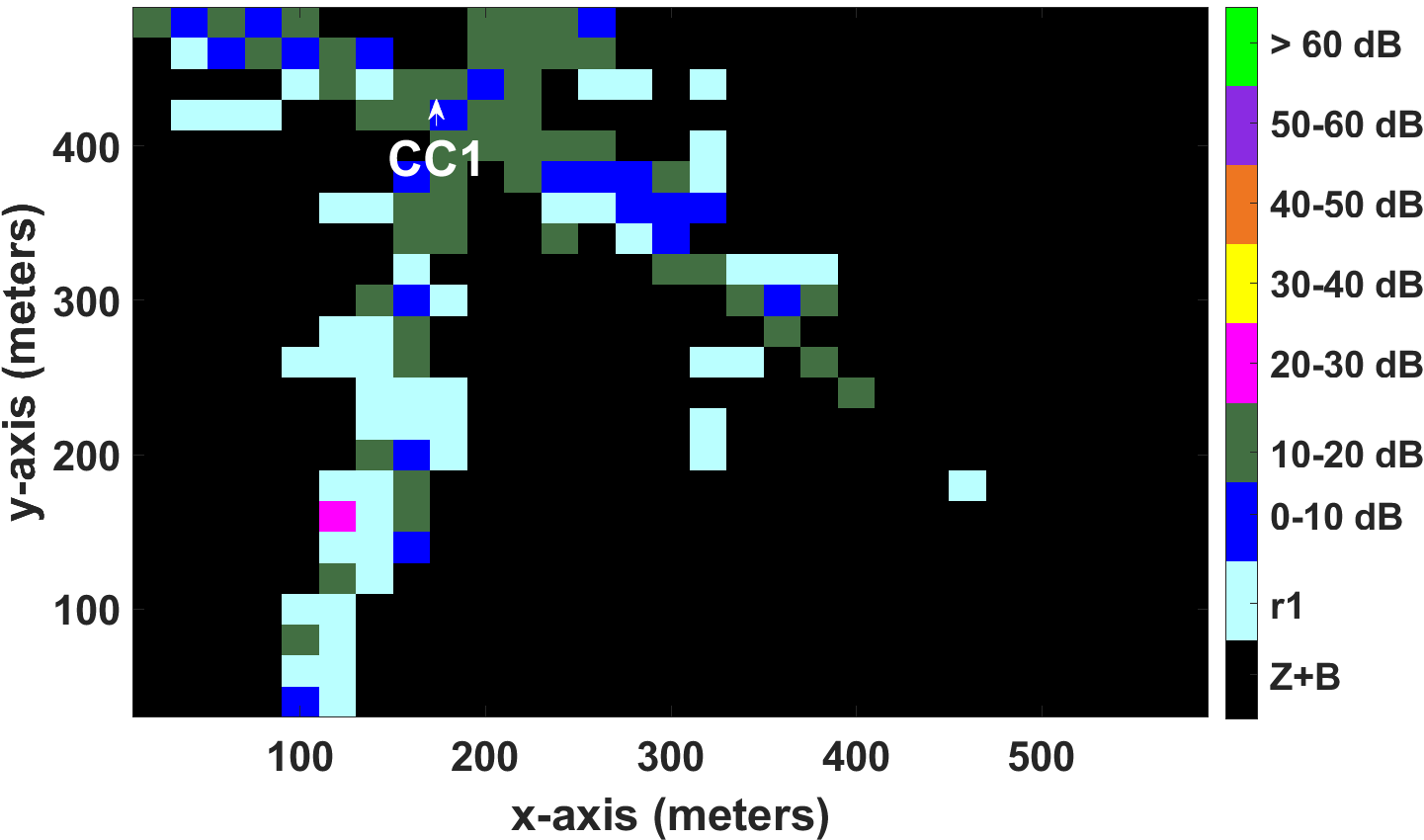}
    \label{fig:CC1_CN_1}    
    }  
    \subfigure[UAV altitude: 30 m (CC1)]{
    \includegraphics[width=0.45\columnwidth]{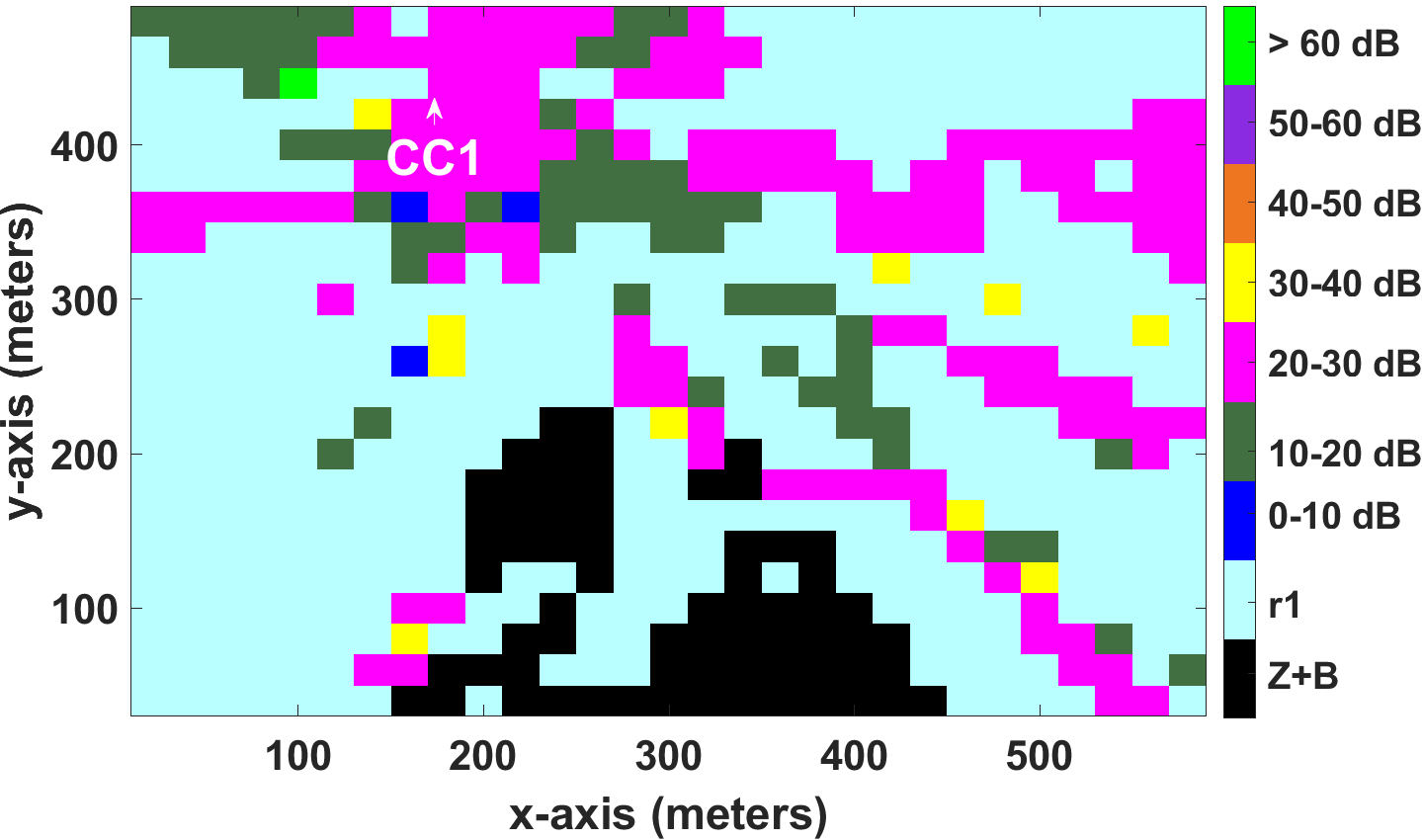}
    \label{fig:CC1_CN_2}
    }
    \subfigure[UAV altitude: 70 m (CC1)]{
    \includegraphics[width=0.45\columnwidth]{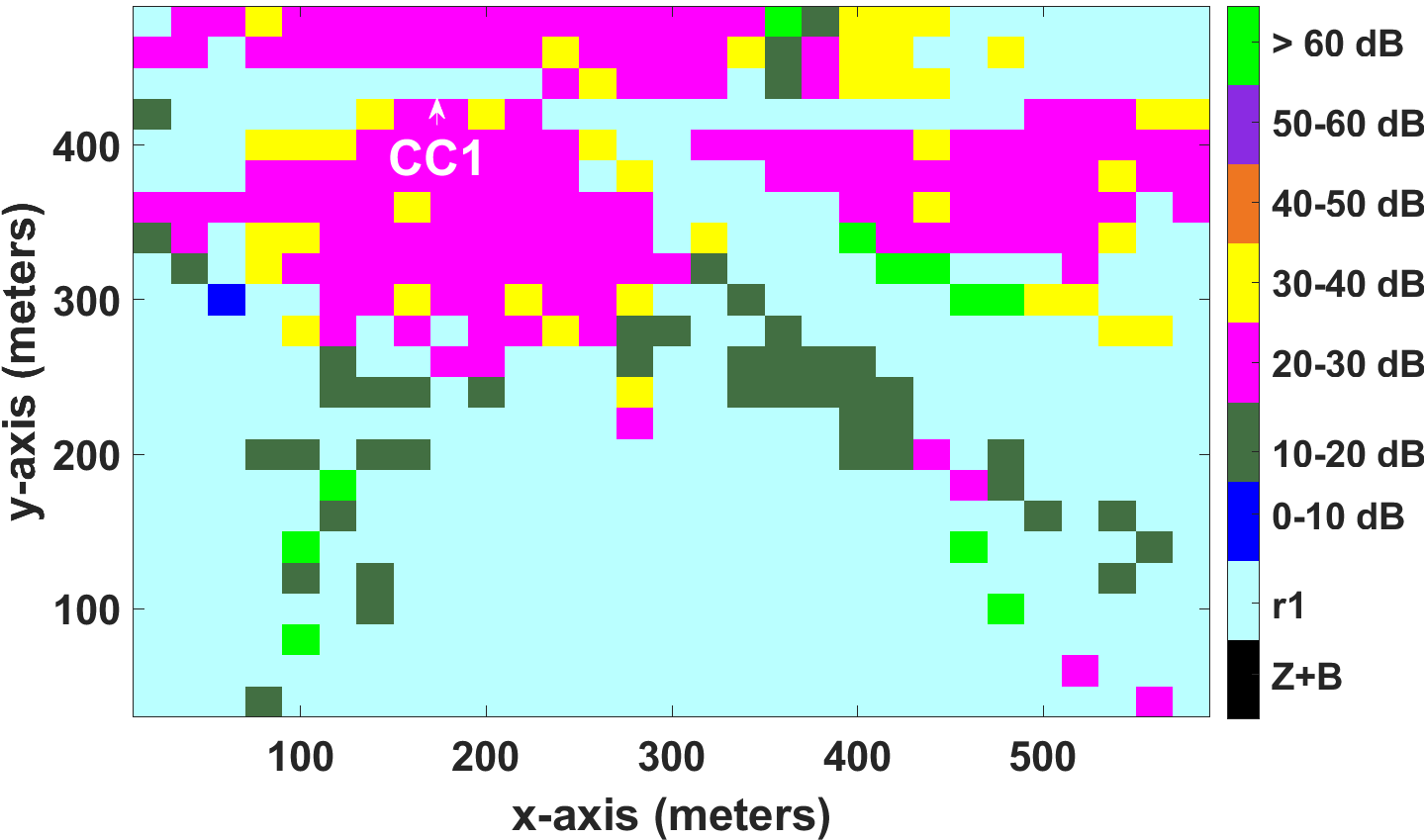}
    \label{fig:CC1_CN_3}
    }
    \subfigure[UAV altitude: 110 m (CC1)]{
    \includegraphics[width=0.45\columnwidth]{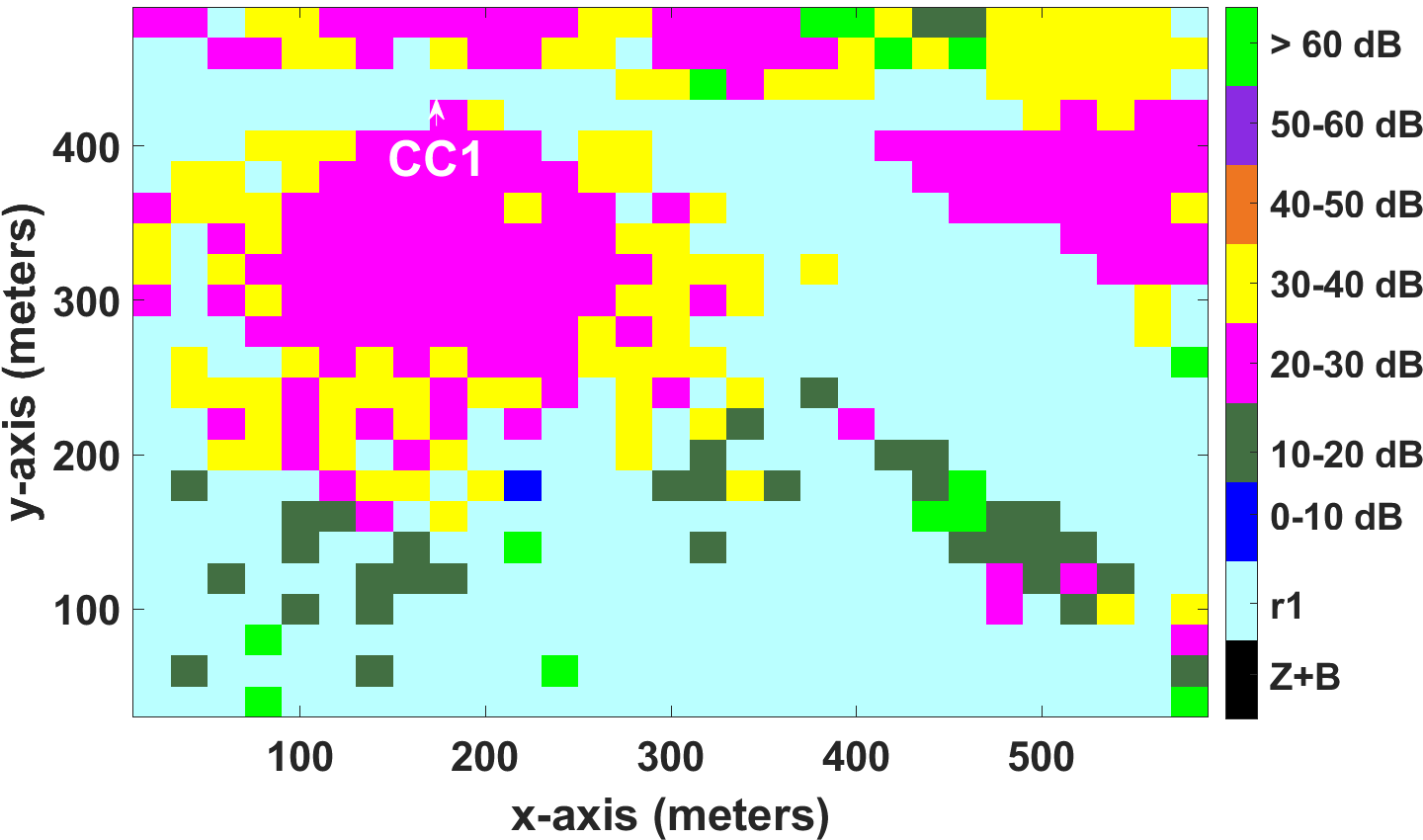}
    \label{fig:CC1_CN_4}
    }
    \subfigure[UAV altitude: 3 m (CC2)]{
    \includegraphics[width=0.45\columnwidth]{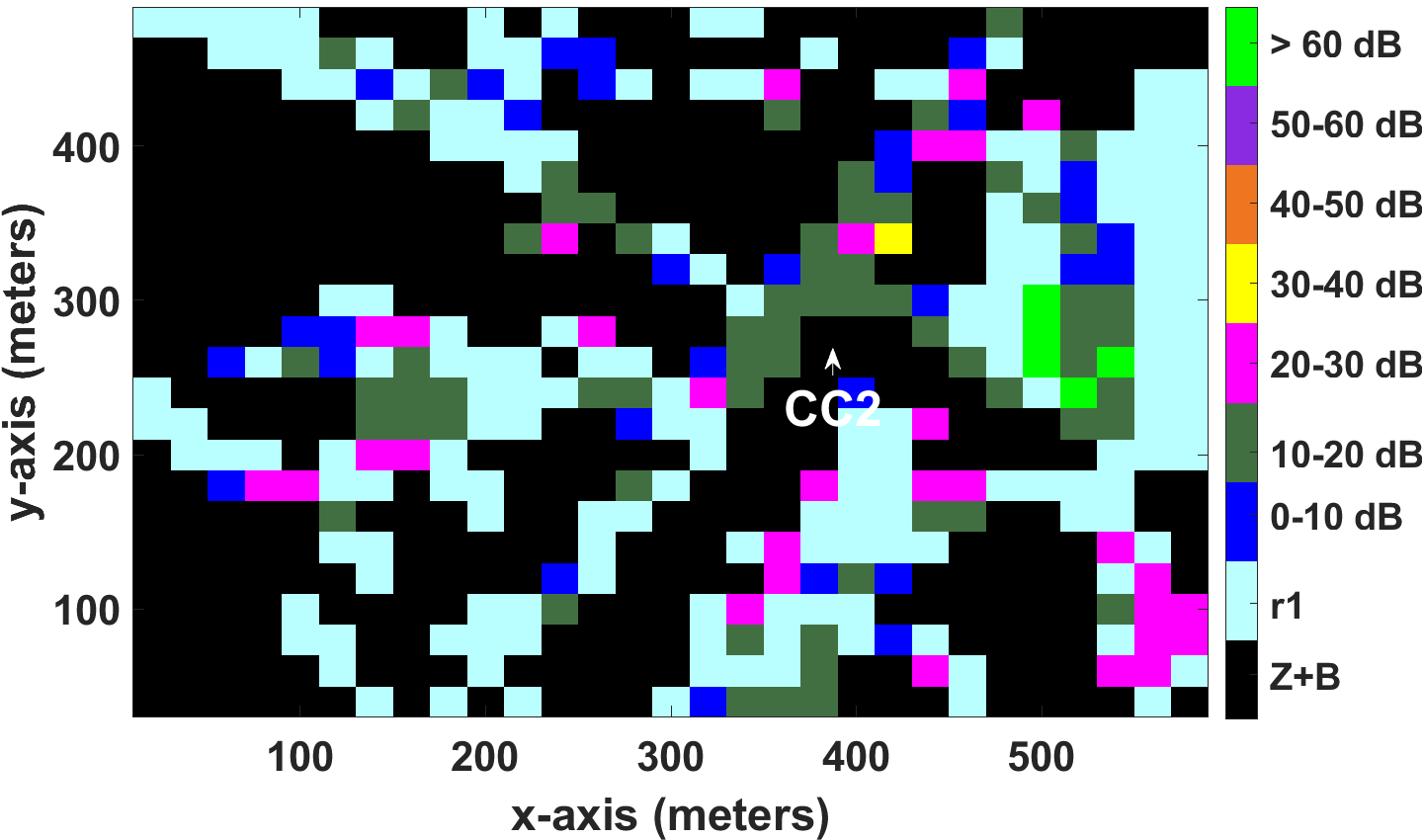}
    \label{fig:CC2_CN_1}
    }
    \subfigure[UAV altitude: 30 m (CC2)]{
    \includegraphics[width=0.45\columnwidth]{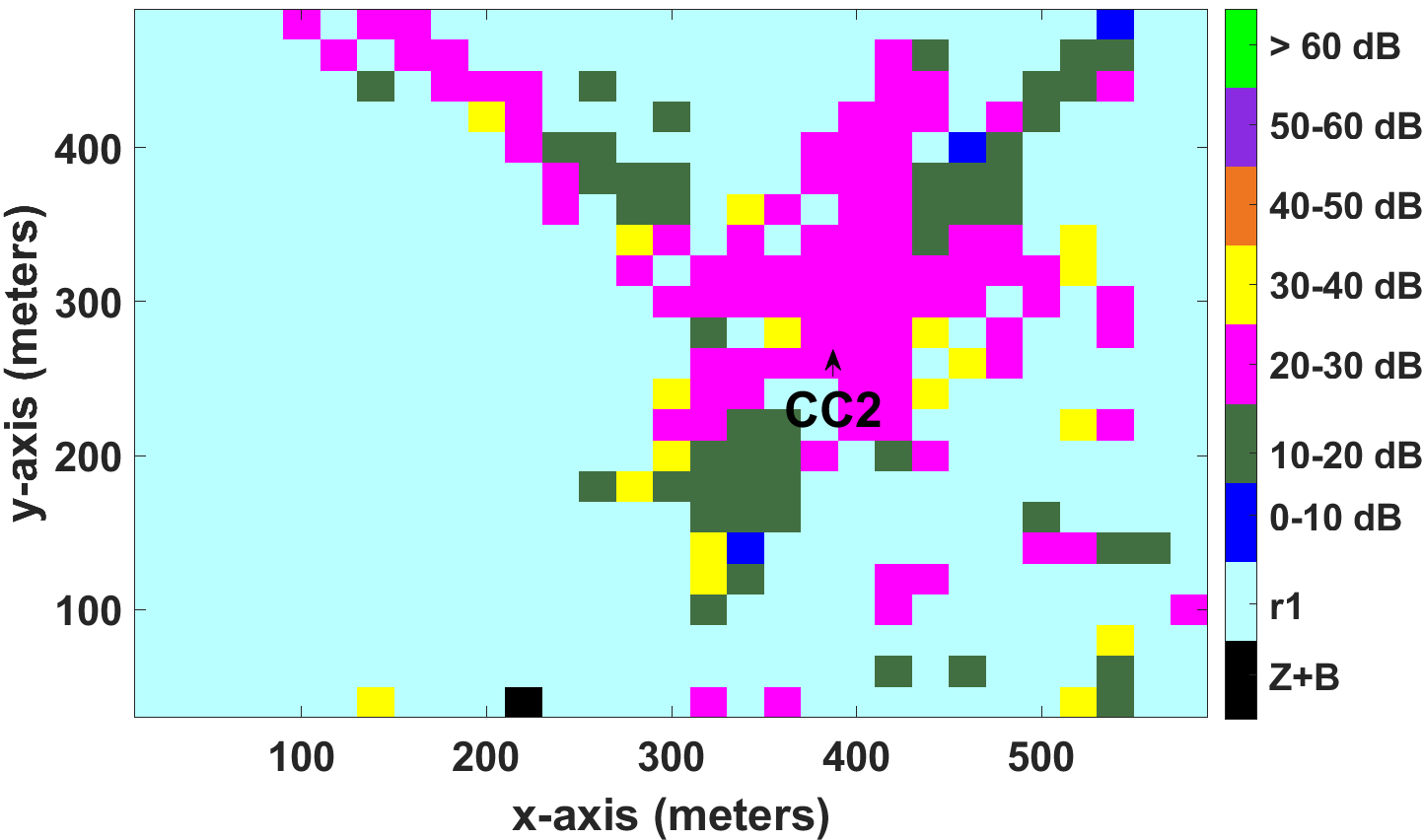}
    \label{fig:CC2_CN_2}
    }
    \subfigure[UAV altitude: 70 m (CC2)]{
    \includegraphics[width=0.45\columnwidth]{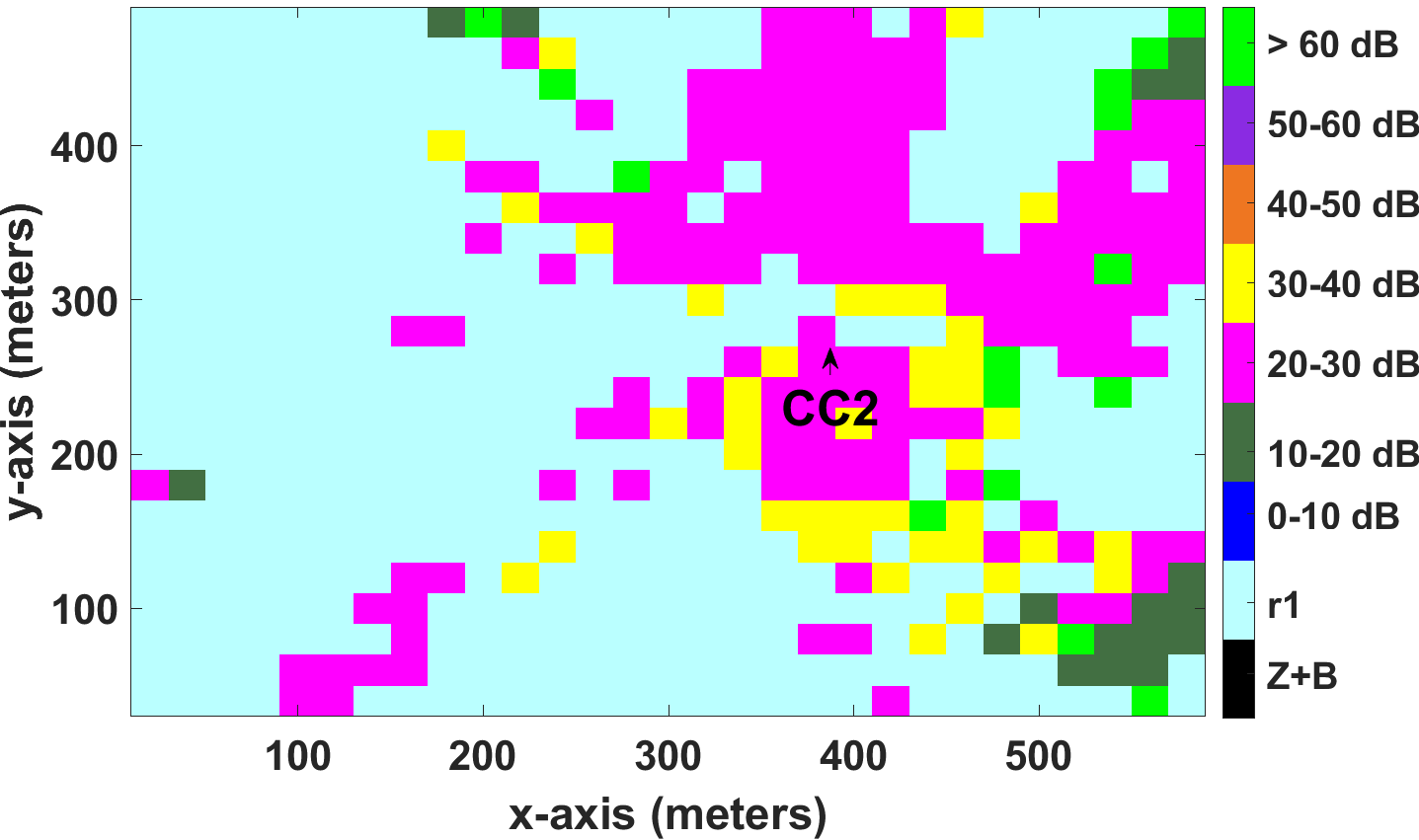}
    \label{fig:CC2_CN_3}
    }
    \subfigure[UAV altitude: 110 m (CC2)]{
    \includegraphics[width=0.45\columnwidth]{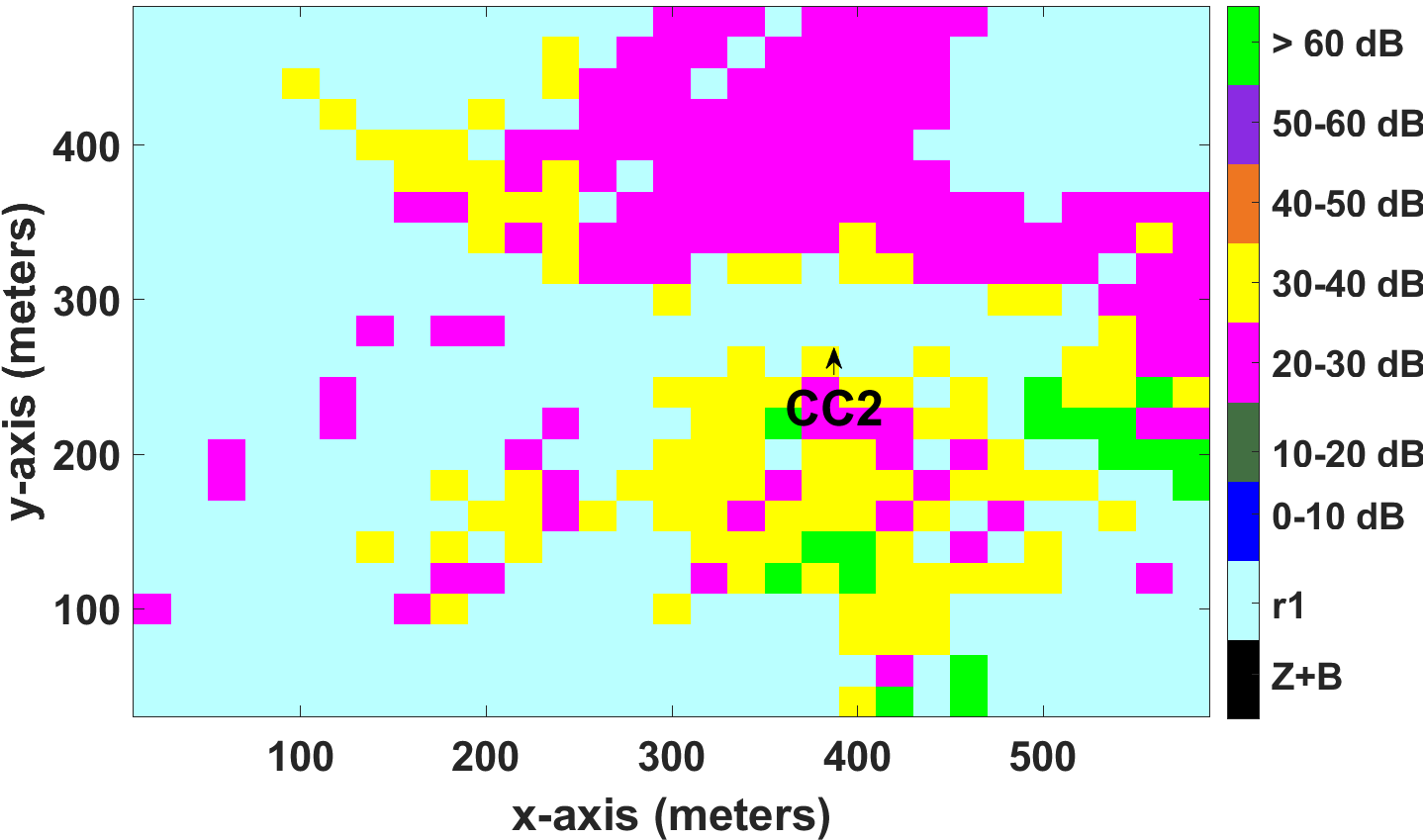}
    \label{fig:CC2_CN_4}
    }
    \subfigure[UAV altitude: 3 m (LW1)]{
    \includegraphics[width=0.45\columnwidth]{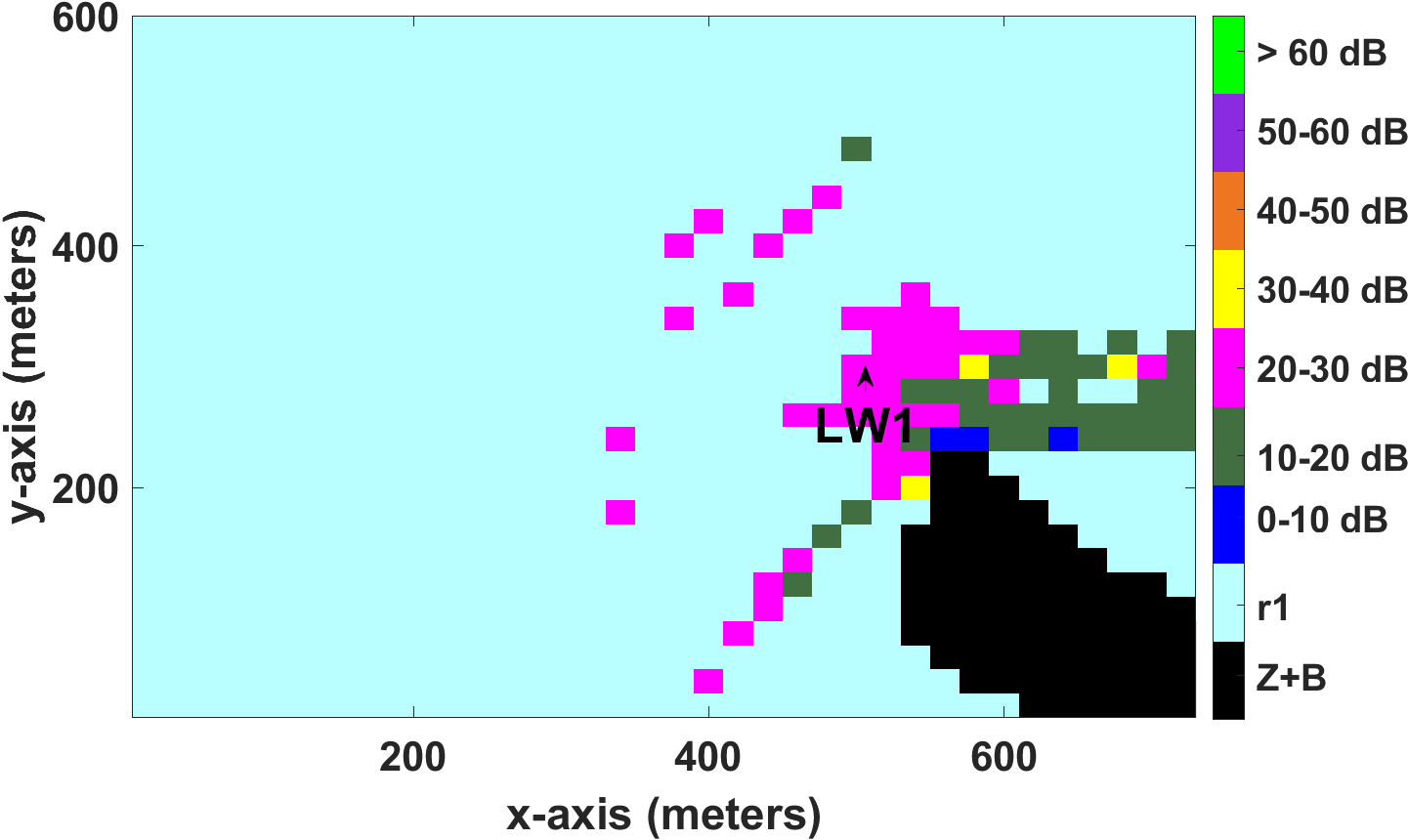}
    \label{fig:LW_CN_1}
    }
    \subfigure[UAV altitude: 30 m (LW1)]{
    \includegraphics[width=0.45\columnwidth]{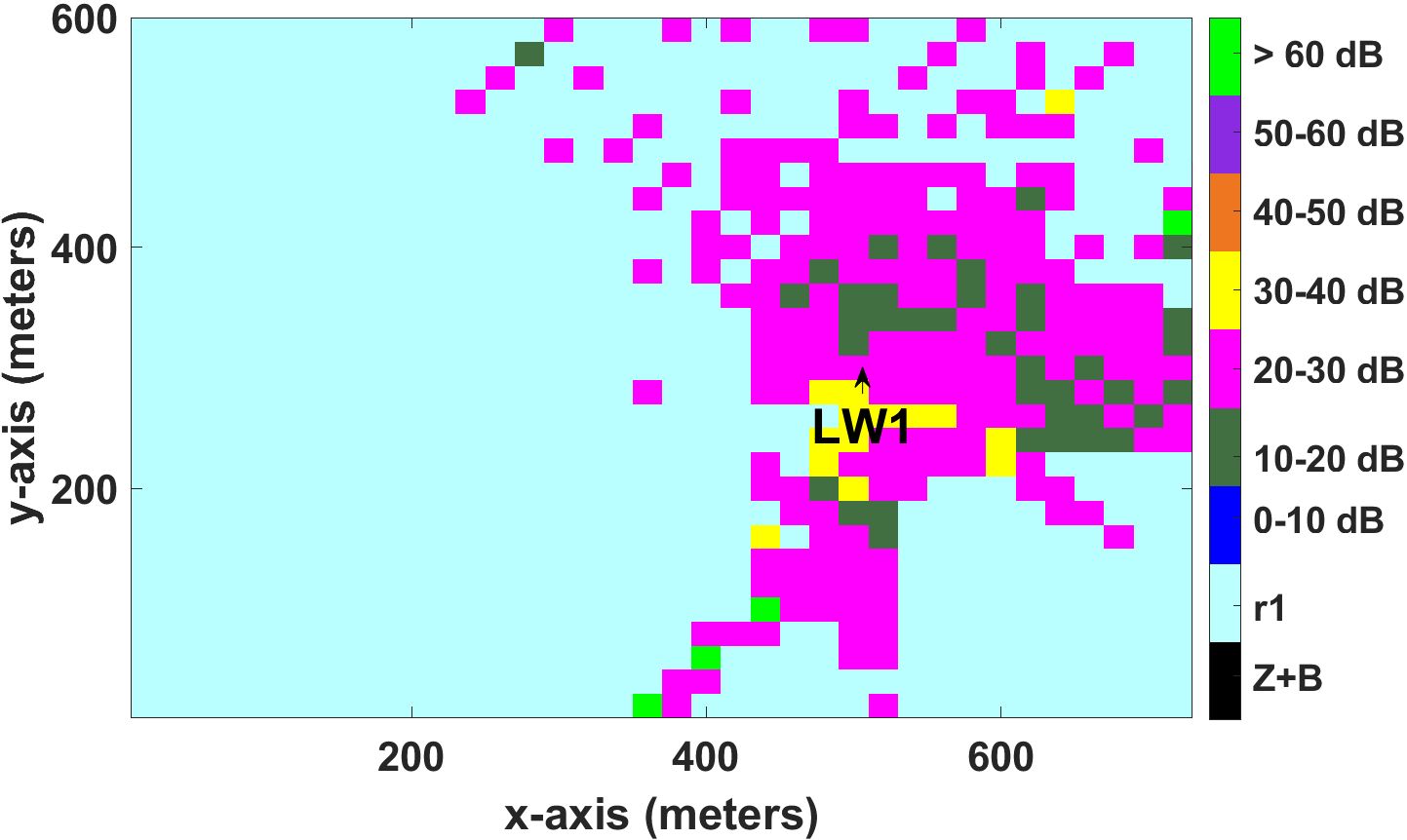}
    \label{fig:LW_CN_2}
    }
    \subfigure[UAV altitude: 70 m (LW1)]{
    \includegraphics[width=0.45\columnwidth]{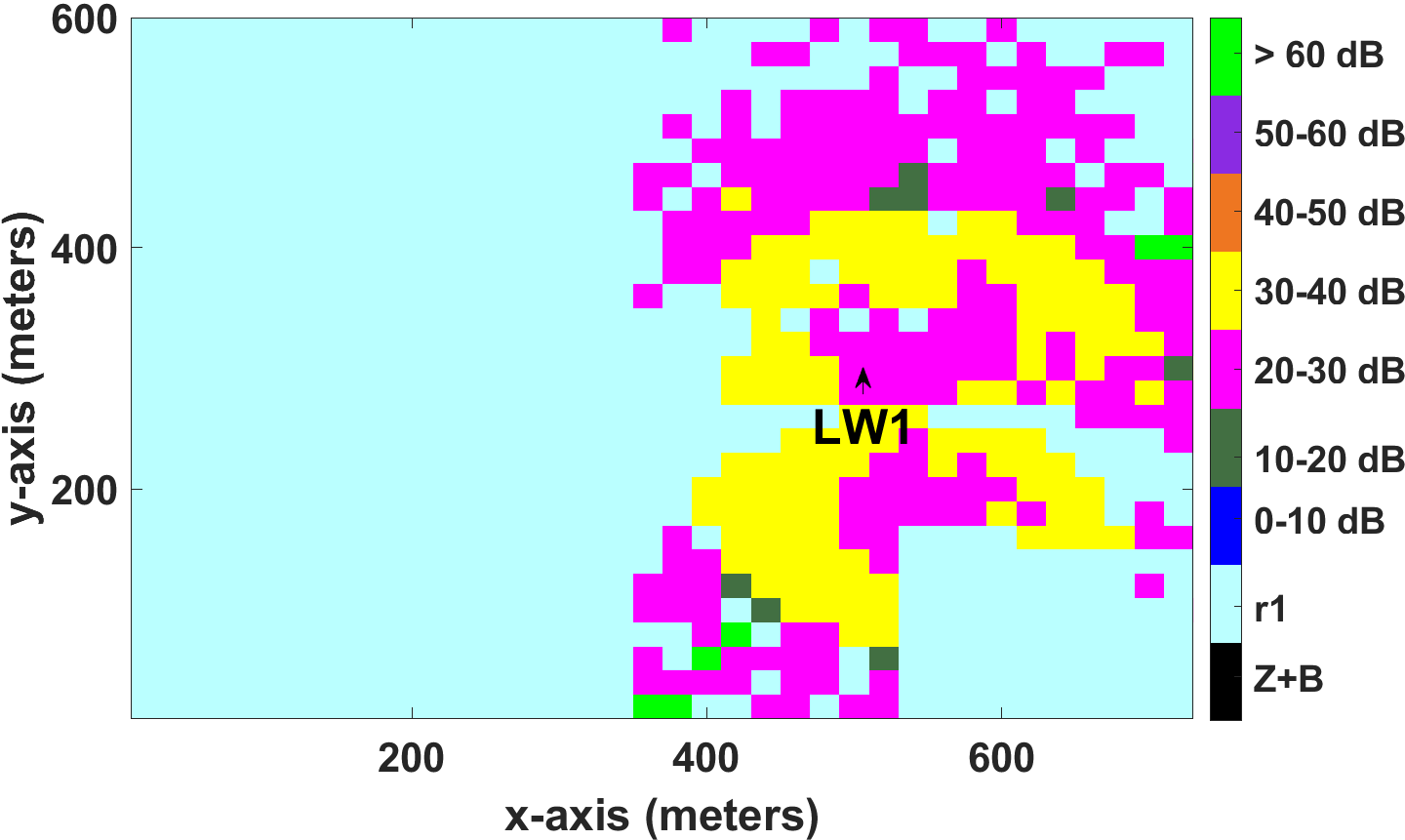}
    \label{fig:LW_CN_3}
    }
    \subfigure[UAV altitude: 110 m (LW1)]{
    \includegraphics[width=0.45\columnwidth]{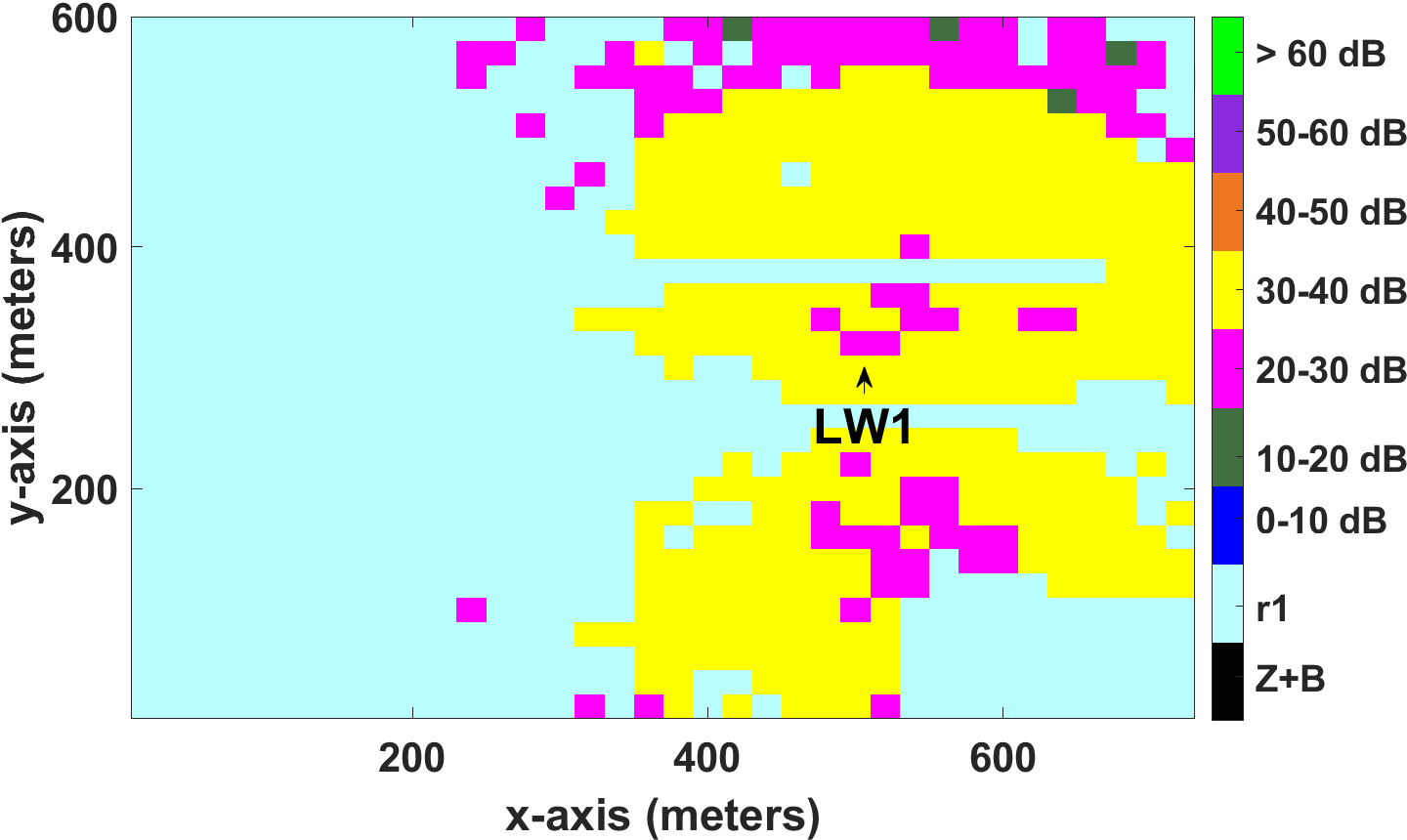}
    \label{fig:LW_CN_4}
    }
    
    \caption{Condition numbers in Centennial Campus with CC1 (a)-(d), CC2 (e)-(h), and in Lake Wheeler with LW1 (i)-(l) as the cellular BS (see Fig.~\ref{fig:my_label}). The condition number is seen to be correlated across different altitudes for a given transmitter.}
    \label{fig:CN_CC1_CC2_LW1}
\end{figure*}

The simulation results on  $\sigma_1/\sigma_2$ with $\sigma_{\rm{Thr}}=\sigma_{s, mean}$ for CC1 are given in Figs.~\ref{fig:CC1_CN_1}-\ref{fig:CC1_CN_4} for four different UE altitudes. Some receiver sites close to the transmitter have condition numbers around $20$~dB for the $3$~m UAV case. The number of receiver locations that have a higher condition number increases as the height of the UAV increases. This is as expected, since fewer multipath components and a stronger LoS dominance are expected at higher altitudes. 
%Moreover, the ratio of receiver locations with high condition numbers tends to increase as the altitude of the receiver increases. 
%This can be from the higher rank distribution with higher altitude cases having high differences among singular values of each order. 
The same trends can be observed with the CC2 transmitter, for which the condition number is shown in Figs.~\ref{fig:CC2_CN_1}-\ref{fig:CC2_CN_4}. The different condition number distribution for CC1 and CC2 are due to the geographical difference for the transmitter-receiver link pairs for those sites. 
In Figs.~\ref{fig:LW_CN_1}-\ref{fig:LW_CN_4}, simulation results on the condition number with LW1 as the transmitter are shown. The receiver locations having a high condition number are sporadically distributed around the center area for the $3$~m UAV case. This can be from reflective paths through the building or vegetation ground area. The areas that have condition numbers up to $40$~dB become wider at higher altitudes. %and having condition numbers greater than 50 dB becomes lesser as the altitude of the UAV increases.

% The distributions of the ratio of the first and the second strongest singular values for $\sigma_{\rm{Thr}}=\sigma_{s, {\rm mean}}$ are shown in Fig.~\ref{fig:CDFs_s1s2}. For all transmitter settings, CDFs show higher \textcolor{red}{$\sigma_1/\sigma_2$} ratios with the higher altitude. Moreover, the corresponding combined portion of the receiver without a link and the receiver of rank 1, $P_{\rm{o}}+P_{\rm{r1}}$, also follows the same pattern as the altitude increases. Further breakdown of $P_{\rm o}$ will be provided later in this section. 
%($\sigma_1/\sigma_4$) 

\begin{figure*}[t!]
    \centering
    \subfigure[CC1, $\sigma_{\rm Thr} = \sigma_{s,{\rm mean}}$]{
    \includegraphics[width=0.45\columnwidth]{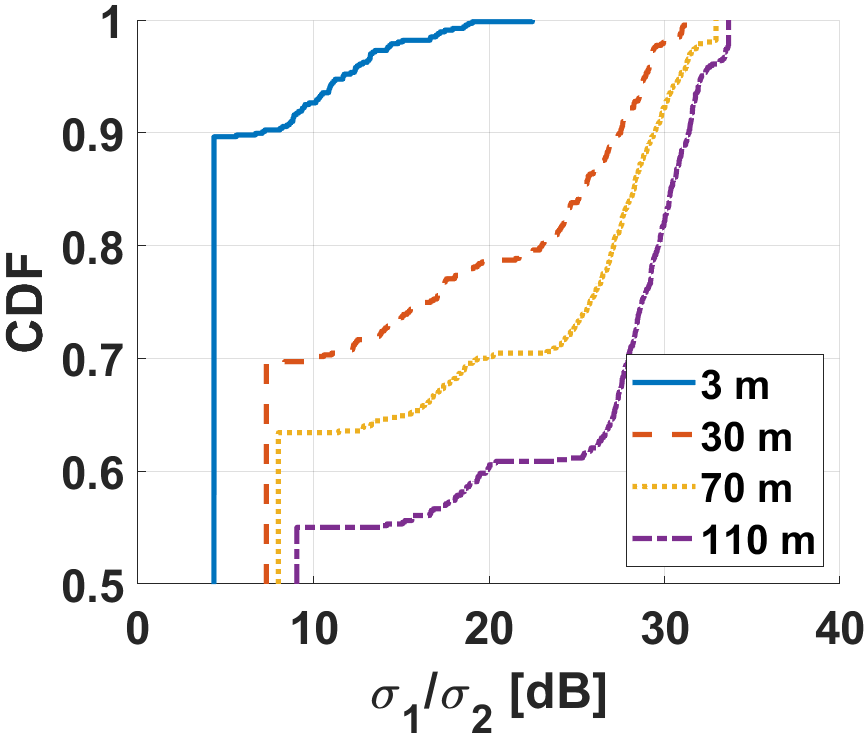}
    \label{fig:CN_CC1}
    }
    \subfigure[CC1, $\sigma_{\rm Thr} = \sigma_{1}/10$]{
    \includegraphics[width=0.45\columnwidth]{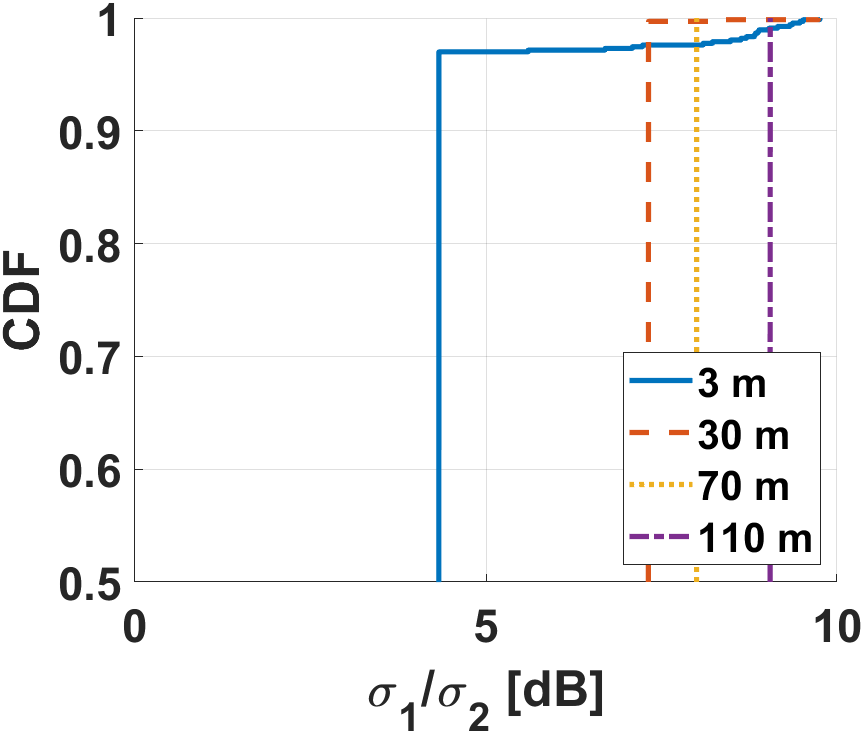}
    \label{fig:CN_CC1_10dB}
    }
    \subfigure[CC1, $\sigma_{\rm Thr} = \sigma_{1}/10^2$]{
    \includegraphics[width=0.45\columnwidth]{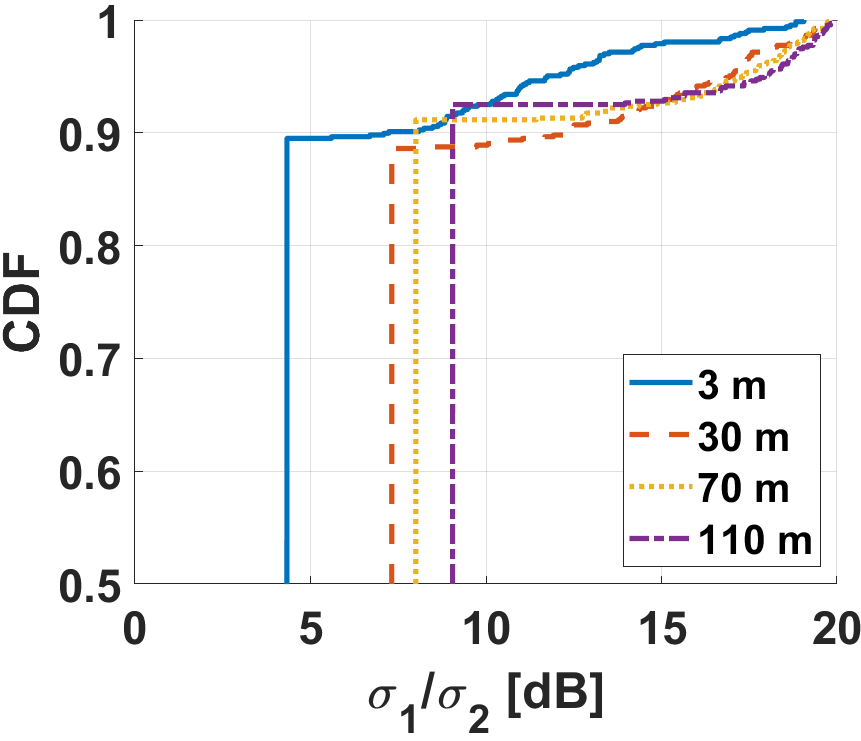}
    \label{fig:CN_CC1_20dB}
    }
    \subfigure[CC1, $\sigma_{\rm Thr} = \sigma_{1}/10^4$]{
    \includegraphics[width=0.45\columnwidth]{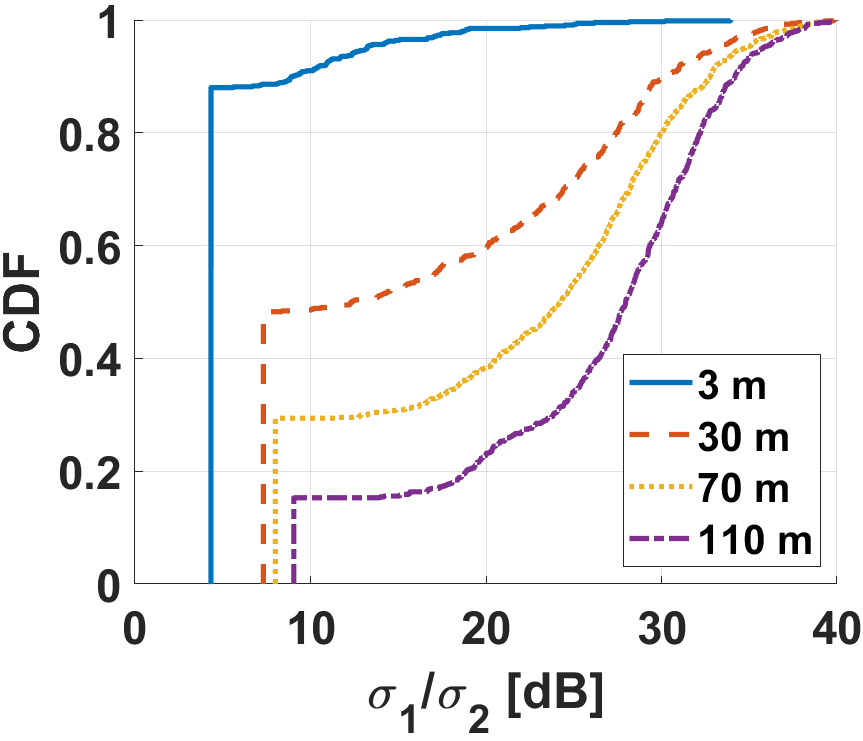}
    \label{fig:CN_CC1_40dB}
    }
    \subfigure[CC2, $\sigma_{\rm Thr} = \sigma_{s,{\rm mean}}$]{
    \includegraphics[width=0.45\columnwidth]{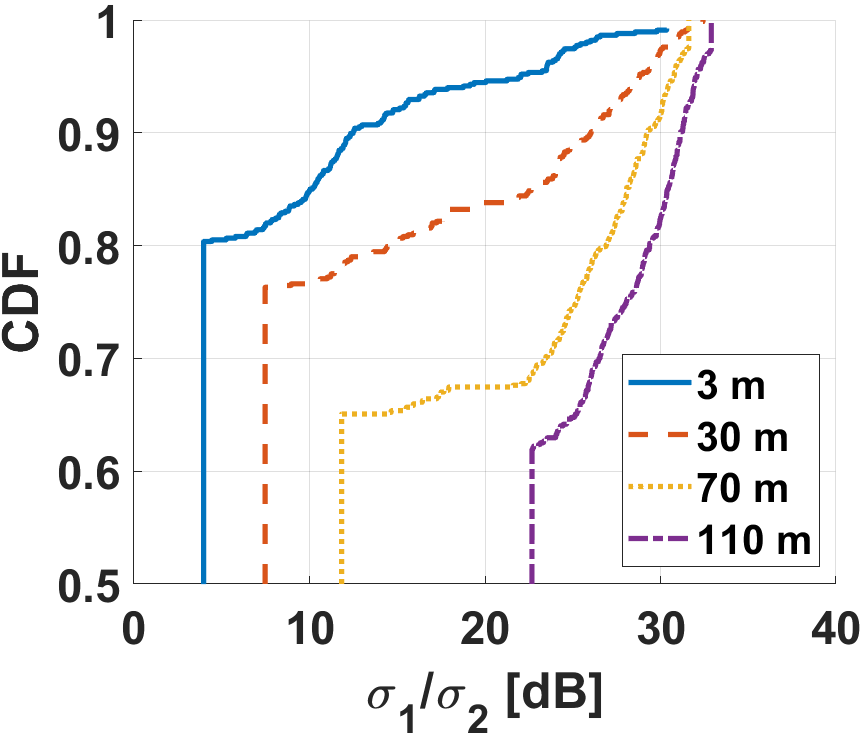}
    \label{fig:CN_CC2}
    }
    \subfigure[CC2, $\sigma_{\rm Thr} = \sigma_{1}/10$]{
    \includegraphics[width=0.45\columnwidth]{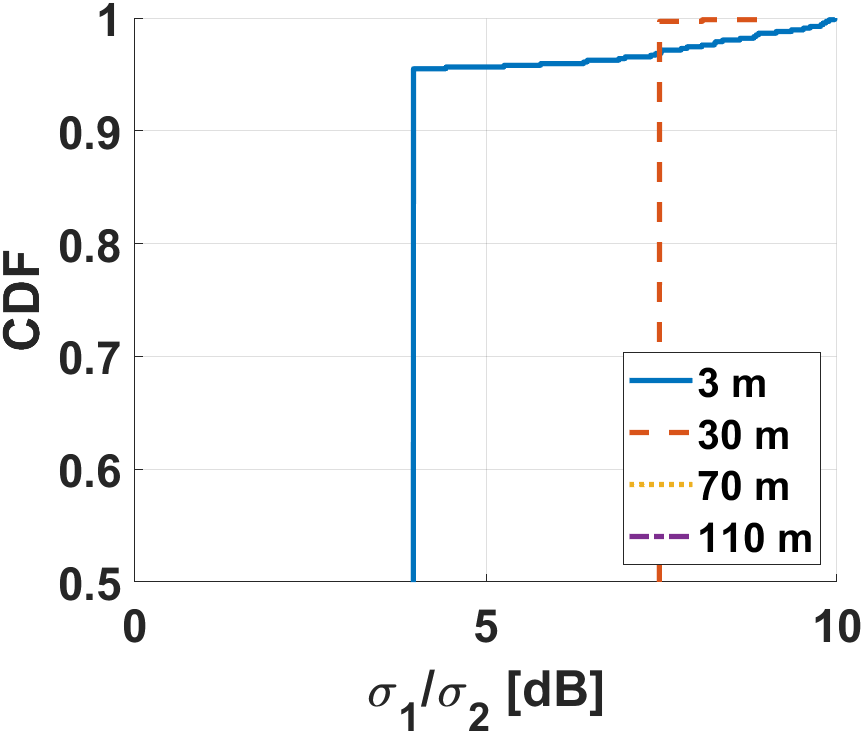}
    \label{fig:CN_CC2_10dB}
    }
    \subfigure[CC2, $\sigma_{\rm Thr} = \sigma_{1}/10^2$]{
    \includegraphics[width=0.45\columnwidth]{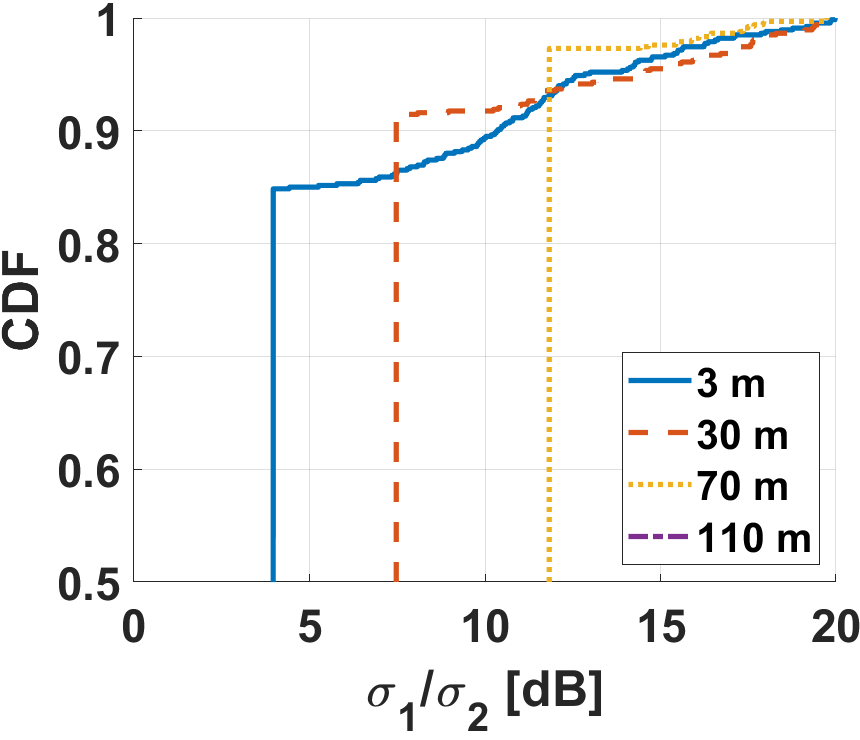}
    \label{fig:CN_CC2_20dB}
    }
    \subfigure[CC2, $\sigma_{\rm Thr} = \sigma_{1}/10^4$]{
    \includegraphics[width=0.45\columnwidth]{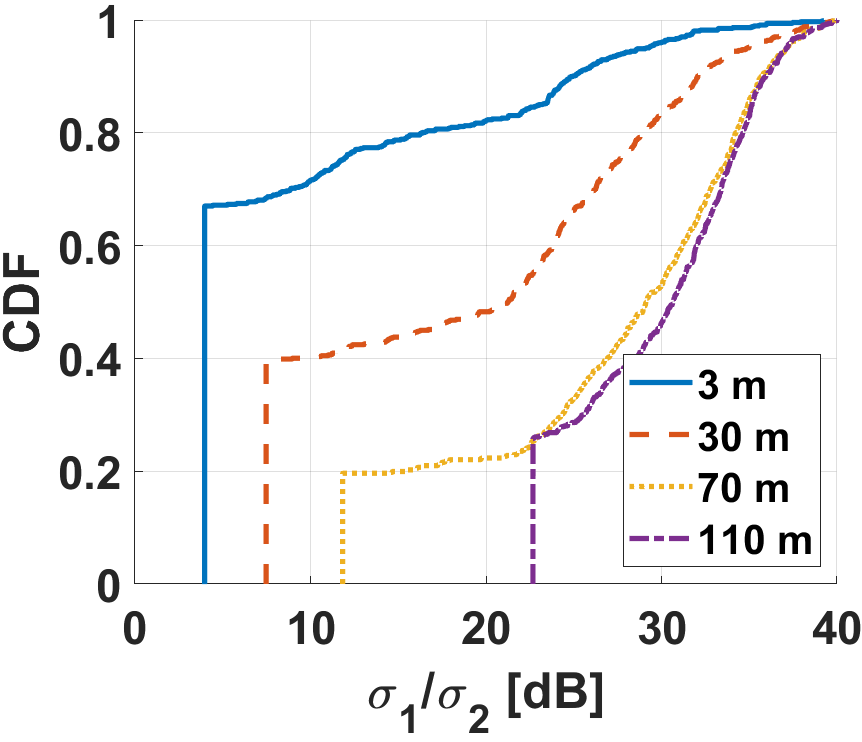}
    \label{fig:CN_CC2_40dB}
    }
    \subfigure[LW1, $\sigma_{\rm Thr} = \sigma_{s,{\rm mean}}$]{
    \includegraphics[width=0.45\columnwidth]{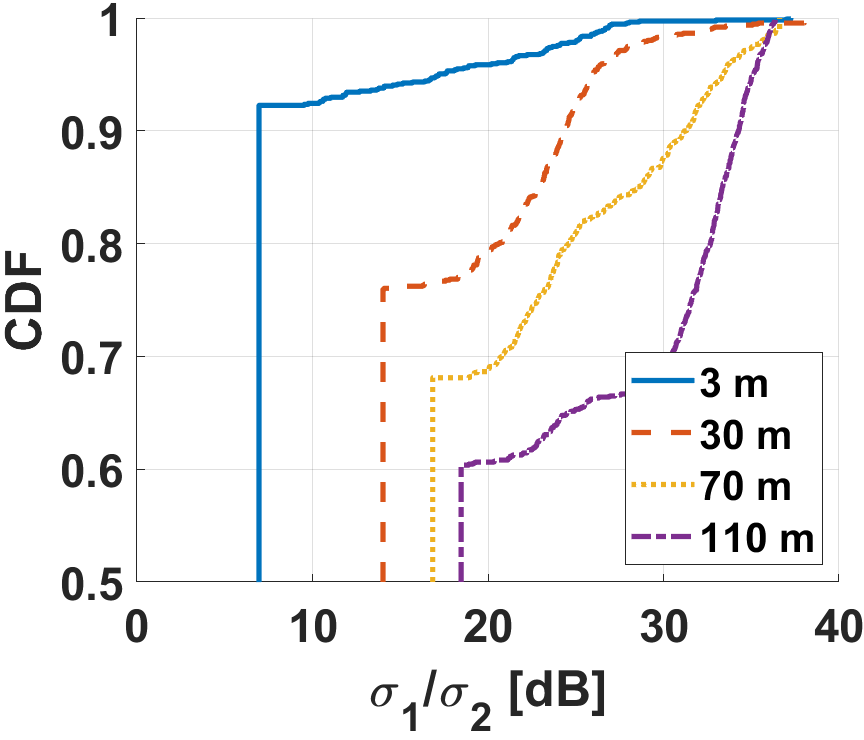}
    \label{fig:CN_LW}
    }   
    \subfigure[LW1, $\sigma_{\rm Thr} = \sigma_{1}/10$]{
    \includegraphics[width=0.45\columnwidth]{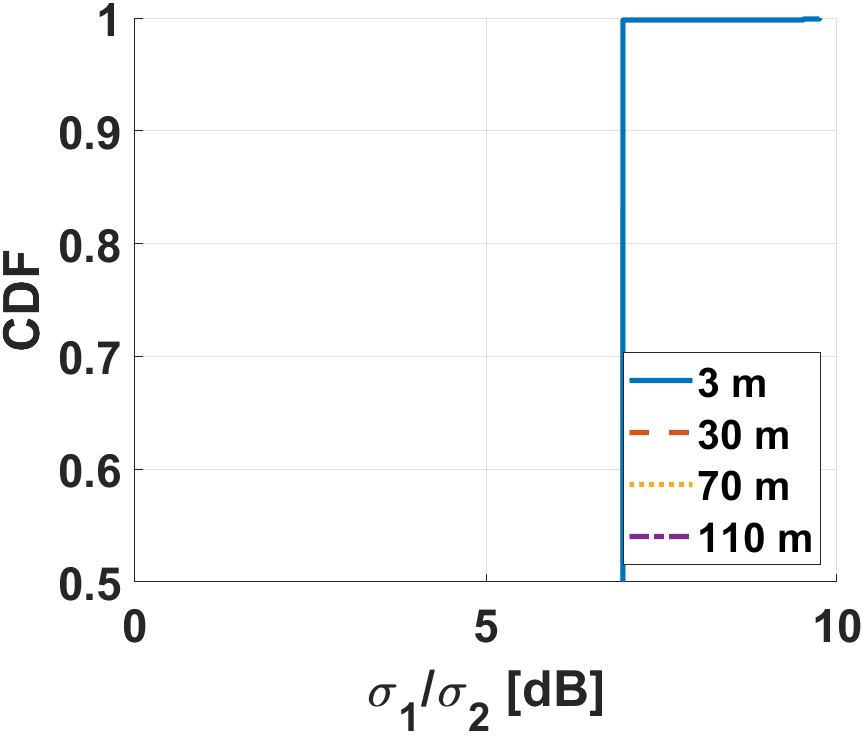}
    \label{fig:CN_LW_10dB}
    }    
    \subfigure[LW1, $\sigma_{\rm Thr} = \sigma_{1}/10^2$]{
    \includegraphics[width=0.45\columnwidth]{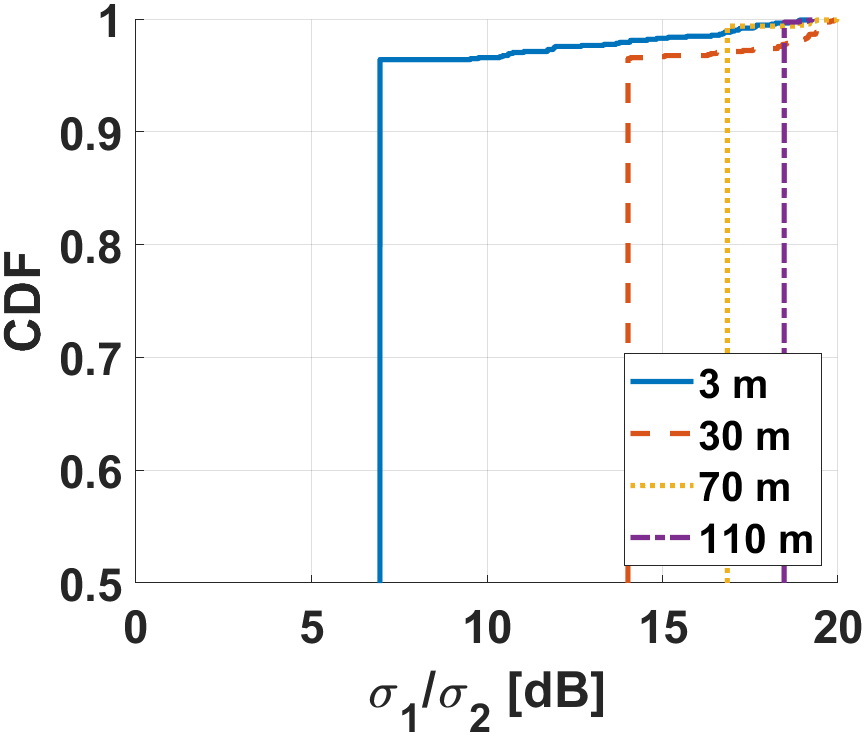}
    \label{fig:CN_LW_20dB}
    }    
    \subfigure[LW1, $\sigma_{\rm Thr} = \sigma_{1}/10^4$]{
    \includegraphics[width=0.45\columnwidth]{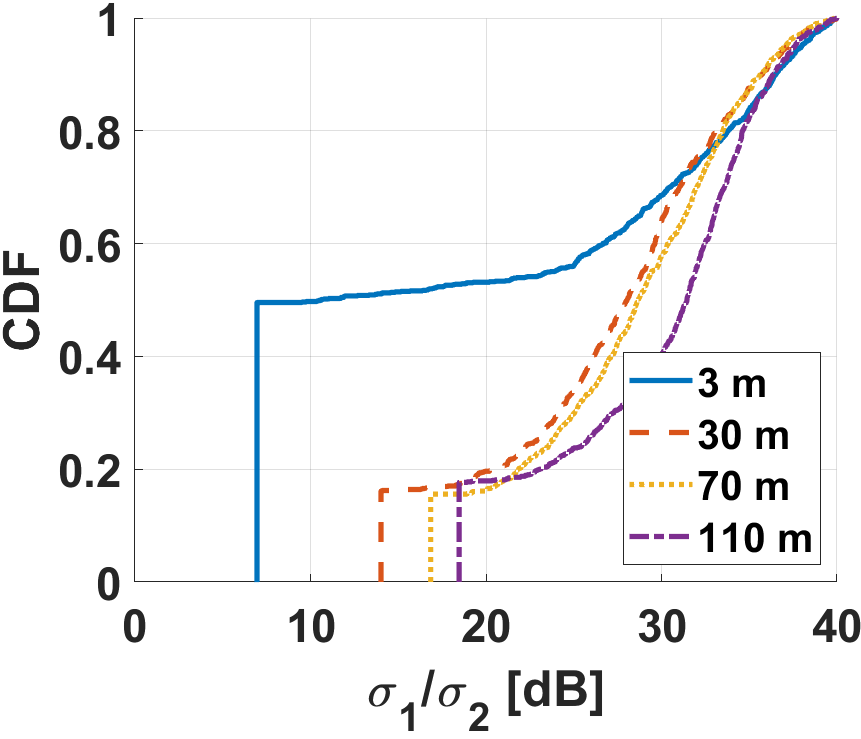}
    \label{fig:CN_LW_40dB}
    }
    % \subfigure[CC1 $\rm{P_o}$ (first left) and $\rm{P_{r1}}$]{
    % \includegraphics[width=0.59\columnwidth]{images/1s/CC1_CDF_bars.png}
    % \label{fig:CDF_bars_CC1}
    % }
    % \subfigure[CC2 $\rm{P_o}$ (first left) and $\rm{P_{r1}}$]{
    % \includegraphics[width=0.59\columnwidth]{images/1s/CC2_CDF_bars.png}
    % \label{fig:CDF_bars_CC2}
    % }
    % \subfigure[LW1 $\rm{P_o}$ (first left) and $\rm{P_{r1}}$]{
    % \includegraphics[width=0.59\columnwidth]{images/1s/LW_CDF_bars.png}
    % \label{fig:CDF_bars_LW1}
    % }
    \vspace{-1mm}
    \caption{CDFs for altitude-dependent $\sigma_1/\sigma_2$ at CC1, CC2, and LW1, for 4 different  thresholding criteria.}
    \label{fig:CDFs CN}    
\end{figure*}

\begin{figure*}
    \centering\vspace{-1mm}
    \subfigure[$\sigma_{\rm{Thr}}=\sigma_{s, \rm{mean}}$]{
    \includegraphics[width=0.45\columnwidth]{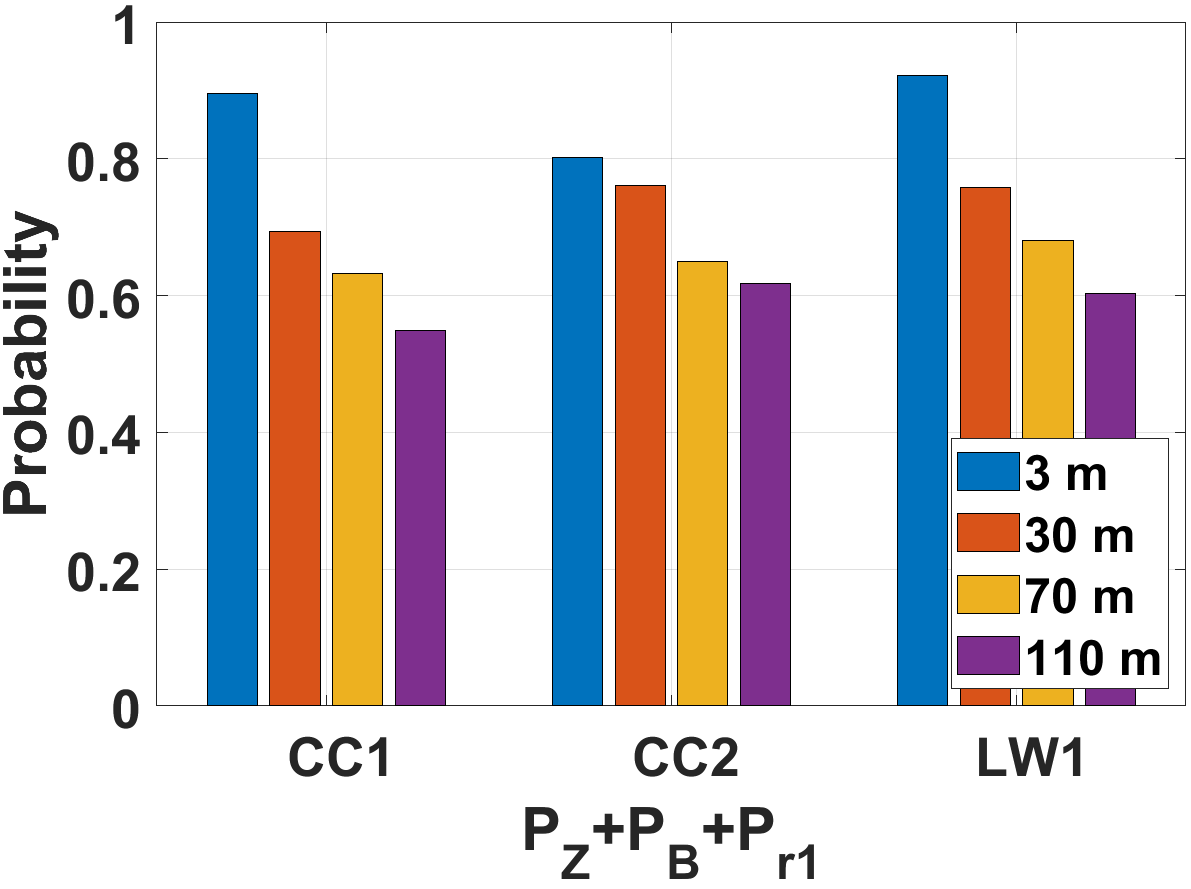}
    \label{fig:CDF_bars_1}
    }
    \subfigure[$\sigma_{\rm{Thr}}=\sigma_1/10$]{
    \includegraphics[width=0.45\columnwidth]{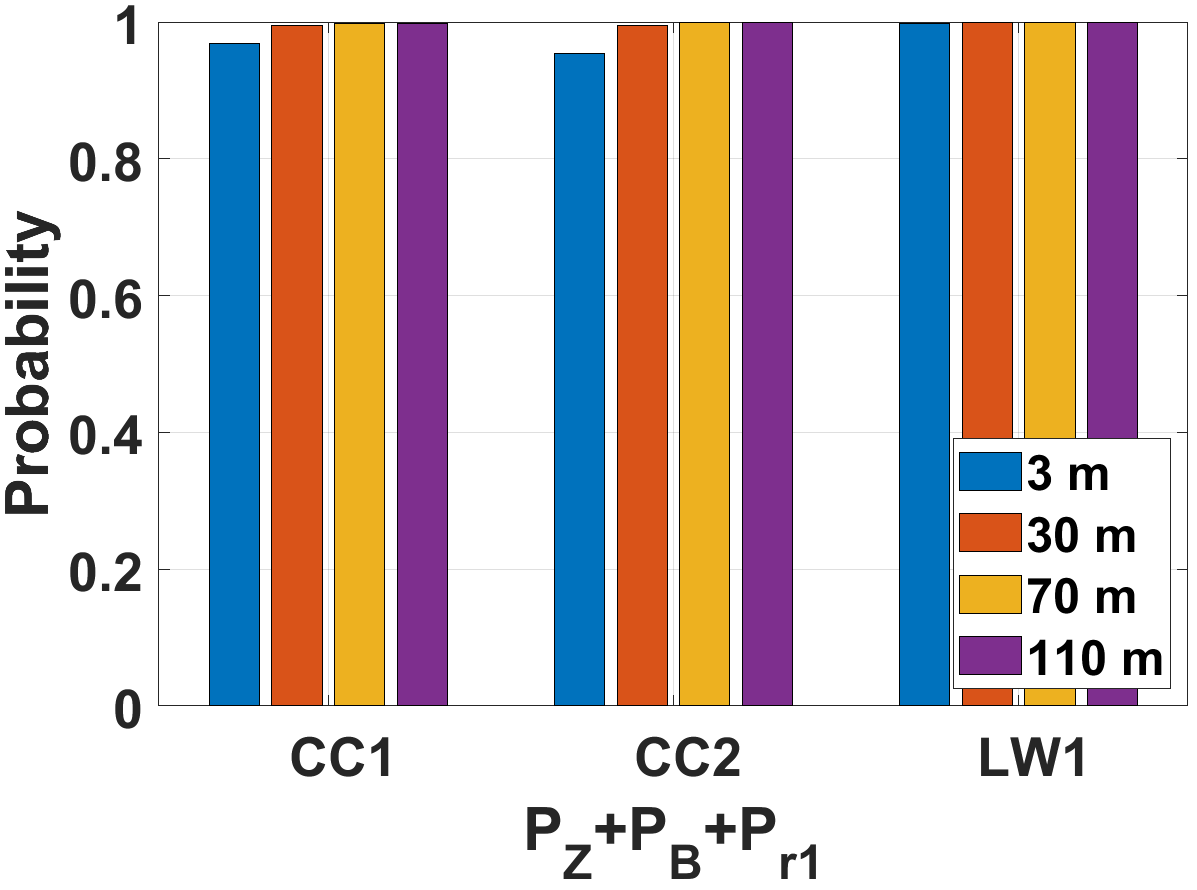}
    \label{fig:CDF_bars_2}
    }
    \subfigure[$\sigma_{\rm{Thr}}=\sigma_1/10^2$]{
    \includegraphics[width=0.45\columnwidth]{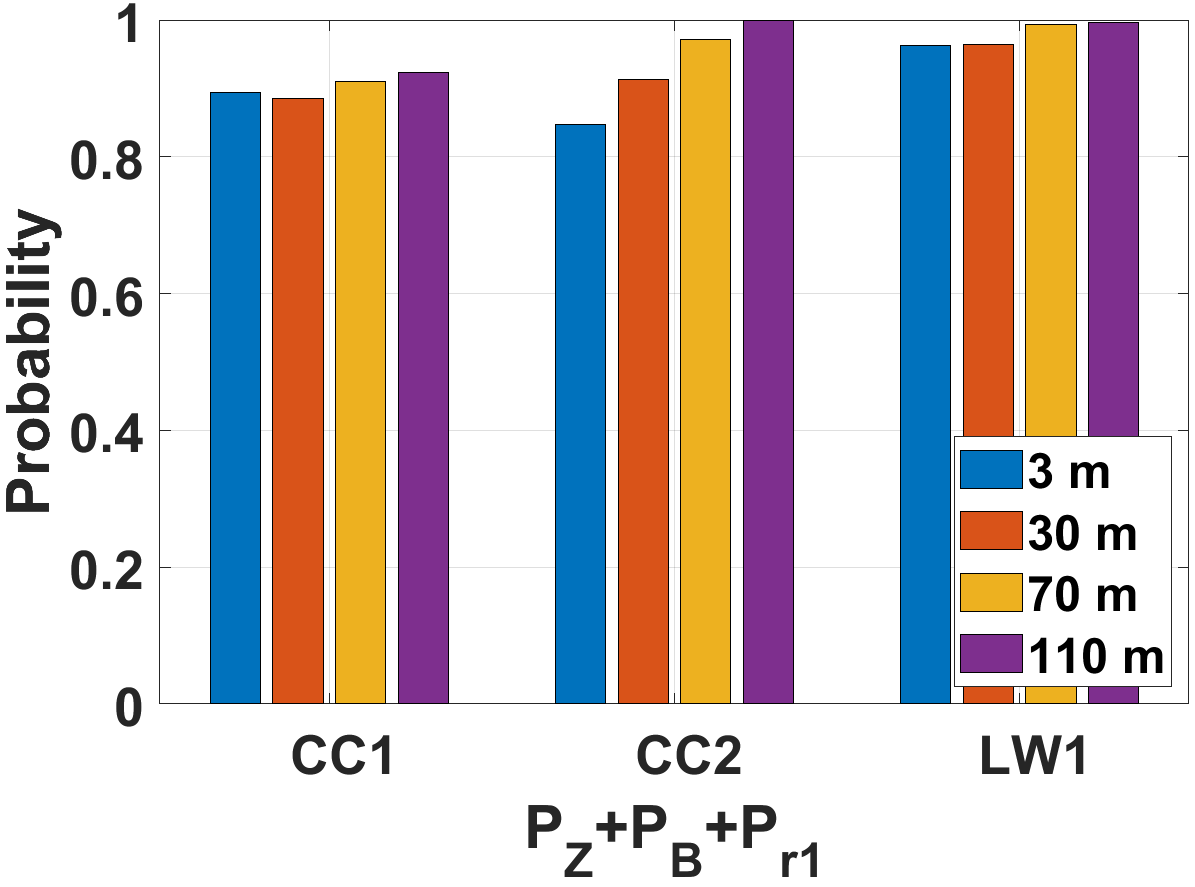}
    \label{fig:CDF_bars_3}
    }\vspace{-1mm}
    \subfigure[$\sigma_{\rm{Thr}}=\sigma_1/10^4$]{
    \includegraphics[width=0.45\columnwidth]{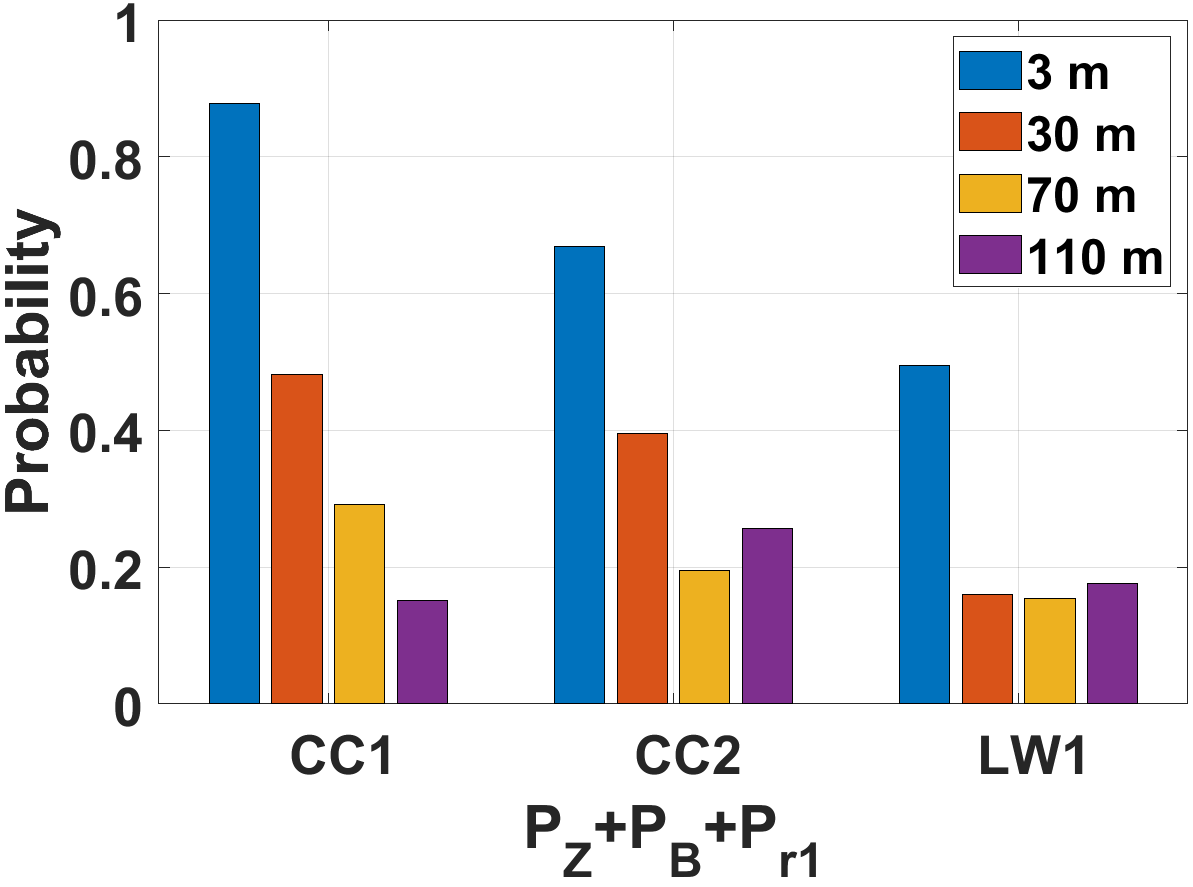}
    \label{fig:CDF_bars_4}
    }
    \caption{Probabilities of $P_Z+P_B$ and ${P_{r1}}$ with different thresholds at CC1, CC2, and LW1.}
    \label{fig:CDFs_Po_r1}\vspace{-3mm}
\end{figure*}

The CDFs of condition number for CC1, CC2, and LW1 with various different thresholds are depicted in Fig.~\ref{fig:CDFs CN}. 
%where $\textnormal{P}_{\rm{o}}$ denotes the case that has no link at a receiver site and $\textnormal{P}_{\rm{r1}}$ represents the case that only have a rank of one. 
Moreover, the distribution of $P_{\boldsymbol{Z}}+P_{\boldsymbol{B}}+P_{\boldsymbol{r1}}$ with different threshold values are shown in Fig.~\ref{fig:CDFs_Po_r1}. As seen for the $3$~m UE height in Fig.~\ref{fig:CDF_bars_1} for $\sigma_{\rm{Thr}}=\sigma_{s, \rm{mean}}$, the combined portion of the receiver locations without a link and with a rank of~1, $P_{\boldsymbol{Z}}+P_{\boldsymbol{B}}+P_{\boldsymbol{r1}}$, is higher than $80\%$  for both CC1 and CC2. In contrast, the same value becomes higher than $90\%$  of receiver sites with the same altitude for LW1. For $\sigma_{\rm{Thr}}=\sigma_{s, \rm{mean}}$, the probability of observing a higher condition number is consistently higher at higher UE altitudes, which implies a lower spatial multiplexing gain. 
%This pattern of curves is continuously observed in the range of 20 dB to 35 dB with the mean threshold setting. 
With the constant threshold settings, the CDFs of the condition number tend to converge to their threshold values. As seen in Fig.~\ref{fig:CDF_bars_2}, for $\sigma_{\rm{Thr}}=\sigma_1/10$, $P_{\boldsymbol{Z}}+P_{\boldsymbol{B}}+P_{\boldsymbol{r1}}$is close to one for most scenarios, which implies either no coverage or a channel rank of~1 across the whole simulation area. 

As $K$ increases, we observe that there are more locations where a channel rank higher than 1 can be observed, especially for the case of $K=10^4$. Condition numbers still tend to increase when the UAV altitude is higher. For $K=10^2$, $P_{\boldsymbol{Z}}+P_{\boldsymbol{B}}+P_{\boldsymbol{r1}}$ is lower than $0.9$ for CC1 and CC2, while it is closer to 0.95 for LW1, for a UE height of $3$~m. These probabilities reduce significantly when $K$ is increased to $10^4$.   
%which is depicted in Fig. \ref{fig:CDF_bars_3}. 
%It can be interpreted that those two constant threshold settings cannot provide sufficient room for condition numbers, which are shown in Figs. \ref{fig:CN_CC1_10dB}-\ref{fig:CN_LW_20dB}. In Figs. \ref{fig:CN_CC1_40dB}-\ref{fig:CN_LW_40dB} and \ref{fig:CDF_bars_4}, this trend is relaxed with the $\sigma_{\rm{Thr}}=\sigma_1/10^4$. However, it has been observed that irregular convergence can be found for the high altitude of the CC2 and LW1 case with this threshold compared to the $\sigma_{\rm{Thr}}=\sigma_{s, \rm{mean}}$ settings.
% The CDFs for $P_{\rm{o}}$ have the same trend as the threshold changed, while it has lower CDFs for $P_{\rm{r1}}$ for $\sigma_{\rm{Thr}}=\sigma_{1}/10^4$ cases. 
Determining the threshold can be an implement-dependent issue for a given environment. It can be interpreted that the mean threshold setting can provide a proper threshold decision among constant threshold settings.
% It can be interpreted that the mean threshold setting can provide a wide range of condition numbers than constant threshold cases.

% \begin{figure*}
%     \centering\vspace{-2mm}
%     \subfigure[CC1]{
%     \includegraphics[width=0.55\columnwidth]{images/1s/CC1_CDF_bars_edited_v2.png}
%     \label{fig:CDF_bars_CC1}
%     }
%     \subfigure[CC2]{
%     \includegraphics[width=0.55\columnwidth]{images/1s/CC2_CDF_bars_edited_v2.png}
%     \label{fig:CDF_bars_CC2}
%     }
%     \subfigure[LW1]{
%     \includegraphics[width=0.55\columnwidth]{images/1s/LW_CDF_bars_edited_v2.png}
%     \label{fig:CDF_bars_LW1}
%     }\vspace{-2mm}
%     \caption{Probabilities of ${P_{\rm{o}}}$ and ${P_{\rm{r1}}}$ with different thresholds at CC1, CC2, and LW1.}
%     \label{fig:CDFs_Po_r1}\vspace{-4mm}
% \end{figure*}

% In Figs. \ref{fig:CDFs CN_CC1_order}, \ref{fig:CDFs CN_CC2_order}, and \ref{fig:CDFs CN_LW_order}, the CDFs for the condition number of various singular value orders for CC1, CC2, LW1 transmitter is given, where the number index right after S denotes the order of the singular values. Through all simulation results, it can be confirmed that the cases of higher-order condition numbers are less distributed. Moreover, condition number with third and fourth order tends to be converged into high condition number in dB, while most cases with second order are distributed in the area of 10 dB to 40 dB.  

\section{Conclusions}
In this paper, we analyze channel rank, condition number, and RF signal coverage over UAV MIMO communication environments at the Centennial Campus of the NC State University and Lake Wheeler Road Field Labs. We adopted an SBR method-based ray tracing scheme to derive channel rank and condition number, which can represent essential information of MIMO systems with given geographic characteristics. We also employed the rounding criteria to decide the singular values of a channel matrix. We have found that the probability of obtaining a channel rank of 3 or 4 at higher UAV altitudes and for rural scenarios is extremely small. We have also observed that the channel rank and condition number tend to be correlated at different UAV altitudes. 
%We confirmed the geographical and statistical distribution of channel rank and condition number over given target areas. 
In future work, we plan to study the prediction of the future MIMO channel rank over the trajectory of a UAV, based on past observations of the channel matrix. Reflecting surfaces can provide channel rank improvements for UAV networks by deliberately generating additional multipath components, as was recently studied in~\cite{IRS_future_work}, which we are planning to study in further detail. 

\bibliographystyle{IEEEtran}
\bibliography{ref}

\begin{thebibliography}{10}
\providecommand{\url}[1]{#1}
\csname url@samestyle\endcsname
\providecommand{\newblock}{\relax}
\providecommand{\bibinfo}[2]{#2}
\providecommand{\BIBentrySTDinterwordspacing}{\spaceskip=0pt\relax}
\providecommand{\BIBentryALTinterwordstretchfactor}{4}
\providecommand{\BIBentryALTinterwordspacing}{\spaceskip=\fontdimen2\font plus
\BIBentryALTinterwordstretchfactor\fontdimen3\font minus
  \fontdimen4\font\relax}
\providecommand{\BIBforeignlanguage}[2]{{%
\expandafter\ifx\csname l@#1\endcsname\relax
\typeout{** WARNING: IEEEtran.bst: No hyphenation pattern has been}%
\typeout{** loaded for the language `#1'. Using the pattern for}%
\typeout{** the default language instead.}%
\else
\language=\csname l@#1\endcsname
\fi
#2}}
\providecommand{\BIBdecl}{\relax}
\BIBdecl

\bibitem{MIMO_performance_metric}
L.~C. Wood and W.~S. Hodgkiss, ``{MIMO} channel models and performance
  metrics,'' in \emph{Proc. IEEE Global Telecommun. Conf. (GLOBECOM)}, 2007,
  pp. 3740--3744.

\bibitem{MIMO_white_paper}
S.~Schindler and H.~Mellein, ``Assessing a {MIMO} channel white paper,''
  \emph{Rohdge and Schwarz App Note}, 2011.

\bibitem{MIMO_adaptation_1}
J.~Yu, F.~Lin, Y.~Teng, and G.~Yue, ``{MIMO-OFDM} transmission adaptation using
  rank,'' in \emph{Proc. Int. Symp. Personal, Indoor and Mobile Radio
  Communications}, 2007, pp. 1--5.

\bibitem{MIMO_adaptation_2}
T.~Jonna, S.~K. Reddy, and J.~K. Milleth, ``Rank and {MIMO} mode adaptation in
  {LTE},'' in \emph{Proc. IEEE Int. Conf. Advanced Networks and
  Telecommunications Systems (ANTS)}, 2013, pp. 1--6.

\bibitem{capacity_analysis_paper}
X.~Hailin, O.~Shan, N.~Zaiping, and Z.~Feng, ``Capacity analysis of high-rank
  line-of-sight {MIMO} channels,'' \emph{J. Systems Engineering and
  Electronics}, vol.~20, no.~4, pp. 706--710, 2009.

\bibitem{vertical_MIMO_paper}
W.~Xie, T.~Yang, X.~Zhu, F.~Yang, and Q.~Bi, ``Measurement-based evaluation of
  vertical separation {MIMO} antennas for base station,'' \emph{IEEE Antennas
  and Wireless Propag. Lett.}, vol.~11, pp. 415--418, 2012.

\bibitem{vertical_MIMO_paper_2}
M.~Lerch and M.~Rupp, ``Measurement-based evaluation of the {LTE MIMO} downlink
  at different antenna configurations,'' in \emph{Proc. Int. ITG Workshop on
  Smart Antennas}, 2013, pp. 1--6.

\bibitem{material_paper}
Z.~Yang, L.~Zhou, G.~Zhao, and S.~Zhou, ``Channel model in the urban
  environment for unmanned aerial vehicle communications,'' in \emph{Proc.
  European Conf. Antennas and Propag. (EuCAP 2018)}, 2018, pp. 1--5.

\bibitem{path_loss_arxiv}
\BIBentryALTinterwordspacing
B.~Galkin, J.~Kibiłda, and L.~A. DaSilva, ``A stochastic geometry model of
  backhaul and user coverage in urban {UAV} networks,'' 2017. [Online].
  Available: \url{https://arxiv.org/abs/1710.03701}
\BIBentrySTDinterwordspacing

\bibitem{RT_survey}
D.~He, B.~Ai, K.~Guan, L.~Wang, Z.~Zhong, and T.~Kürner, ``The design and
  applications of high-performance ray-tracing simulation platform for {5G} and
  beyond wireless communications: A tutorial,'' \emph{IEEE Commun. Surveys \&
  Tutorials}, vol.~21, no.~1, pp. 10--27, 2019.

\bibitem{RT_mode_paper}
A.~O. Kaya, W.~Trappe, L.~J. Greenstein, and D.~Chizhik, ``Predicting {MIMO}
  performance in urban microcells using ray tracing to characterize the
  channel,'' \emph{IEEE Trans. Wireless Commun.}, vol.~11, no.~7, pp.
  2402--2411, 2012.

\bibitem{lte_book}
S.~Sesia, I.~Toufik, and M.~Baker, \emph{{LTE-the UMTS long term evolution:
  from theory to practice}}.\hskip 1em plus 0.5em minus 0.4em\relax John Wiley
  \& Sons, 2011.

\bibitem{ray_tracing_access_paper}
Z.~Yun and M.~F. Iskander, ``Ray tracing for radio propagation modeling:
  Principles and applications,'' \emph{IEEE Access}, vol.~3, pp. 1089--1100,
  2015.

\bibitem{sbr_paper}
H.~Ling, R.-C. Chou, and S.-W. Lee, ``Shooting and bouncing rays: calculating
  the {RCS} of an arbitrarily shaped cavity,'' \emph{IEEE Trans. Antennas and
  Propagation}, vol.~37, no.~2, pp. 194--205, 1989.

\bibitem{aerpawWebsite}
\BIBentryALTinterwordspacing
``{AERPAW} - aerial experimentation and research platform for advanced
  wireless.'' [Online]. Available: \url{https://aerpaw.org}
\BIBentrySTDinterwordspacing

\bibitem{matlab_ray_tracing}
\BIBentryALTinterwordspacing
``{Matlab Ray Tracing Tool}.'' [Online]. Available:
  \url{https://www.mathworks.com/help/antenna/ref/rfprop.raytracing.html}
\BIBentrySTDinterwordspacing

\bibitem{osm}
\BIBentryALTinterwordspacing
``{OpenStreetMap}.'' [Online]. Available: \url{https://www.openstreetmap.org}
\BIBentrySTDinterwordspacing

\bibitem{osm_buildings}
\BIBentryALTinterwordspacing
``{OpenStreetMap Buildings}.'' [Online]. Available:
  \url{https://osmbuildings.org}
\BIBentrySTDinterwordspacing

\bibitem{brweb}
\BIBentryALTinterwordspacing
{ITU-R-REC-P.2040}. [Online]. Available:
  \url{https://www.itu.int/rec/R-REC-P.2040/en}
\BIBentrySTDinterwordspacing

\bibitem{tse2005fundamentals}
D.~Tse and P.~Viswanath, \emph{Fundamentals of wireless communication}.\hskip
  1em plus 0.5em minus 0.4em\relax Cambridge University Press, 2005.

\bibitem{IRS_future_work}
\BIBentryALTinterwordspacing
E.~Ozturk, C.~K. Anjinappa, F.~Erden, I.~Guvenc, H.~Dai, and A.~Bhuyan,
  ``Channel rank improvement in urban drone corridors using passive intelligent
  reflectors,'' 2021. [Online]. Available:
  \url{https://arxiv.org/abs/2108.02179}
\BIBentrySTDinterwordspacing

\end{thebibliography}

\end{document}